\documentclass[a4paper,11pt]{article}
\pdfoutput=1 

\usepackage{jcappub} 

\usepackage[T1]{fontenc} 
\usepackage{multirow}
\usepackage{times}

\title{\boldmath Quasinormal Modes and Hawking Radiation Sparsity of GUP corrected Black 
Holes in Bumblebee Gravity with Topological Defects}

 \author{Dhruba Jyoti Gogoi}
 \author{and Umananda Dev Goswami}
 \affiliation{Department of Physics, Dibrugarh University,\\
Dibrugarh 786004, Assam, India}

\emailAdd{moloydhruba@yahoo.in}
\emailAdd{umananda2@gmail.com}

\abstract{We have obtained the Generalized Uncertainty Principle (GUP) corrected de 
Sitter and anti-de Sitter black hole solutions in bumblebee gravity with a 
topological defect. We have calculated the scalar, electromagnetic and 
gravitational quasinormal modes for the both vanishing and non-vanishing 
effective cosmological constant using Pad\'e averaged sixth order WKB 
approximation method. Apart from this, the time evolutions for all three 
perturbations are studied, and quasinormal modes are calculated using the time 
domain profile. We found that the first order and second order GUP parameters 
$\alpha$ and $\beta$, respectively have opposite impacts on the quasinormal 
modes. The study also finds that the 
presence of a global monopole can decrease the quasinormal frequencies and the 
decay rate significantly. On the other hand, Lorentz symmetry violation has 
noticeable impacts on the quasinormal frequencies and the decay rate. We 
have studied the greybody factors, power spectrum and sparsity of the black 
hole with the vanishing effective cosmological constant for all the three 
perturbations. The presence of Lorentz symmetry breaking and the GUP 
parameter $\alpha$ decrease, while other GUP parameter $\beta$ and the presence 
of global monopole increase the probability of Hawking radiation to reach the 
spatial infinity. The presence of Lorentz violation can make the black holes 
less sparse, while the presence of a global monopole can increase the sparsity 
of the black holes. Moreover, we have seen that the black hole area 
quantization rule is modified by the presence of Lorentz symmetry breaking.}

\begin{document}
\maketitle
\flushbottom

\section{Introduction}
Although the recent discovery of Gravitational Waves (GWs) has provided  
strong support to the General Relativity (GR), GR is not renormalizable at UV 
scale, and hence it describes the gravitation at classical level only. On the 
other hand, the Standard Model (SM) of particle physics describes particles 
and other three fundamental interactions at the quantum level. As both GR and 
SM are the most successful field theories in describing nature, the unification 
of these two theories is one of the prime pursuits of physicists to describe 
all fundamental interactions of nature at the quantum level so that we will
understand nature to the most deeper level. In the quest of this unification 
goal, some theories of Quantum Gravity (QG) have already been introduced, whose
direct test would be possible only at the Planck scale ($\approx 10^{19}$ GeV).
As this energy scale is far more beyond the reach of current experiments, so 
direct tests of those QG models are not possible at present and nor will be in the 
near future. However, there are some effects of QG models, such as the 
breaking of Lorentz symmetry \cite{kost2004}, which may be observed at the 
current low energy scales \cite{casana2018}. The violation of Lorentz symmetry 
arises as a possibility in the context of loop quantum gravity, noncommutative 
field theories, string theory, standard-model extension (SME) etc. Especially,
SME is a most general field theoretical framework that includes the fields of
SM and GR with the terms in the lagrangian containing information about the
Lorentz symmetry violation. The effects of Lorentz symmetry breaking in the 
gravitational sector of SME were studied in Ref.s \cite{bluhm2005,bailey2006, bailey2009,tso2011,kost2009,maluf2013,maluf2014, santos2015}. In the case of 
GWs, the Lorentz violation was studied in Ref.s \cite{kost2016, kost2016_2}.
The simplest extended gravitational field theories with the spontaneous breaking
 of Lorentz and diffeomorphism are usually referred to as bumblebee models 
\cite{kost2004} in which breaking of Lorentz symmetry takes place due to the 
nonzero vacuum expectation value of a single vector field, known as the 
bumblebee field. It should be noted that the diffeomorphism violation is 
always associated with the local Lorentz symmetry violation. Recently a 
spherically symmetric exact vacuum solution to 
Einstein field equations in the presence of a spontaneous breaking of Lorentz 
symmetry due to nonzero vacuum expectation value of the bumblebee field has 
been obtained in Ref.\ \cite{casana2018}. This study also explored three 
classical tests viz., the advance of perihelion, bending of light and Shapiro's 
time delay, and showed that corrections from Lorentz violation were present 
even in the absence of a massive gravitational source and the Lorentz violation 
background deforms the spacetime. In another study, particle motion in Snyder 
noncommutative spacetime structures was studied in the presence of Lorentz 
violation \cite{kumar2021}.

In a recent study, it was seen that the Lorentz symmetry breaking reduces the
greybody factor of black holes in Generalized Uncertainty Principle (GUP) 
modified bumblebee gravity \cite{kanzi2019}. In a different study, the photon 
orbits of Kerr-Sen-like black holes in bumblebee gravity have been 
investigated, where the authors examined the effects of charge, Lorentz 
violation parameter and plasma as a dispersive medium \cite{jha2022}. The 
impact of a topological defect on black holes in bumblebee gravity was studied 
for the first time in Ref.\ \cite{gullu}. In that work, the authors studied 
the black hole horizon, temperature and the photon sphere extensively and 
found that the radius of the shadow of black holes increases with an increase in 
the global monopole parameter. Their study also shows an increasing effect of 
the deflection angle coming from the Lorentz symmetry breaking parameter and 
the global monopole. The exact traversable wormhole solution in bumblebee gravity has been 
obtained in Ref. \cite{Jusufi19}. In this work, the authors studied the energy conditions of 
the wormhole and deflection angle of light in explicit form. They found that the bumblebee 
wormhole solutions support the normal matter wormhole geometries under some certain conditions. 

In this work, we shall study the quasinormal modes of GUP corrected black 
holes in the presence of a topological defect in bumblebee gravity. The 
quasinormal modes are basically some complex numbers that are related to the 
emission of GWs from the compact and massive perturbed objects in the 
universe \cite{Vishveshwara, Press, Chandrasekhar_qnms}. The real part of the 
quasinormal modes is related to the emission frequency, while the imaginary 
part is connected to its damping. In recent times, the properties of 
GWs and quasinormal modes of black holes were studied extensively in different 
modified gravity theories \cite{Ma, gogoi1, gogoi2, Liang_2017, qnm_bumblebee, gogoi3, Graca, Zhang2, lopez2020, Liang2018, Hu, hemawati2022, gogoi4}. In 
Ref.\ \cite{qnm_bumblebee}, the quasinormal modes of black holes in bumblebee 
gravity were studied, and it was shown that the Lorentz violation has a 
significant impact on the quasinormal frequencies. In another recent study, 
the quasinormal modes of GUP corrected Schwarzschild black holes have been 
studied and it was seen that the GUP correction could change the quasinormal 
frequencies and the decay rates of GWs \cite{gupbh}. Being inspired by this 
study, we have considered the GUP correction in black holes in bumblebee 
gravity. To have a better comparison, we have considered three different 
perturbations viz., scalar, electromagnetic and gravitational perturbations to 
calculate the quasinormal modes of black holes. To obtain the quasinormal 
modes with higher accuracy, we have implemented Pad\'e averaged WKB 
approximation method upto sixth order. WKB approximation method 
provides a good approximation to the quasinormal modes, however, in 
some cases the method may be deficient to calculate quasinormal 
modes \cite{Daghigh2012}. For a comparison of our results in time
domain, we have studied the time evolution of the perturbation profiles and 
obtained the quasinormal modes from the time domain analysis. Moreover, we 
have also studied the aspect of the sparsity of black holes along with greybody 
factors and Hawking radiation power spectrum. 
The greybody factor of a black hole is related to the quantum nature of a black hole and 
it has been widely studied for black holes in different modified theories. In Ref. 
\cite{Boonserm2018}, greybody factors of a black hole in dRGT massive gravity have been studied 
using rigorous bound and the matching technique. In another work, the nonlinear electrodynamic 
effects on the black hole shadow, deflection angle, quasinormal modes and greybody factors have 
been studied in details for a magnetically charged black hole in `double-logarithmic' nonlinear 
electrodynamics \cite{Okyay2022}. The greybody factors in black 
strings have been studied in Ref. \cite{Boonserm2019} in dRGT massive 
gravity theory where the authors calculated the rigorous bounds on 
greybody factors. The light rays in a Kazakov-Solodukhin black hole 
has been studied using Gauss-Bonnet theorem and the greybody bounds 
have been calculated in details in a recent study \cite{Javed2022}. In 
another important work, quasinormal modes and greybody factors have 
been also studied in $f(R)$ gravity minimally coupled to cloud of 
strings in $(2+1)$ dimensions, where the authors found explicit 
analytical results for decay rate, reflection coefficient, greybody 
factors and temperature of the black hole \cite{Jusufi18}. The scalar 
perturbations of a single-horizon regular black hole has been studied 
in Ref. \cite{Daghigh2020} where the regular black hole solution was 
obtained using polymer quantization inspired by loop quantum gravity. 
The quantum corrections like GUP and non-commutativity are also 
implemented in black holes to study the thermodynamic behaviours 
\cite{Jusufi16}. GUP effects on Hawking temperature of a black hole in 
warp Dvali-Gabadadze-Porrati (DGP) gravity model have been studied in 
Ref. \cite{Jusufi17}, where it was found that the mass, angular 
momentum of vector and scalar particles impact the Hawking temperature 
of a black hole, and it is connected with the type of the particle 
emitted from the black hole.

Finally, we perform a study on 
the area of the black holes with the help of the adiabatic invariance method. Since, 
the area of black holes is connected with the GW echo 
\cite{cardoso2019, datta2021, coates2022}, entropy spectrum and power spectrum 
of black holes, we believe the results can be further used to study the 
impacts of Lorentz violation on different aspects of black holes.

One may note that as the Lorentz symmetry breaking carries the 
possibilities of beyond standard model physics, it has been an 
objective of many experimental searches \cite{nw1, Seifert2010}. Apart 
from bumblebee gravity, such symmetry breaking also arises in 
different theories including string theory \cite{nw3} and 
noncommutative geometry \cite{nw4}. Such theories with a spontaneously 
broken symmetry can give rise to different topological defects such as 
domain wall, cosmic string or monopole solutions etc. Hence, global 
monopoles are a possibility in bumblebee gravity. In such theories, the 
topological defects like global monopoles arise during the phase 
transitions in the early universe via Kibble mechanism 
\cite{Kibble1976} and since isolated defects are stable in nature, one 
may expect that they can still exist in the present universe. The 
presence of such relics can have some important implications in 
different aspects of astrophysics and cosmology including structure 
formation and inflation \cite{Guth1981,Durrer2002}. In this work, we 
consider one such type of relics known as global monopole and study 
its implications in black hole properties such as quasinormal modes, 
sparsity etc. If one can have some direct or indirect evidence of the 
existence of such a monopole, they will contribute significantly to 
the understanding of beyond standard model physics. In other words, 
such an evidence might represent the first field observed to break 
Lorentz symmetry, which is not described by the standard model. Apart 
from the bumblebee field and global monopole, we have considered 
another ingredient known as GUP. GUP quantum corrections are expected 
for a black hole in quantum regime or in such length scale. In such an SME 
theory, GUP has been implemented previously \cite{kanzi2019}. GUP is 
based on a momentum-dependent modification in the standard dispersion 
relation which is conjectured to violate the Lorentz invariance 
\cite{LGUP1}. In another work, it has been shown that the GUP 
deformation parameter is connected with the violation of Lorentz 
invariance \cite{LGUP2}. Hence for a study of the black holes in 
bumblebee gravity, inclusion of global monopoles and GUP corrections 
are necessary for a better understanding of the system.

The work is organized as follows. In the next section \ref{section2}, we have 
included a very short review on the field equations of bumblebee gravity and 
obtained a GUP corrected black hole solution around the global monopole in the
bumblebee gravity. The black hole solutions with non-vanishing $\Lambda_{eff}$ 
are obtained in section \ref{section4}. Here we have discussed the GUP 
corrected de Sitter and anti-de Sitter black hole solutions. We have obtained 
the quasinormal modes of black holes in section \ref{section5}. In this 
section, we have tried to show the dependency of the quasinormal modes on 
different parameters of black holes. We have studied the time evolution of 
three different perturbations in section \ref{section6}. The possibility of 
experimental detection of quasinormal modes has been studied in brief 
in the section \ref{section6.5}. The sparsity of black 
holes is studied in section \ref{section7}. Finally, we have summarized our 
results and findings in section \ref{section8}. Throughout the paper we have 
considered $ G=c=k_B=\hbar=l_p=1$, where $l_p$ is the Planck length.

\section{GUP Corrected Black Hole solutions in Bumblebee Gravity with topological defects} \label{section2}
Here we shall briefly review the simplest bumblebee gravity model and then 
derive the GUP corrected black hole solution in this gravity. In general, 
bumblebee models are vector or tensor theories that include a potential term 
which provides nonzero vacuum expectation values in the configuration of the 
fields. This behaviour of the fields affects the dynamics of other fields 
coupled to them, maintaining the conservation laws and geometric structures as 
required by pseudo-Riemannian manifold used in GR \cite{kost2004}.     

The Lagrangian density of the simplest bumblebee gravity model for the
bumblebee field coupled to gravity in a torsion-free spacetime in the presence 
of the global monopole is given by \cite{casana2018}

\begin{align}
\mathcal{L}_{B} & =\sqrt{-g}\left[\frac{1}{2\kappa}(R- 2\Lambda)+\frac{\xi}{2\kappa}\mathcal{B}^{\mu}\mathcal{B}^{\nu}R_{\mu\nu}-\frac{1}{4}\mathcal{B}_{\mu\nu}^{2}-V\!\!\left(\mathcal{B}^{\mu}\mathcal{B}_{\mu} \pm b^2 \right)\right]+\mathcal{L}_{M},\label{lagrange_bumble}
\end{align}
where $\kappa = 8\pi$, $\Lambda$ is the cosmological constant, 
$\mathcal{B}_{\mu}$ is the bumblebee field with the field strength tensor
$\mathcal{B}_{\mu\nu}=\partial_{\mu}\mathcal{B}_{\nu}-\partial_{\nu}\mathcal{B}_{\mu}$. $V\!\!\left(\mathcal{B}^{\mu}\mathcal{B}_{\mu} \pm b^2 \right)$
is the potential with $b^2$ as a real positive number, for which the 
spontaneous Lorentz violation takes place and $\xi$ is the coupling constant 
for a non-minimal gravity-bumblebee field interaction. $\mathcal{L}_{M}$ is 
basically the Lagrangian density of matter. In our case, we shall consider this 
to be for the global monopole. First, we shall calculate the field equations
of the theory for the vanishing cosmological constant i.e.\ $\Lambda = 0$. The 
corresponding field equations associated with the theory can be obtained from 
the Lagrangian (\ref{lagrange_bumble}) by varying its action with respect to 
the metric $g_{\mu\nu}$, which are given by
\begin{equation}
G_{\mu\nu} = \kappa\left(T_{\mu\nu}^{B}+T_{\mu\nu}^{M}\right), \label{eq_of_motion}
\end{equation}
where $G_{\mu\nu}$ is the standard Einstein's tensor. It is seen the 
energy-momentum tensor part of these field equations is a sum of two tensors: 
$T_{\mu\nu}^{B}$ and $T_{\mu\nu}^{M}$. The tensor $T_{\mu\nu}^{B}$ is that 
part of energy momentum tensor which depends on the bumblebee field and is 
given by
\begin{align}
T_{\mu\nu}^{B}\equiv & -\mathcal{B}_{\mu\sigma}\mathcal{B}_{\phantom{\sigma}\nu}^{\sigma}-\frac{1}{4}g_{\mu\nu}\mathcal{B}_{\alpha\beta}^{2}-g_{\mu\nu}V\!\left(\mathcal{B}^{\mu}\mathcal{B}_{\mu}\right)+4V^{\prime}\mathcal{B}_{\mu}\mathcal{B}_{\nu}\nonumber \\
 & +\frac{\xi}{\kappa}\left(\frac{1}{2}g_{\mu\nu}\mathcal{B}^{\alpha}\mathcal{B}^{\beta}R_{\alpha\beta}-\mathcal{B}_{\nu}\mathcal{B}^{\alpha}R_{\alpha\mu}-\mathcal{B}_{\mu}\mathcal{B}^{\alpha}R_{\alpha\nu}\right)\nonumber \\
 & +\frac{\xi}{\kappa}\left(\frac{1}{2}\nabla_{\alpha}\nabla_{\mu}\!\left(\mathcal{B}^{\alpha}\mathcal{B}_{\nu}\right)+\frac{1}{2}\nabla_{\alpha}\nabla_{\nu}\!\left(\mathcal{B}_{\mu}\mathcal{B}^{\alpha}\right)\right)\nonumber \\
 & +\frac{\xi}{\kappa}\left(-\frac{1}{2}\nabla^{\lambda}\nabla_{\lambda}\!\left(\mathcal{B}_{\mu}\mathcal{B}_{\nu}\right)-\frac{1}{2}g_{\mu\nu}\nabla_{\alpha}\nabla_{\beta}\!\left(\mathcal{B}^{\alpha}\mathcal{B}^{\beta}\right)\right),\label{T_mn^B}
\end{align}
where the prime denotes the derivative with respect to the argument, and the 
usual matter part $T_{\mu\nu}^M$ for our case is considered to be of the form:
\begin{equation}
T_{\mu}^{\left(M\right)\nu} = \text{diag}\!\left(\frac{\eta^{2}}{r^{2}},\frac{\eta^{2}}{r^{2}},0,0\right). \label{T_mn^M}
\end{equation}
Here, the parameter $\eta$ is a constant quantity which is connected to the 
global monopole charge. The second set of field equations can be obtained by 
considering the variation of the Lagrangian \eqref{lagrange_bumble} action 
with respect to the bumblebee field and it is given by
\begin{equation}
\nabla^{\mu}\mathcal{B}_{\mu\nu}=\mathcal{J}_{\nu}\label{EOM_2},
\end{equation}
where $\mathcal{J}_{\nu}=\mathcal{J}_{\nu}^{\mathcal{B}}+\mathcal{J}_{\nu}^{M}$, in which  $\mathcal{J}_{\nu}^{M}$ acts as a source term for the bumblebee field, while $\mathcal{J}_{\nu}^{\mathcal{B}}=2V^{\prime}\mathcal{B}_{\nu}-\frac{\xi}{\kappa}\mathcal{B}^{\mu}R_{\mu\nu}$ is the self interaction bumblebee field 
current.Taking trace of equation (\ref{eq_of_motion}) we get,
\begin{align}
-\frac{1}{\kappa}R= & -4V\!\left(\mathcal{B}^{\mu}\mathcal{B}_{\mu}\right)+4V^{\prime}\mathcal{B}_{\mu}\mathcal{B}^{\mu}\nonumber \\
 & +\frac{\xi}{\kappa}\left(-\frac{1}{2}\nabla^{\lambda}\nabla_{\lambda}\!\left(\mathcal{B}_{\mu}\mathcal{B}^{\mu}\right)-\nabla_{\alpha}\nabla_{\beta}\!\left(\mathcal{B}^{\alpha}\mathcal{B}^{\beta}\right)\right)+T^{M}.\label{R}
\end{align}
Now, using equation \eqref{R} in equation \eqref{eq_of_motion} we can have,
\begin{align}
\frac{1}{\kappa}R_{\mu\nu} & = T_{\mu\nu}^{M}-\frac{1}{2}g_{\mu\nu}T^{M}-\mathcal{B}_{\mu\sigma}\mathcal{B}_{\phantom{\sigma}\nu}^{\sigma}-\frac{1}{4}g_{\mu\nu}\mathcal{B}_{\alpha\beta}^{2}-g_{\mu\nu}V\!\left(\mathcal{B}^{\mu}\mathcal{B}_{\mu}\right)+4V^{\prime}\mathcal{B}_{\mu}\mathcal{B}_{\nu}\nonumber \\
 & +\frac{\xi}{\kappa}\left(\frac{1}{2}g_{\mu\nu}\mathcal{B}^{\alpha}\mathcal{B}^{\beta}R_{\alpha\beta}-\mathcal{B}_{\nu}\mathcal{B}^{\alpha}R_{\alpha\mu}-\mathcal{B}_{\mu}\mathcal{B}^{\alpha}R_{\alpha\nu}\right)\nonumber \\
 & +\frac{\xi}{\kappa}\left(\frac{1}{2}\nabla_{\alpha}\nabla_{\mu}\!\left(\mathcal{B}^{\alpha}\mathcal{B}_{\nu}\right)+\frac{1}{2}\nabla_{\alpha}\nabla_{\nu}\!\left(\mathcal{B}_{\mu}\mathcal{B}^{\alpha}\right)\right)\nonumber \\
 & +\frac{\xi}{\kappa}\left(-\frac{1}{2}\nabla^{\lambda}\nabla_{\lambda}\!\left(\mathcal{B}_{\mu}\mathcal{B}_{\nu}\right)-\frac{1}{2}g_{\mu\nu}\nabla_{\alpha}\nabla_{\beta}\!\left(\mathcal{B}^{\alpha}\mathcal{B}^{\beta}\right)\right)\nonumber \\
 & +\frac{1}{2}g_{\mu\nu}\left(4V\!\left(\mathcal{B}^{\alpha}\mathcal{B}_{\alpha}\right)-4V^{\prime}\mathcal{B}^{\alpha}\mathcal{B}_{\alpha}\right)\nonumber \\
 & -\frac{1}{2}g_{\mu\nu}\frac{\xi}{\kappa}\left(-\frac{1}{2}\nabla^{\lambda}\nabla_{\lambda}\!\left(\mathcal{B}^{\alpha}\mathcal{B}_{\alpha}\right)-\nabla_{\alpha}\nabla_{\beta}\!\left(\mathcal{B}^{\alpha}\mathcal{B}^{\beta}\right)\right).
\label{R_mn}
\end{align}

As mentioned above, the potential in the Lagrangian \eqref{lagrange_bumble} 
generates the nonzero vacuum expectation value for the field $\mathcal{B}_\mu$
since for $V\!\!\left(\mathcal{B}^{\mu}\mathcal{B}_{\mu} \pm b^2 \right) = 0$ 
it is required that
\begin{equation}
\mathcal{B}_{\mu}\mathcal{B}^{\mu}\pm b^{2}= 0.\label{Condition}
\end{equation}
One can see that the solution of equation \eqref{Condition} gives a nonzero 
vacuum expectation value $\left\langle \mathcal{B}_{\mu}\right\rangle =b_{\mu},$
where $b_{\mu}$ is a vector, which is a function of spacetime coordinates and
hence $b^{\mu}b_{\mu} = \pm\, b^2$. This nonzero background vector $b_{\mu}$   
spontaneously violates the Lorentz symmetry. For the completeness it needs to be 
mentioned that $\pm$ signs in front of $b^2$ decide whether the background
field $b_{\mu}$ is timelike or spacelike respectively \cite{casana2018, kumar2021, kanzi2019, gullu, qnm_bumblebee}. For a static and spherically symmetric 
black hole solution with the Lorentz symmetry violation in the bumblebee field
we consider the Birkhoff metric as an ansatz: 
\begin{equation}
g_{\mu\nu}=  \text{diag}\!\left(-e^{2\gamma},e^{2\rho},r^{2},r^{2}\sin^{2}\theta\right),\label{Bhirkoff}
\end{equation}
where $\gamma$ and $\rho$ are functions of $r$ and we fix the bumblebee field 
in its vacuum expectation value \cite{Bertolami2005}, i.e.
\begin{equation}
\mathcal{B}_{\mu}= b_{\mu}\label{B_equal_b}
\end{equation}
These lead the radial background field as given by
\begin{equation}
b_{r}(r)= \left|b\right|e^{\rho},\label{b_r}
\end{equation}
where $b_r(r)$ is the radial component of $b_{\mu}$. Using this component it is
possible to write the equation \eqref{R_mn} as the extended Einstein's equations
in vacuum in the form as
\begin{align}
0 & =  \bar{R}_{\mu\nu}\nonumber \\
 & =  R_{\mu\nu}-\kappa\left(T_{\mu\nu}^{M}-\frac{1}{2}g_{\mu\nu}T^{M}\right)\\
 & -\frac{\xi}{2}g_{\mu\nu}b^{\alpha}b^{\beta}R_{\alpha\beta}+\xi b_{\nu}b^{\alpha}R_{\alpha\mu}+\xi b_{\mu}b^{\alpha}R_{\alpha\nu}\nonumber \\
 & -\frac{\xi}{2}\nabla_{\alpha}\nabla_{\mu}\!\left(b^{\alpha}b_{\nu}\right)-\frac{\xi}{2}\nabla_{\alpha}\nabla_{\nu}\!\left(b_{\mu}b^{\alpha}\right)\nonumber \\
& +\frac{\xi}{2}\nabla^{\lambda}\nabla_{\lambda}\!\left(b_{\mu}b_{\nu}\right),
\label{Einstein_eqns}
\end{align}
where the trace of the energy-momentum tensor is given by
\begin{equation}
T^{M}=  2\,\frac{\eta^{2}}{r^{2}}.\label{Trace_of_T_mn}
\end{equation}
The components of equation (\ref{Einstein_eqns}) can be written explicitly as
\begin{align}
\bar{R}_{tt} & = \left(1+\frac{\lambda}{2}\right)\!R_{tt}+\frac{\lambda}{r}\left(\partial_{r}\rho+\partial_{r}\gamma\right)e^{2\left(\gamma-\rho\right)} = 0,\label{barR_tt}\\
\bar{R}_{rr} & = \left(1+\frac{3\lambda}{2}\right)\!R_{rr} = 0,\label{barR_rr}\\
\bar{R}_{\theta\theta} & = \left(1+\lambda\right)R_{\theta\theta}-\lambda\left(\frac{1}{2}r^{2}e^{-2\rho}R_{rr}+1\right)+\eta^{2} = 0,\label{barR_theta}\\
\bar{R}_{\phi\phi} & = \sin^{2}\!\theta\bar{R}_{\theta\theta} = 0,\label{barR_phi}
\end{align}
where $\lambda=\xi b^{2}$. Solving these equations \eqref{barR_tt}, 
\eqref{barR_rr}, \eqref{barR_theta} and \eqref{barR_phi} we obtain,
\begin{align}
e^{2\rho} & = \left(1+\lambda\right)\left(1+\eta^{2}-\frac{\rho_{0}}{r}\right)^{-1}\!\!\!,\label{rho_func}\\
e^{2\gamma} & = 1+\eta^{2}-\frac{\rho_{0}}{r}.\label{gamma_func}
\end{align}
Thus, with these results the Lorentz symmetry breaking spherically 
symmetric solution for the field equations can be given as
\begin{equation}
ds^{2} = -\left(1-\mu-\frac{2M}{r}\right)dt^{2}+\left(1+\lambda\right)\left(1-\mu-\frac{2M}{r}\right)^{-1}\!\!\!\!dr^{2}
 +r^{2}d\theta^{2}+r^{2}\sin^{2}\!\theta\,d\phi^{2}.\label{solution}
\end{equation}
Here, we have chosen $\rho_{0}=2M,$ where $M$ is the mass of the black hole 
and $\mu= -\,\eta^2,$ the global monopole term. The metric (\ref{solution}) will
give the Lorentz symmetry breaking standard spherically symmetric solution 
when $\eta=0$ \cite{casana2018}, and the usual Schwarzschild metric for both
$\eta=0$ and  $\lambda\rightarrow0$.

The prediction of the existence of a minimal length from different approaches 
of QG such as the black hole physics and the string theory has many important
physical implications at very high energy scales. For example, at Planck scale
the Schwarzschild radius of the corresponding black hole becomes comparable to 
the Compton wavelength. Thus the existence of a minimal length demands for the 
GUP, which is defined as \cite{Anacleto_01, Gangopadhyay, Ali_gup, Tawfik, Tawfik2}
\begin{equation}\label{gup_eq}
\Delta x\Delta p\geq \frac{1}{2}\left( 1-\alpha \Delta p +\beta(\Delta p)^2 \right),
\end{equation}
where $\alpha$ and $\beta$ are dimensionless positive parameters. It is clear 
from the above relation that for $\alpha=\beta=0$ one can easily recover the
Heisenberg's uncertainty principle, i.e.\ $\Delta x\Delta p\geq \frac{1}{2}$.
Following Ref.\ \cite{Anacleto_01}, we can obtain a bound for the massless 
particles as 
\begin{equation}
E\Delta x\geq \frac{1}{2},
\end{equation}
which changes the equation \eqref{gup_eq} to
\begin{equation}
\label{rdgup}
{\cal E}\geq E\left[1-\frac{\alpha}{2(\Delta x)}+ \frac{\beta}{2(\Delta x)^2}+\cdots    \right],
\end{equation}
where $\cal E$ denotes the GUP corrected energy. 
Assuming  $\Delta p\sim E\sim M$ and $ \Delta x\sim r_h=\frac{2M}{(1-\mu)}$, we
can write the above equation \eqref{rdgup} in terms of mass as the following 
mass relation: 
\begin{equation}
{\cal M}=M_{gup}\geq M\left( 1- \frac{\alpha (1-\mu)}{4M}+\frac{\beta (1-\mu)^2}{8M^2} \right).
\end{equation}
This mass relation leads to the GUP corrected event horizon as given by
\begin{equation}
r_{hgup} = \frac{2M_{gup}}{(1-\mu)}\geq r_h\left( 1- \frac{\alpha}{2r_h}+\frac{\beta}{2r^2_h} \right).
\end{equation}
Finally, the metric for a GUP corrected black hole in the bumblebee gravity 
with topological defects can be obtained from the metric \eqref{solution} by
replacing the black hole mass $M$ with the corresponding GUP corrected mass
$M_{gup}$ as given by
\begin{equation}\label{flat_GUP_BH_sol}
ds^{2}= -\left(1-\mu-\frac{2M_{gup}}{r}\right)\!dt^{2}+\left(1+\lambda\right)\left(1-\mu-\frac{2M_{gup}}{r}\right)^{-1}\!\!\!\!dr^{2} +r^{2}d\theta^{2}+r^{2}\sin^{2}\!\theta\,d\phi^{2}
\end{equation}
and the event horizon of this black hole would be
\begin{equation}
r_{hgup} = \frac{2M_{gup}}{(1-\mu)} = r_h\! \left( 1- \frac{\alpha (1-\mu)}{4M}+\frac{\beta (1-\mu)^2}{8M^2} \right).
\end{equation}
\section{Black holes with non-vanishing $\Lambda_{eff}$}\label{section4}
In presence of non-vanishing cosmological constant, the metric functions which 
satisfy the modified field equations in absence of magnetic monopole are given 
by \cite{maluf2021}
\begin{equation}
g_{tt} =1 -\frac{2 M}{r} -\frac{r^2}{3} (1+\lambda) \Lambda _{eff}
\end{equation}
and
\begin{equation}
g_{rr} =(1+\lambda)\left( 1 -\frac{2 M}{r}  -\frac{r^2}{3} (1+\lambda) \Lambda _{eff}\right)^{-1}\!\!\!\!.
\end{equation}
Here $\Lambda _{eff} = \dfrac{\kappa \lambda}{\xi},$ the effective cosmological
 constant. Now, in presence of global monopole, these metric functions are 
found to be of the form:
\begin{equation}\label{gtt_cosmological_const}
g_{tt} =1 -\frac{2 M}{r} -\mu -\frac{r^2}{3} (1+\lambda) \Lambda _{eff}
\end{equation}
and
\begin{equation}\label{grr_cosmological_const}
g_{rr} =(1+\lambda)\left( 1 -\frac{2 M}{r} -\mu -\frac{r^2}{3} (1+\lambda) \Lambda _{eff}\right)^{-1}\!\!\!\!.
\end{equation}
Depending on the value of $\Lambda_{eff}$, the above solutions can be 
classified into anti-de Sitter and de Sitter solutions which are discussed 
below.

\subsection{GUP corrected Anti de Sitter Black Hole}
In case of anti-de Sitter (AdS) space ($\Lambda_{eff}<0$), there is a unique 
horizon of the black hole defined by the metric functions 
\eqref{gtt_cosmological_const} and \eqref{grr_cosmological_const}, which is 
given by
\begin{equation}
r_{h(AdS)} = - \dfrac{(1-\mu)}{A^{1/3}(M)} - \dfrac{A^{1/3}(M)}{(1+\lambda) \Lambda_{eff}},
\end{equation}
where $A(M) = \sqrt{(\lambda +1)^3 \Lambda_{eff} ^3 \left((\mu -1)^3+9 (\lambda +1) \Lambda_{eff}  M^2\right)}+3 (\lambda +1)^2 \Lambda_{eff} ^2 M.$ Now the 
GUP corrected metric functions can be given by
\begin{equation}
g_{tt(gup)} =1 -\frac{2 M_{AdS(gup)}}{r} -\mu -\frac{r^2}{3} (1+\lambda) \Lambda_{eff} 
\end{equation}
and
\begin{equation}
g_{rr(gup)} =(1+\lambda)\left( 1 -\frac{2 M_{AdS(gup)}}{r} -\mu -\frac{r^2}{3} (1+\lambda) \Lambda_{eff} \right)^{-1}\!\!\!\!\!.
\end{equation}
Here, the GUP corrected mass term is obtained by
\begin{align}
{\cal M}_{AdS} & = M_{AdS(gup)} \\ \notag &\geq M\left( 1- \frac{\alpha}{2 \left(- \dfrac{(1-\mu)}{A^{1/3}(M)} - \dfrac{A^{1/3}(M)}{(1+\lambda) \Lambda_{eff}}\right)}+\frac{\beta}{2 \left(- \dfrac{(1-\mu)}{A^{1/3}( M)} - \dfrac{A^{1/3}(M)}{(1+\lambda) \Lambda_{eff}}\right)^2} \right).
\end{align}
Hence the GUP corrected event horizon is given by
\begin{equation}
\bar{r}_{h(AdS)} =  - \dfrac{(1-\mu)}{A^{1/3}(M_{AdS(gup)})} - \dfrac{A^{1/3}(M_{AdS(gup)})}{(1+\lambda) \Lambda_{eff}}\geq r_{h(AdS)}\left( 1- \frac{\alpha}{2r_{h(AdS)}}+\frac{\beta}{2r^2_{h(AdS)}} \right).
\end{equation}
The above equations give the GUP corrected anti-de Sitter black hole in 
bumblebee gravity with the topological defects.
\subsection{GUP corrected de Sitter Black Hole}
In case of de Sitter black holes $\Lambda_{eff}>0.$ In this case there are two 
horizons of the black hole defined by the metric functions 
\eqref{gtt_cosmological_const} and \eqref{grr_cosmological_const}. One is the 
inner horizon or the event horizon given by
\begin{equation}
r_{h(dS)} = \frac{2 \sqrt{(1-\mu ) } \sin\zeta(M) }{\sqrt{(\lambda +1) \Lambda_{eff} }},
\end{equation}
and other is the cosmological horizon as given by
\begin{equation}
r_{c} = \frac{2 \sqrt{(1-\mu )} \sin \left(\frac{\pi }{3}-\zeta(M) \right)}{\sqrt{(\lambda +1) \Lambda_{eff}}},
\end{equation}
where $\zeta(M) = \frac{1}{3} \sin ^{-1}\!\left(3 M \sqrt{\frac{(\lambda +1) \Lambda_{eff} }{(1-\mu )^3}}\right).$ However, one should note that the black hole 
can have two horizons when $g^{rr}=0$ has two real roots that impose a 
constraint on the cosmological constant $\Lambda_{eff}$ given by
\begin{equation}\label{Lambda_bound}
0<\Lambda_{eff} <\frac{(1-\mu )^3}{9 (\lambda +1)},
\end{equation}
where $\mu<1$ and $\lambda>-1$ must be satisfied. When $\Lambda_{eff}$ approaches $\frac{(1-\mu )^3}{9 (\lambda +1)}$, the two horizons approach each other and finally 
at $\Lambda_{eff} =\frac{(1-\mu )^3}{9 (\lambda +1)}$, we have $r_{h(dS)} = r_c$ i.e.\ a single horizon of the black hole.

In this case, the GUP corrected metric functions can be given by
\begin{equation}\label{ds_gtt}
g_{tt(gup)} =1 -\frac{2 M_{dS(gup)}}{r} -\mu -\frac{r^2}{3} (1+\lambda) \Lambda_{eff} 
\end{equation}
and
\begin{equation}\label{ds_grr}
g_{rr(gup)} =(1+\lambda)\left( 1 -\frac{2 M_{dS(gup)}}{r} -\mu -\frac{r^2}{3} (1+\lambda) \Lambda_{eff} \right)^{-1}\!\!\!\!.
\end{equation}
Here, for the event horizon the GUP corrected mass term is given by
\begin{align}\label{ds_mgup}
{\cal M}_{dS} &= M_{dS(gup)} \\ \notag &\geq M\left( 1- \frac{\alpha \sqrt{(\lambda +1) \Lambda_{eff} }}{4 \sqrt{(1-\mu ) } \sin\zeta(M)}+\frac{\beta (\lambda +1) \Lambda_{eff}}{8  (1-\mu) \sin^2\zeta(M)} \right).
\end{align}
Thus, the GUP corrected event horizon is given by
\begin{equation}\label{ds_rgup}
\bar{r}_{h(dS)} =  \frac{2 \sqrt{(1-\mu ) } \sin\zeta(M_{dS(gup)}) }{\sqrt{(\lambda +1) \Lambda_{eff} }}\geq r_{h(dS)}\left( 1- \frac{\alpha}{2r_{h(dS)}}+\frac{\beta}{2r^2_{h(dS)}} \right).
\end{equation}
The equations \eqref{ds_gtt} and \eqref{ds_grr} define the GUP corrected 
de Sitter black hole in bumblebee gravity with a global monopole.
\section{Quasinormal modes}\label{section5}
We have obtained the black hole solutions in bumblebee gravity with topological 
defects in the previous section and implemented GUP corrections to the solutions. In 
this this section, we shall study the quasinormal modes obtained from these black 
holes for three different types of perturbations viz., massless scalar perturbation, 
electromagnetic perturbation and gravitational perturbation. In case of the 
quasinormal modes of a black hole obtained from the perturbation of the test field 
i.e.\ of scalar field or electromagnetic field, it is considered that the test field has 
negligible reaction on the spacetime. The Schr\"odinger like wave equations are 
derived from the conservation relations of the test fields on the black hole 
spacetime. For a scalar field it will be a Klein Gordon type equation and for 
electromagnetic fields, it will be the Maxwell equations. The Schr\"odinger like wave 
equations for gravitational perturbations can be obtained by introducing perturbation 
to the spacetime metric and the field equations. In this work, we have considered the 
axial gravitational perturbation and calculated the quasinormal modes associated with it.

Now, considering only the axial perturbations, the perturbed metric can be given by \cite{lopez2020}
\begin{equation}\label{pert_metric}
ds^2 = -\, |g_{tt}|\, dt^2 + r^2 \sin^2\!\theta\, (d\phi - a_1\, dt - a_2\, dr - a_3\, d\theta)^2 + g_{rr}\, dr^2 + r^2 d\theta^2,
\end{equation}
here the parameters $a_1, a_2$ and $a_3$ are the functions of time $t$, radial 
coordinate $r$ and polar angle $\theta$. These parameters define the perturbation 
introduced to the metric. $g_{tt}$ and $g_{rr}$ being independent of $t$ and 
$\theta$, are the zeroth order terms or the background terms.
\subsection{Scalar Quasinormal Modes}
First we consider a massless scalar field around the GUP corrected bumblebee 
quantum black holes defined previously. Since, we assumed that the reaction of 
the scalar field on the spacetime is negligible, it is possible to reduce the 
perturbed metric equation \eqref{pert_metric} to the following form:
\begin{equation}
ds^2 = -\,|g_{tt}|\, dt^2 + g_{rr}\, dr^2 +r^2 d \Omega^2
\end{equation}
 In this case, it is possible to describe the quasinormal modes of the black holes by the Klein Gordon equation in curved spacetime given by
\begin{equation}\label{scalar_KG}
\square \Phi = \dfrac{1}{\sqrt{-g}} \partial_\mu (\sqrt{-g} g^{\mu\nu} \partial_\nu \Phi) = 0.
\end{equation}
Using the spherical harmonics, we may decompose the scalar field in the 
following form:
\begin{equation}
\Phi(t,r,\theta, \phi) = \dfrac{1}{r} \sum_{l,m} \psi_l(t,r) Y_{lm}(\theta, \phi),
\end{equation}
where $\psi_l(t,r)$ is the radial time dependent wave function, and $l$ and $m$ are 
the indices of the spherical harmonics. Using this equation in equation 
\eqref{scalar_KG}, we get,
\begin{equation}
\partial^2_{r_*} \psi(r_*)_l + \omega^2 \psi(r_*)_l = V_s(r) \psi(r_*)_l,  
\end{equation}
where $r_*$ is the tortoise coordinate defined by
\begin{equation}\label{tortoise}
\dfrac{dr_*}{dr} = \sqrt{g_{rr}\, |g_{tt}^{-1}|}
\end{equation}
and $V_s(r)$ is the effective potential of the field, which can be obtained as
\begin{equation}\label{Vs}
V_s(r) = |g_{tt}| \left( \dfrac{l(l+1)}{r^2} +\dfrac{1}{r \sqrt{|g_{tt}| g_{rr}}} \dfrac{d}{dr}\sqrt{|g_{tt}| g_{rr}^{-1}} \right).
\end{equation}
Here $l$ is known as multipole moment of the quasinormal modes of the black hole.

\subsection{Electromagnetic Quasinormal Modes}
In case of electromagnetic perturbation, we need to use the tetrad 
formalism \cite{chandrasekhar, lopez2020} in which a basis say $e^\mu_{a}$ is defined 
associated with the metric $g_{\mu\nu}$. The basis should satisfy,
\begin{align}
e^{(a)}_\mu e^\mu_{(b)} &= \delta^{(a)}_{(b)} \notag \\
e^{(a)}_\mu e^\nu_{(a)} &= \delta^{\nu}_{\mu} \notag \\
e^{(a)}_\mu &= g_{\mu\nu} \eta^{(a)(b)} e^\nu_{(b)}\notag \\
g_{\mu\nu} &= \eta_{(a)(b)}e^{(a)}_\mu e^{(b)}_\nu = e_{(a)\mu} e^{(a)}_\nu.
\end{align}
In terms of these basis the tensor fields can be expressed as
\begin{align*}
S_\mu &= e^{(a)}_\mu S_{(a)}, \\ 
S_{(a)} &= e^\mu_{(a)} S_\mu, \\
P_{\mu\nu} &=  e^{(a)}_\mu e^{(b)}_\nu P_{(a)(b)}, \\
P_{(a)(b)} &= e^\mu_{(a)} e^\nu_{(b)} P_{\mu\nu}.
\end{align*}
One should note that in the tetrad formalism the covariant derivative in the 
actual coordinate system is replace with the intrinsic derivative in the 
tetrad frame as shown below \cite{chandrasekhar, lopez2020}:
\begin{align}
K_{(a)(b)|(c)} &\equiv e^{\lambda}_{(c)} K_{\mu\nu;\lambda} e^\mu_{(a)} e^\nu_{(b)} \notag \\
&= K_{(a)(b),(c)} - \eta^{(m)(n)} (\gamma _{(n)(a)(c)} K_{(m)(b)} + \gamma _{(n)(b)(c)} K_{(a)(m)}),
\end{align}
where the Ricci rotation coefficients are given by $\gamma_{(c)(a)(b)} \equiv e^\mu_{(b)} e_{(a)\nu;\mu} e^\nu_{(c)}.$ The vertical rule and the comma denote 
the intrinsic and directional derivative respectively in the tetrad basis. 
Now for the electromagnetic perturbation in the tetrad formalism, the Bianchi 
identity of the field strength $F_{[(a)(b)(c)]} = 0$ gives
\begin{align}
\left( r \sqrt{|g_{tt}|}\, F_{(t)(\phi)}\right)_{,r} + r \sqrt{g_{rr}}\, F_{(\phi)(r), t} &=0, \label{em1} \\
\left( r \sqrt{|g_{tt}|}\, F_{(t)(\phi)}\sin\theta\right)_{,\theta} + r^2 \sin\theta\, F_{(\phi)(r), t} &=0. \label{em2}
\end{align}
The conservation equation is
\begin{equation}
\eta^{(b)(c)}\! \left( F_{(a)(b)} \right)_{|(c)} =0.
\end{equation}
The above equation can be further written as
\begin{equation} \label{em3}
\left( r \sqrt{|g_{tt}|}\, F_{(\phi)(r)}\right)_{,r} +  \sqrt{|g_{tt}| g_{rr}}\, F_{(\phi)(\theta),\theta} + r \sqrt{g_{rr}}\, F_{(t)(\phi), t} = 0.
\end{equation}
Differentiating equation \eqref{em3} w.r.t.\ $t$ and using equations \eqref{em1} and \eqref{em2}, we get,
\begin{equation}\label{em4}
\left[ \sqrt{|g_{tt}| g_{rr}^{-1}} \left( r \sqrt{|g_{tt}|}\, \mathcal{F} \right)_{,r} \right]_{,r} + \dfrac{|g_{tt}| \sqrt{g_{rr}}}{r} \left( \dfrac{\mathcal{F}_{,\theta}}{\sin\theta} \right)_{,\theta}\!\! \sin\theta - r \sqrt{g_{rr}}\, \mathcal{F}_{,tt} = 0,
\end{equation}
where we have considered $\mathcal{F} = F_{(t)(\phi)} \sin\theta.$ Using the 
Fourier decomposition $(\partial_t \rightarrow -\, i \omega)$ and field 
decomposition $\mathcal{F}(r,\theta) = \mathcal{F}(r) Y_{,\theta}/\sin\theta,$ where $Y(\theta)$ is the Gegenbauer function and it satisfies the following 
relation,
\begin{equation}
\sin\theta\, \dfrac{d}{d\theta} \left( \dfrac{1}{\sin\theta} \dfrac{d}{d\theta} \dfrac{Y_{,\theta}}{\sin\theta} \right) = -l(l+1) \dfrac{Y_{,\theta}}{\sin\theta},
\end{equation}
we can write equation \eqref{em4} in the following form:
\begin{equation}\label{em5}
\left[ \sqrt{|g_{tt}| g_{rr}^{-1}} \left( r \sqrt{|g_{tt}|}\, \mathcal{F} \right)_{,r} \right]_{,r} + \omega^2 r \sqrt{g_{rr}}\, \mathcal{F} - |g_{tt}| \sqrt{g_{rr}} r^{-1} l(l+1)\, \mathcal{F} = 0.
\end{equation}
Now, finally using the tortoise coordinate from equation \eqref{tortoise} and redefining $\psi_e \equiv r \sqrt{|g_{tt}|}\, \mathcal{F}$, equation \eqref{em5} can be written in the Schr\"odinger like form given by
\begin{equation}
\partial^2_{r_*} \psi_e + \omega^2 \psi_e = V_e(r) \psi_e,
\end{equation}
where the potential is given by
\begin{equation}\label{Ve}
V_e(r) = |g_{tt}|\, \dfrac{l(l+1)}{r^2}. 
\end{equation}
\subsection{Gravitational Quasinormal Modes}
Here, we shall consider the axial gravitational perturbation and find out a Schr\"odinger 
like equation with an effective potential $V_g$ which will be helpful to calculate 
the quasinormal modes in the next part of the study. It is shown in 
Ref.\ \cite{chen19} that in case of axial perturbation, the axial components of 
perturbed energy-momentum tensor for an anisotropic fluid are zero, which gives us a privilege to write in tetrad formalism,
\begin{equation}
R_{(a)(b)} = 0.
\end{equation}
The $\theta \, \phi$ and $r\, \phi$ components of this equation give \cite{lopez2020},
\begin{align}
\left[ r^2 \sqrt{|g_{tt}| g_{rr}^{-1}}\, (a_{2, \theta} - a_{3,r}) \right]_{,r} = r^2 \sqrt{|g_{tt}|^{-1} g_{rr}}\, (a_{1,\theta} - a_{3,t})_{,t}, \label{g1}\\
\left[ r^2 \sqrt{|g_{tt}| g_{rr}^{-1}}\, (a_{3,r} - a_{2,\theta}) \sin^3\theta  \right]_{,\theta} = \dfrac{r^4 \sin^3\theta}{\sqrt{|g_{tt}| g_{rr}}}\, (a_{1,r} - a_{2,t})_{,t}. \label{g2}
\end{align}
Now, using $\mathcal{F}_g (r, \theta) = \mathcal{F}_g (r) Y(\theta),$ where $Y(\theta)$ satisfies $\frac{d}{d\theta} \left(\sin^{-3}\theta \frac{dY}{d\theta} \right) = -\,\lbrace l(l+1) - 2 \rbrace Y \sin^{-3}\theta$ and we simplify equations \eqref{g1} and \eqref{g2} to obtain,
\begin{equation}
\partial^2_{r_*} \psi_g + \omega^2 \psi_g = V_g(r) \psi_g,
\end{equation}
where we used $\psi_g r = \mathcal{F}_g$ and $r_*$ is the tortoise coordinate defined in equation \eqref{tortoise}. The effective potential in this expression is given by
\begin{equation}\label{Vg}
V_g(r) = |g_{tt}| \left[ \dfrac{2}{r^2} \left( \dfrac{1}{g_{rr}} - 1 \right) + \dfrac{l(l+1)}{r^2} - \dfrac{1}{r \sqrt{|g_{tt}| g_{rr}}} \left( \dfrac{d}{dr} \sqrt{|g_{tt}| g_{rr}^{-1}} \right) \right].
\end{equation}
We shall use this expression of potential to calculate the quasinormal modes of 
gravitational perturbation.

\begin{figure}[!h]
\centerline{
   \includegraphics[scale = 0.3]{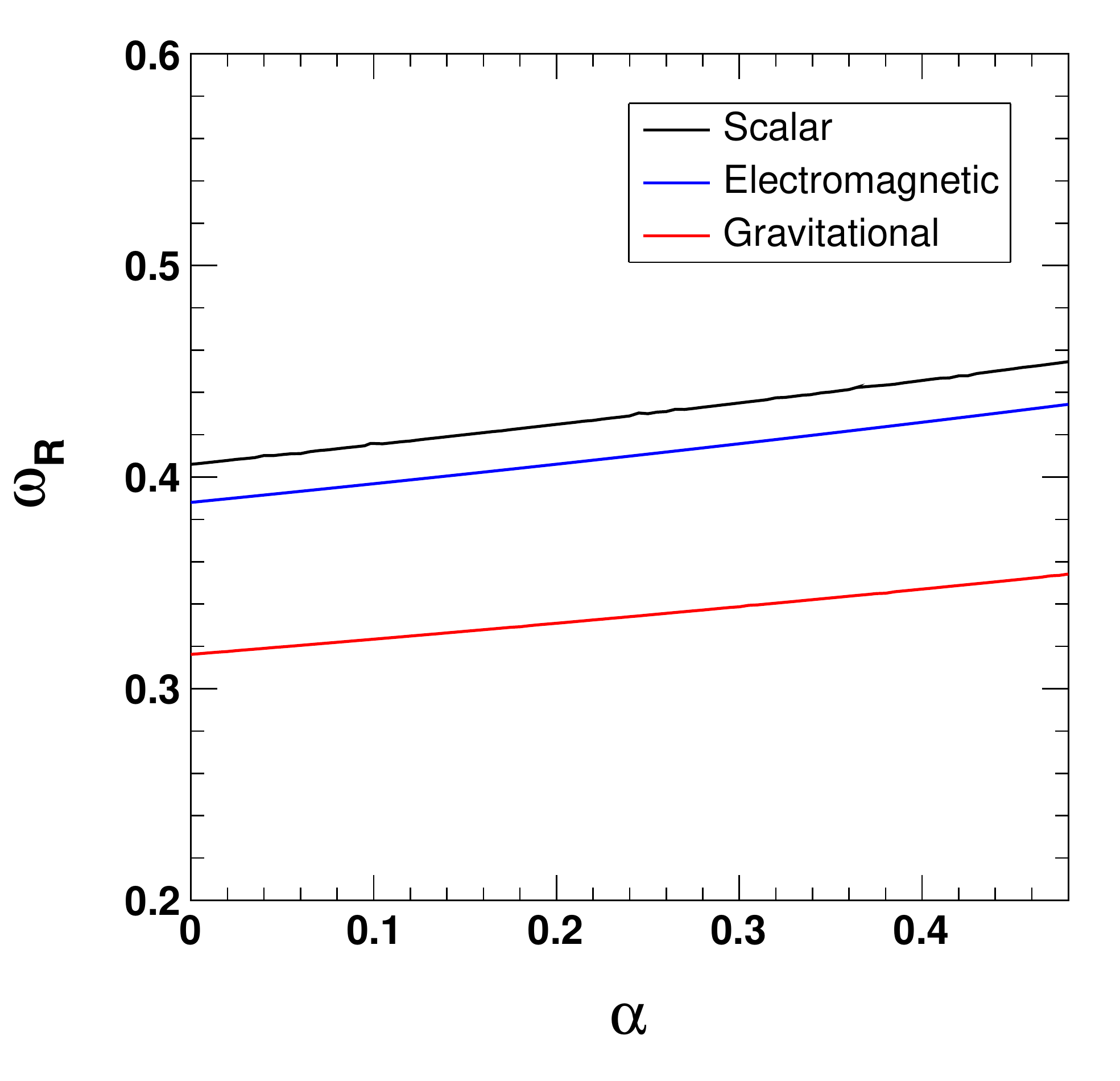}\hspace{1cm}
   \includegraphics[scale = 0.3]{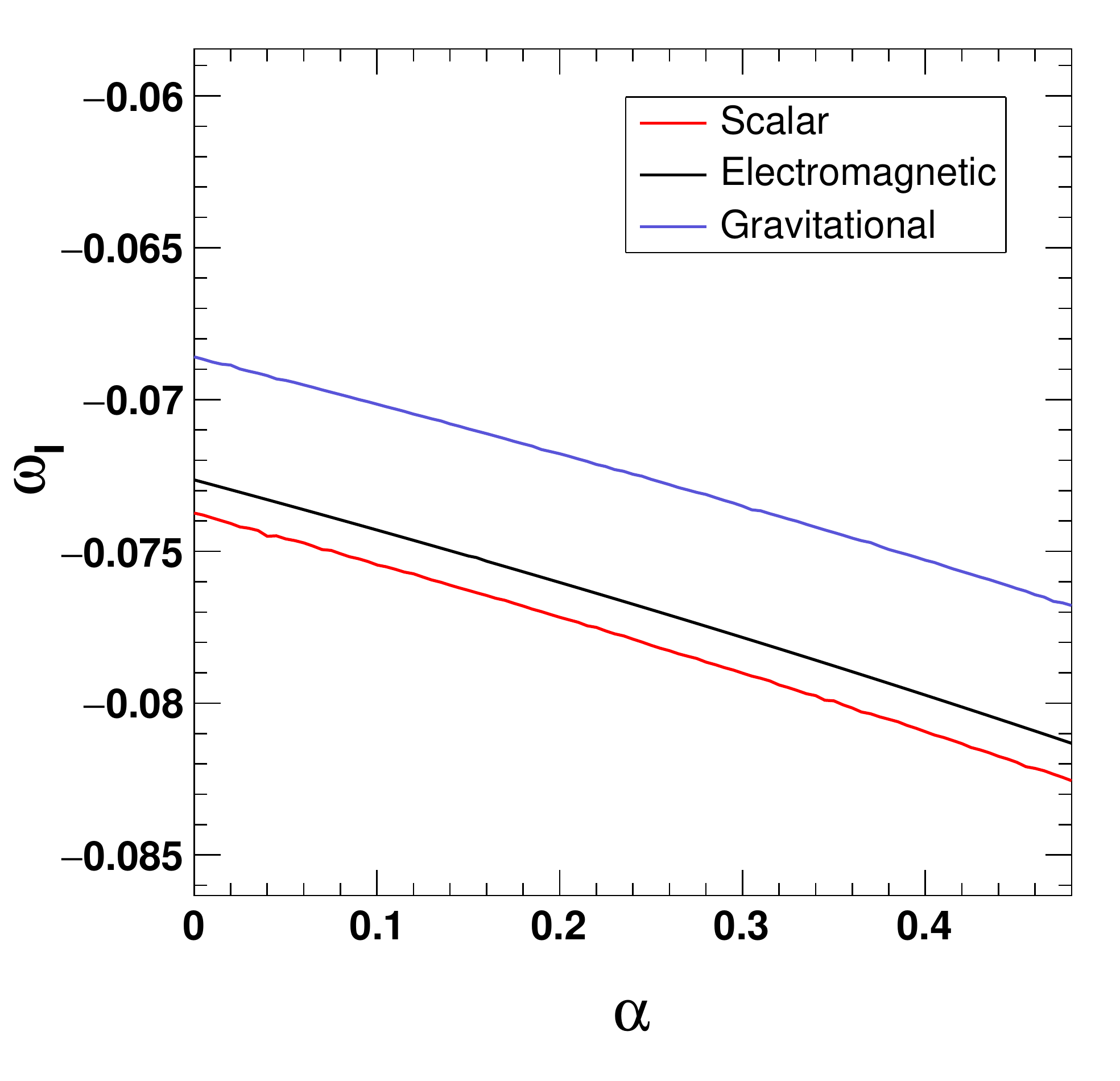}}
\vspace{-0.2cm}
\caption{Variation of quasinormal modes w.r.t.\ $\alpha$ for the black hole defined by the metric \eqref{flat_GUP_BH_sol} with $\lambda=0.1, \beta = 0.1, M=1, \mu=0.1$, $n=0$ and $l=2.$}
\label{fig_qnm01}
\end{figure}

\begin{figure}[!h]
\centerline{
   \includegraphics[scale = 0.3]{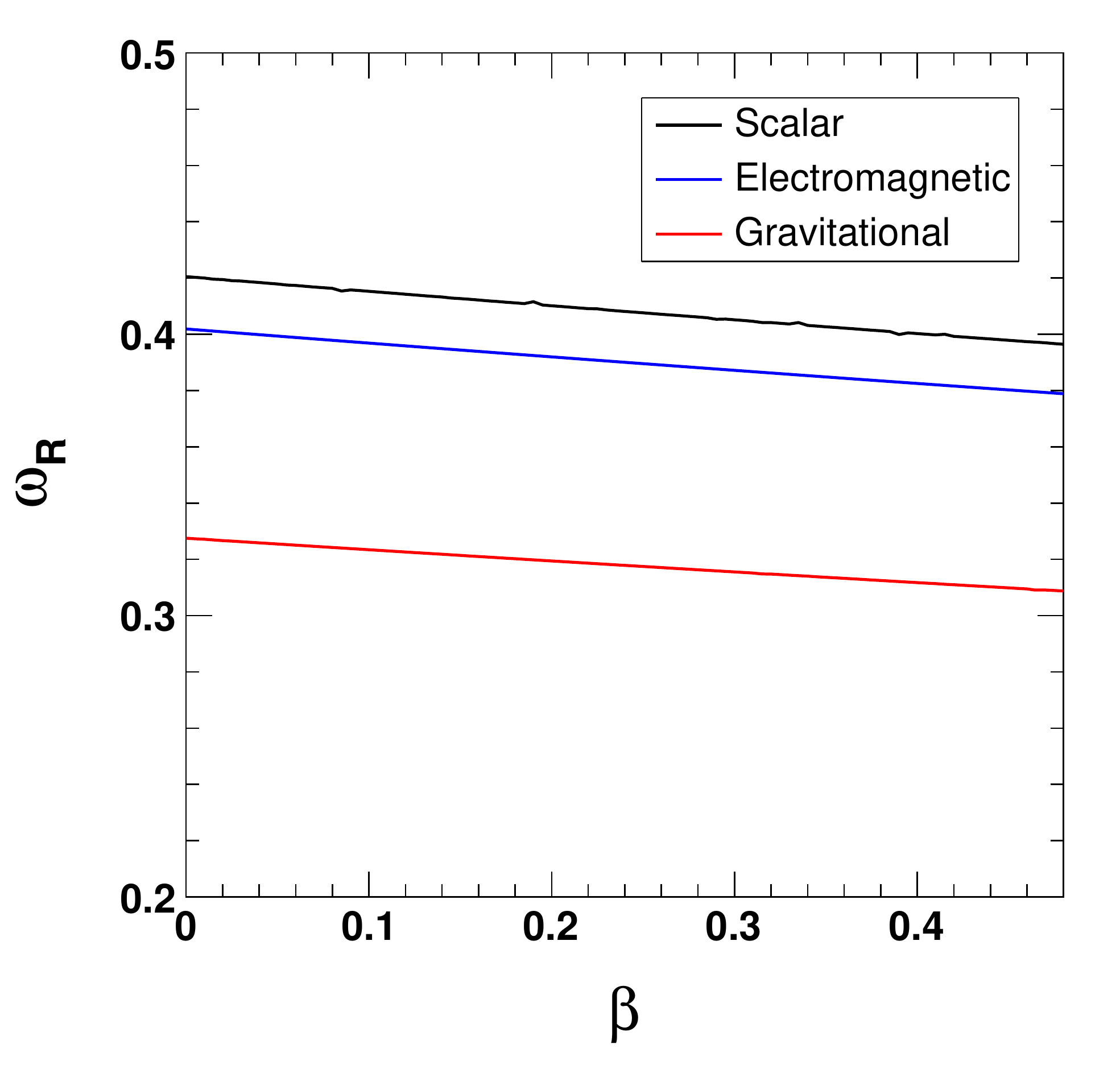}\hspace{1cm}
   \includegraphics[scale = 0.3]{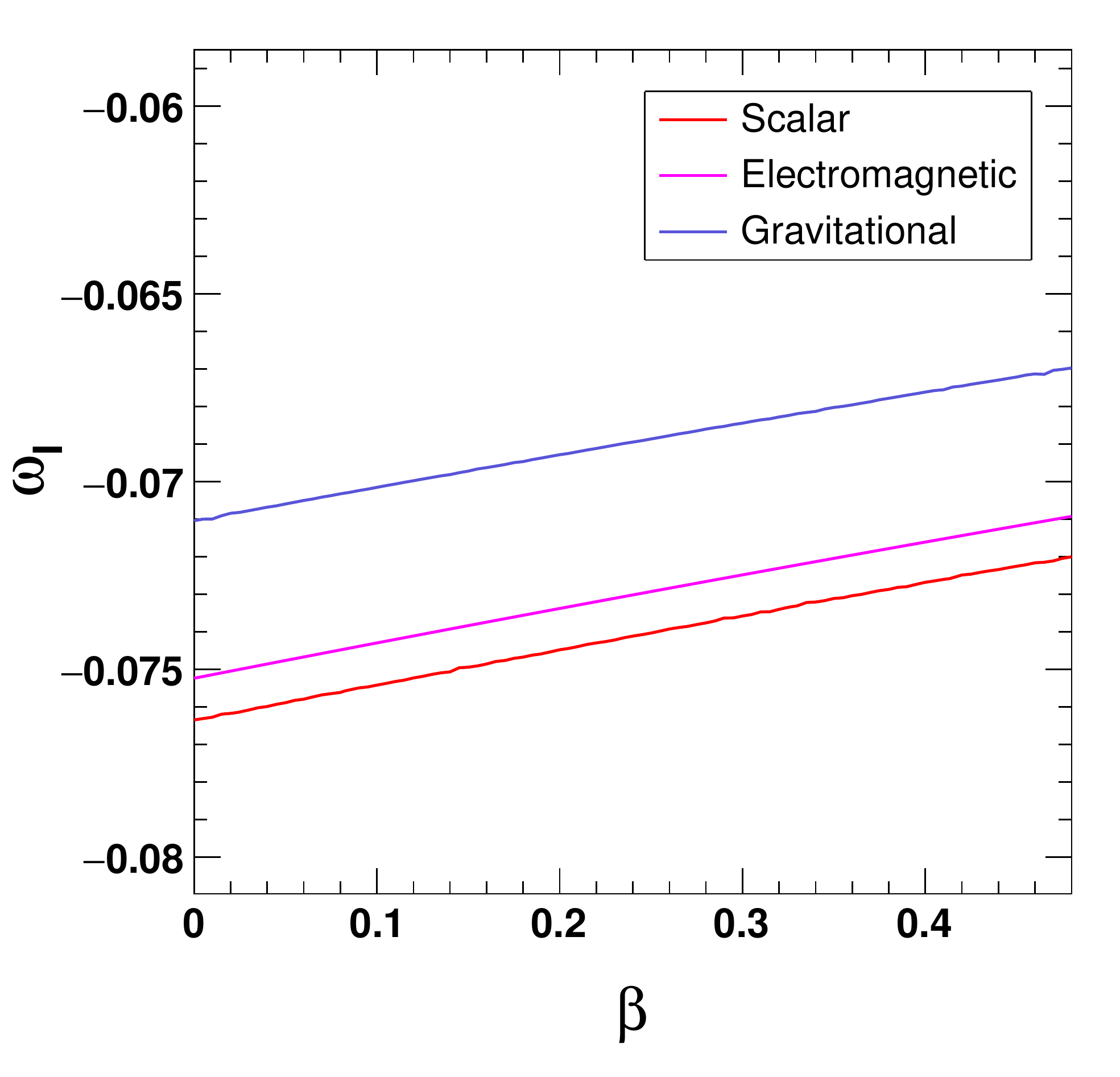}}
\vspace{-0.2cm}
\caption{Variation of quasinormal modes w.r.t.\ $\beta$ for the black hole defined by the metric \eqref{flat_GUP_BH_sol} with $\lambda=0.1, \alpha = 0.1, M=1, \mu=0.1$, $n=0$ and $l=2.$}
\label{fig_qnm02}
\end{figure}

\begin{figure}[!h]
   \centerline{
   \includegraphics[scale = 0.3]{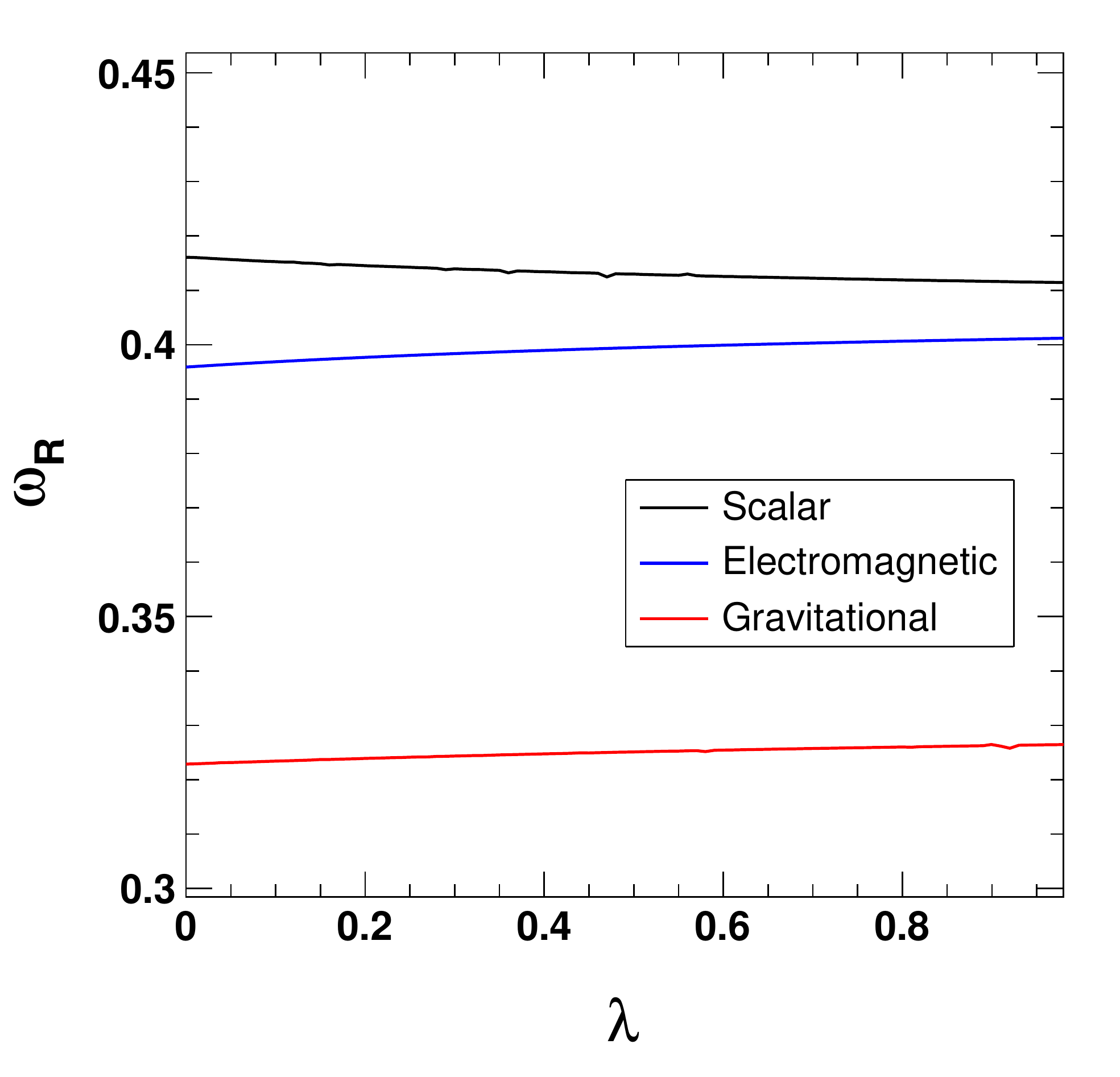}\hspace{1cm}
   \includegraphics[scale = 0.3]{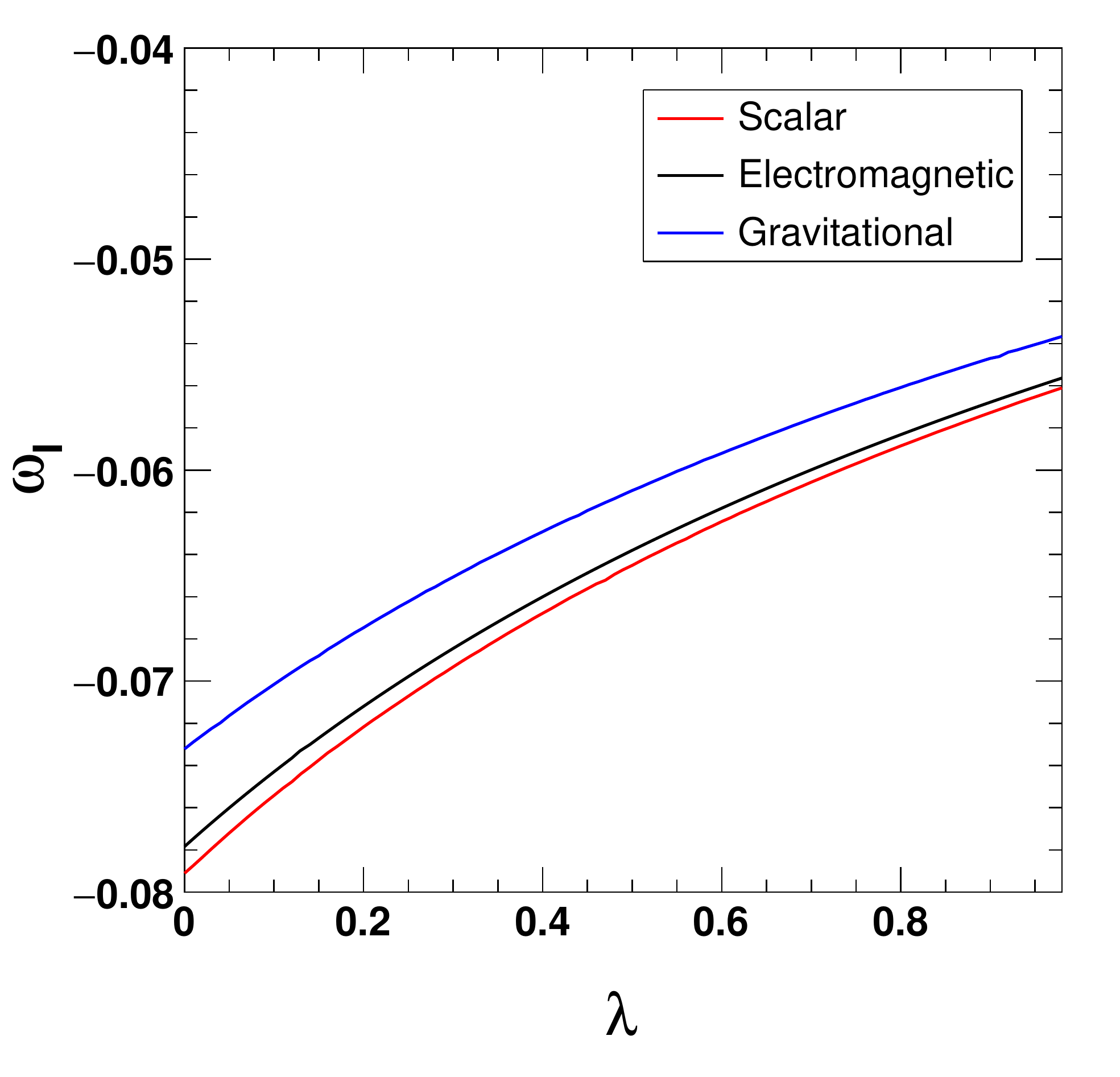}}
\vspace{-0.2cm}
\caption{Variation of quasinormal modes w.r.t.\ $\lambda$ for the black hole defined by the metric \eqref{flat_GUP_BH_sol} with $\alpha=0.1, \beta = 0.1, M=1, \mu=0.1$, $n=0$ and $l=2.$}
\label{fig_qnm03}
\end{figure}

\begin{figure}[!h]
   \centerline{
   \includegraphics[scale = 0.3]{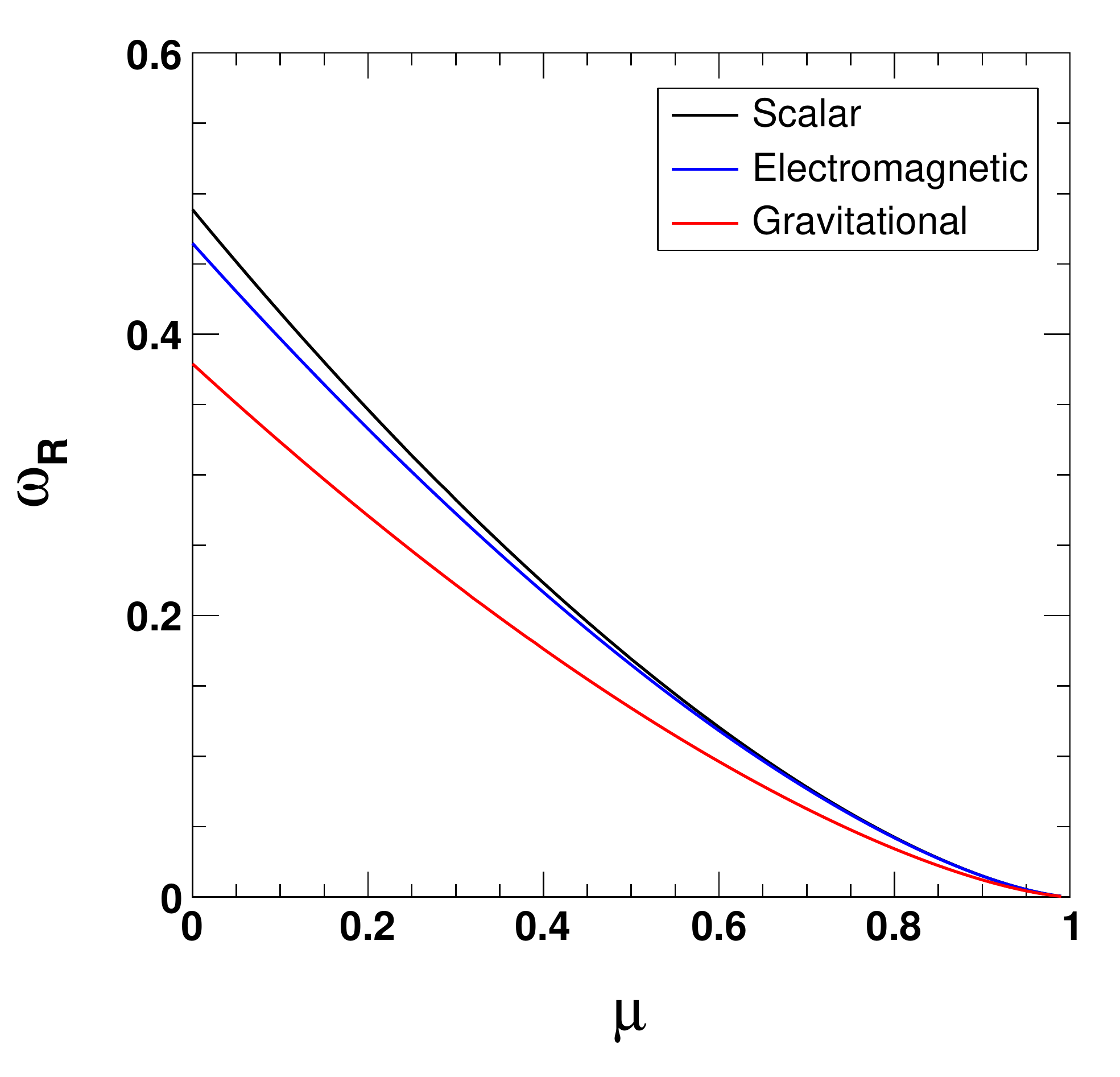}\hspace{1cm}
   \includegraphics[scale = 0.3]{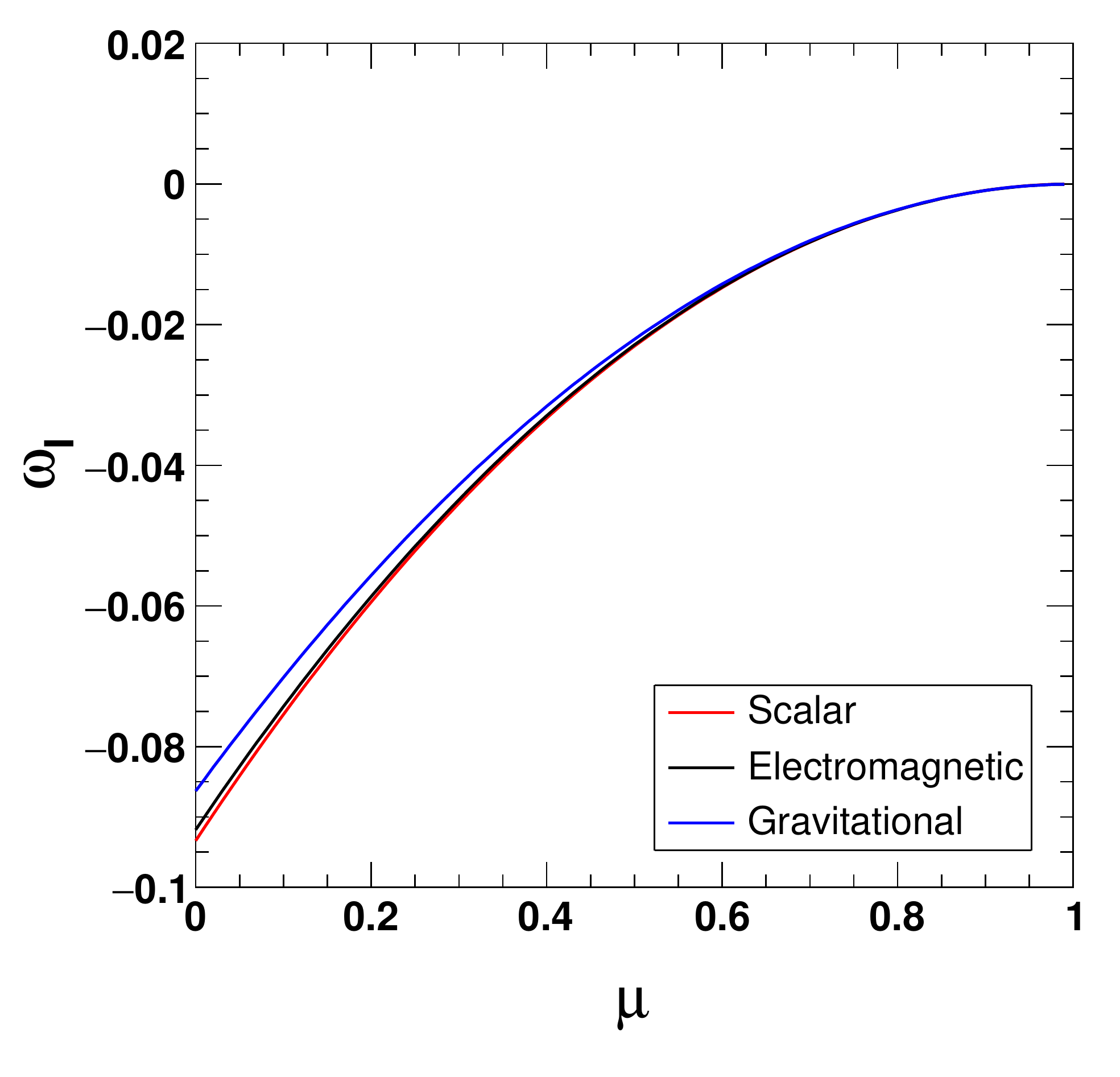}}
\vspace{-0.2cm}
\caption{Variation of quasinormal modes w.r.t.\ $\mu$ for the black hole defined by the metric \eqref{flat_GUP_BH_sol} with $\alpha=0.1, \beta = 0.1, M=1, \lambda=0.1$, $n=0$ and $l=2.$}
\label{fig_qnm04}
\end{figure}

\begin{figure}[!h]
\centerline{
   \includegraphics[scale = 0.3]{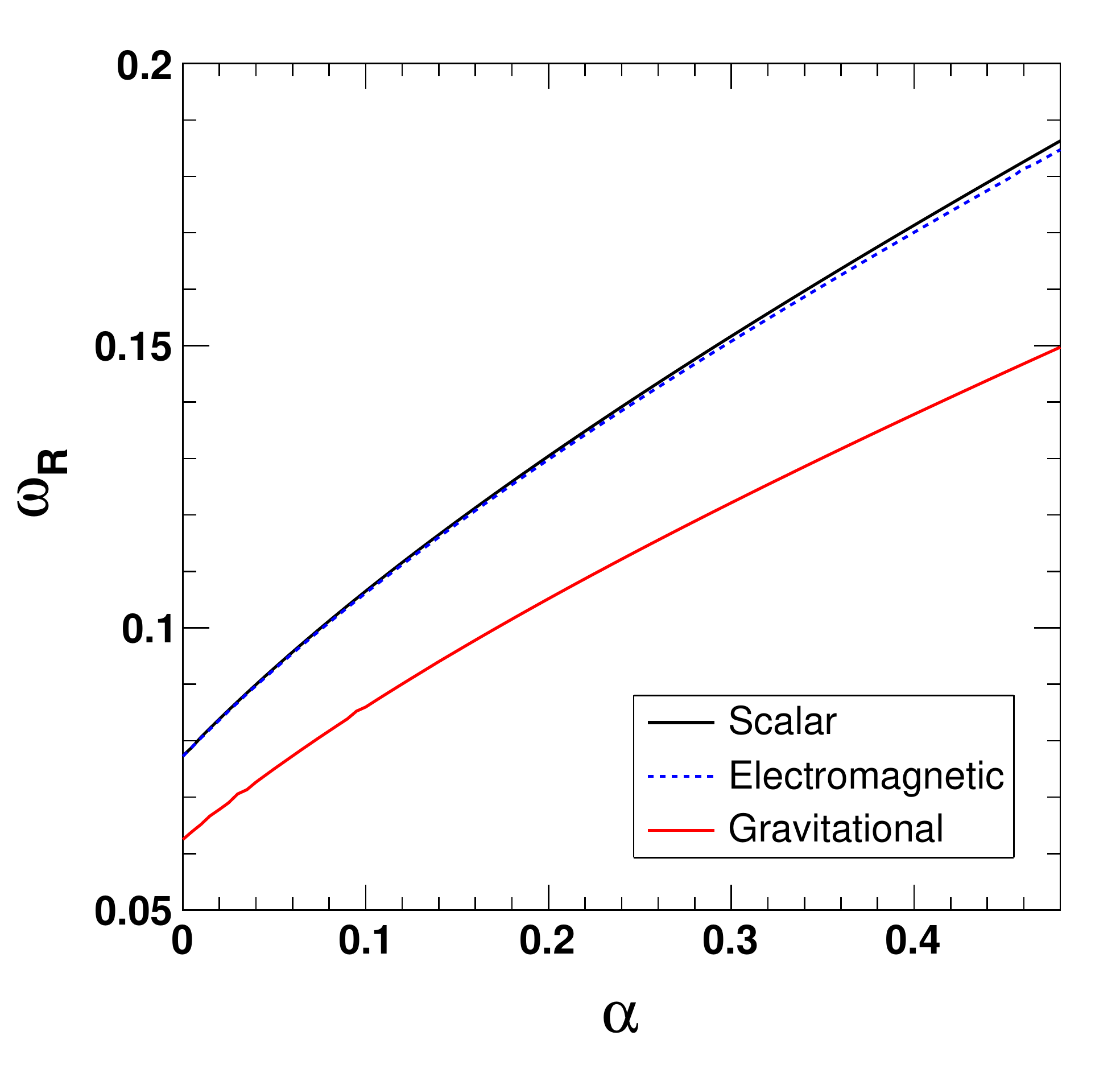}\hspace{1cm}
   \includegraphics[scale = 0.3]{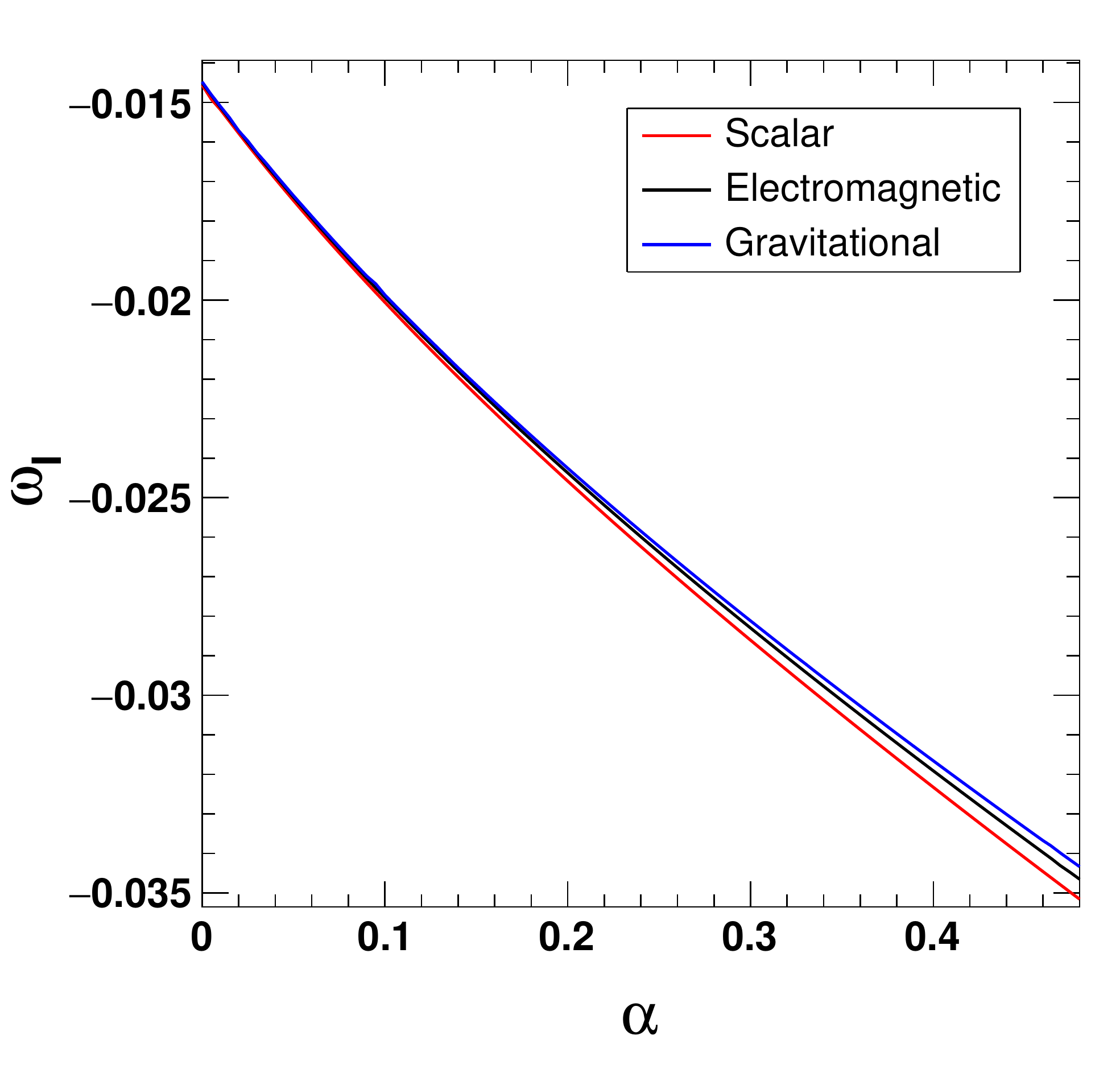}}
\vspace{-0.2cm}
\caption{Variation of quasinormal modes w.r.t.\ $\alpha$ for GUP corrected de Sitter black hole in bumblebee gravity with $\lambda=0.1, \beta = 0.1, M=1, \mu=0.1, \Lambda_{eff} = 0.07$, $n=0$ and $l=2.$}
\label{fig_qnm05}
\end{figure}

\begin{figure}[!h]
\centerline{
   \includegraphics[scale = 0.3]{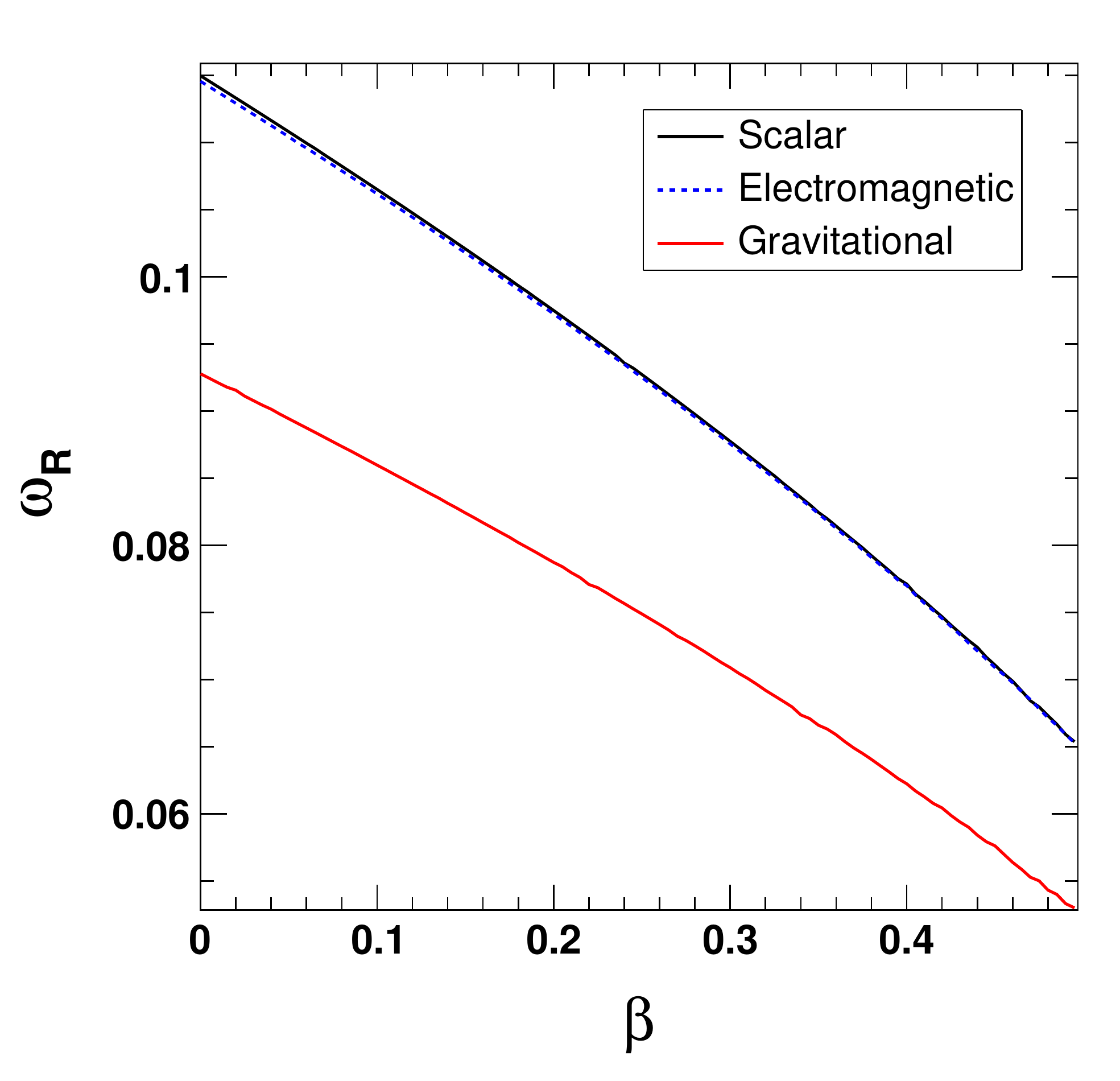}\hspace{1cm}
   \includegraphics[scale = 0.3]{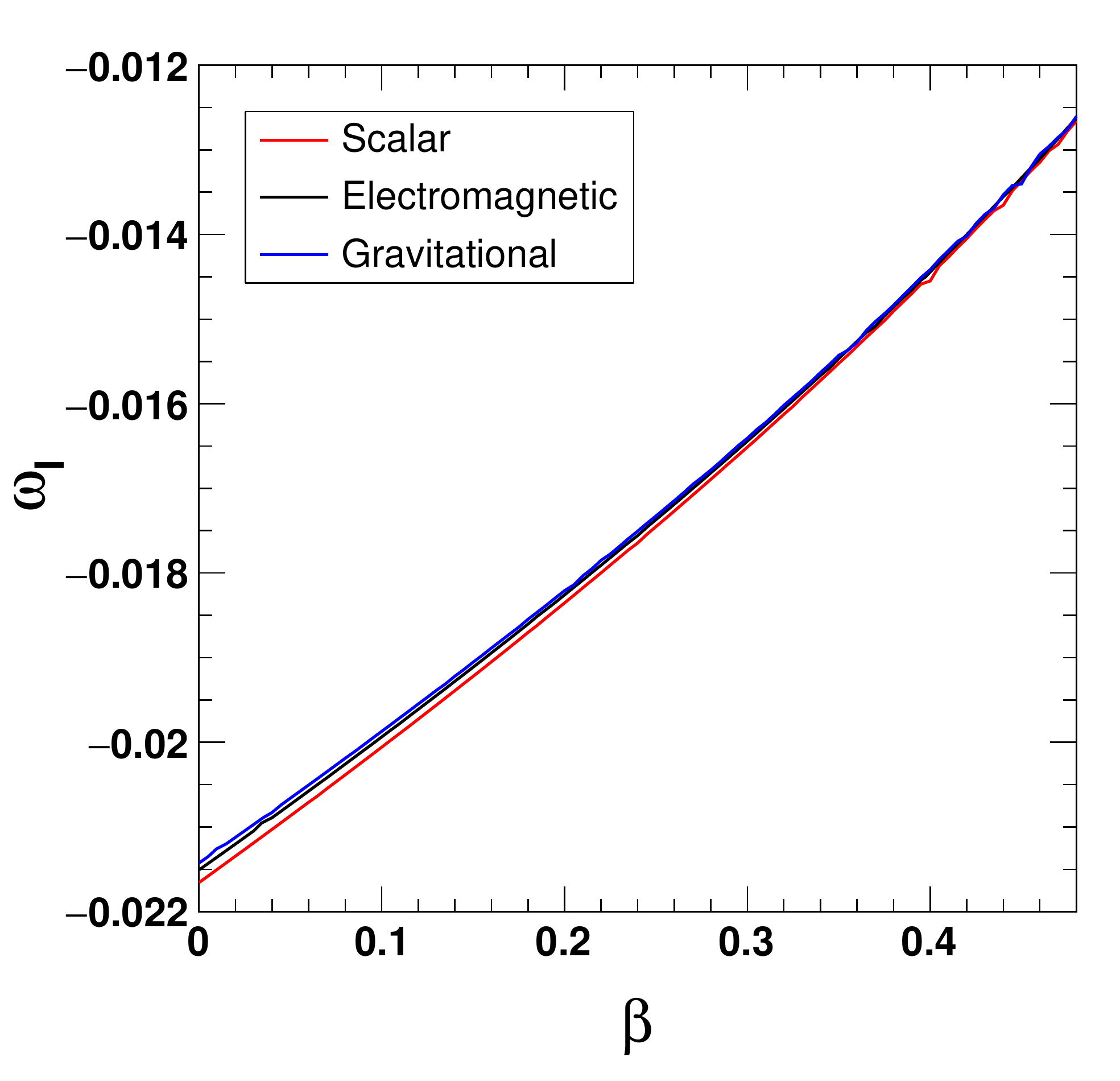}}
\vspace{-0.2cm}
\caption{Variation of quasinormal modes w.r.t.\ $\beta$ for GUP corrected de Sitter black hole in bumblebee gravity with $\lambda=0.1, \alpha = 0.1, M=1, \mu=0.1, \Lambda_{eff} = 0.07$, $n=0$ and $l=2.$}
\label{fig_qnm06}
\end{figure}

\begin{figure}[!h]
   \centerline{
   \includegraphics[scale = 0.3]{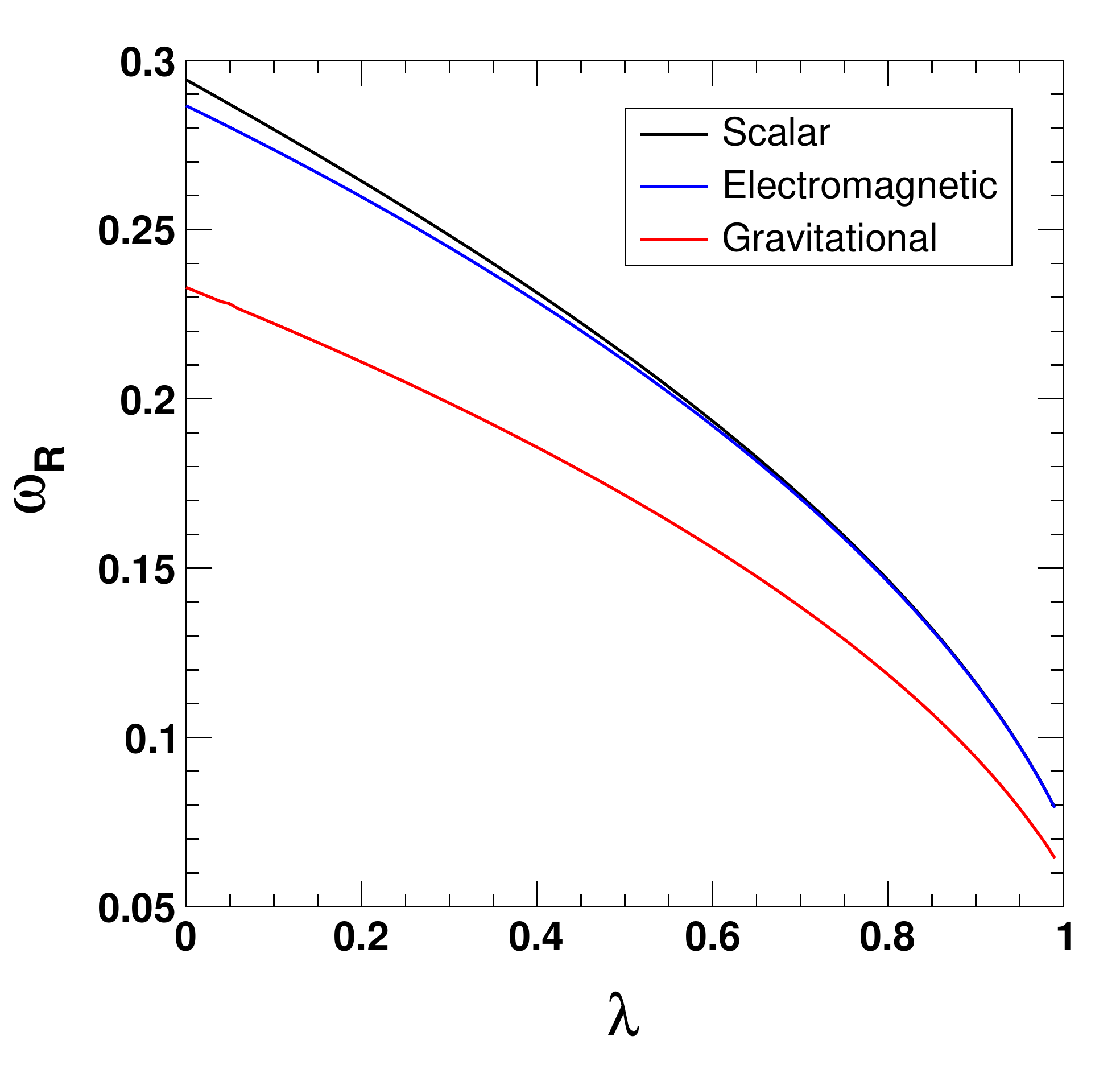}\hspace{1cm}
   \includegraphics[scale = 0.3]{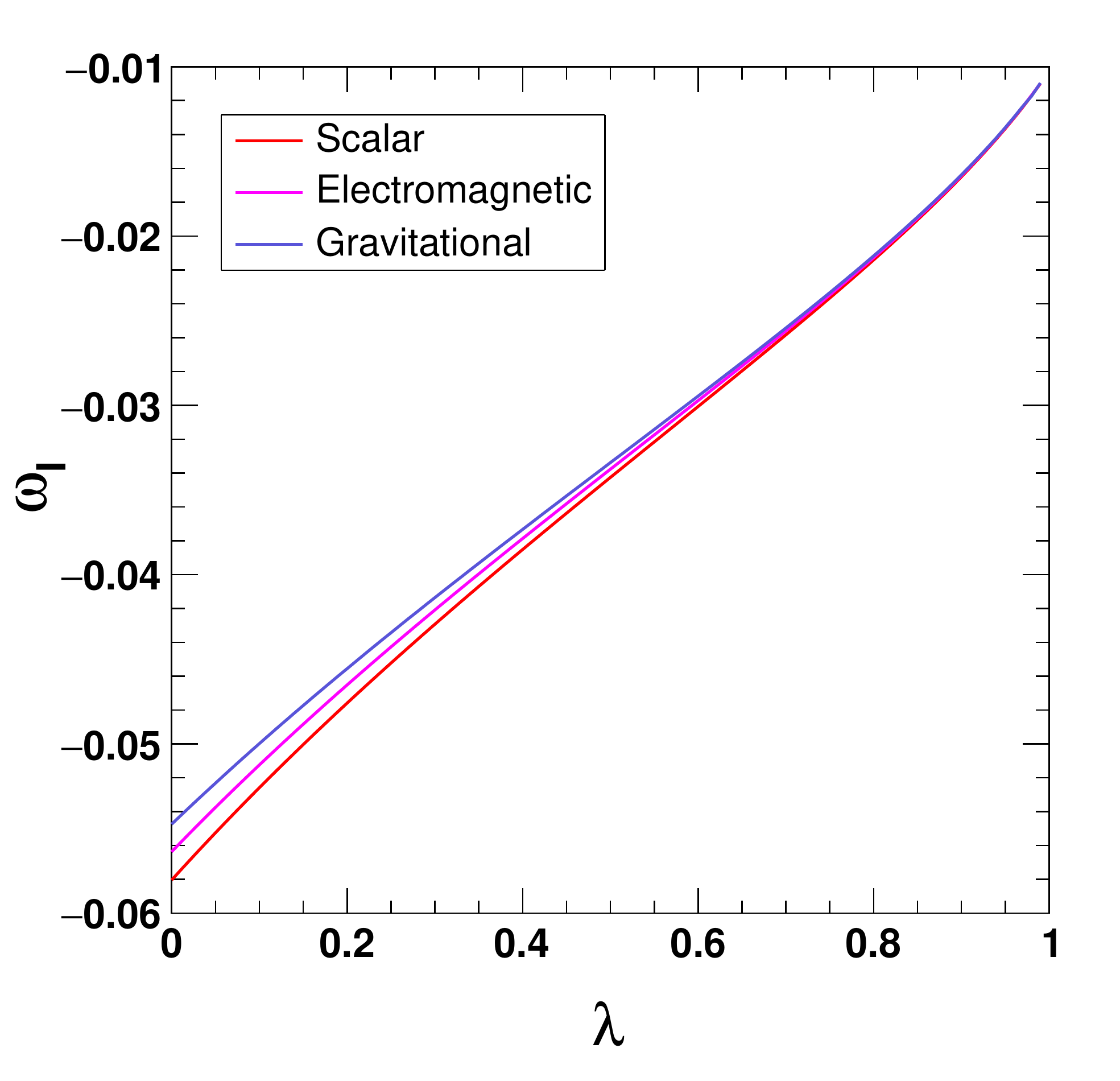}}
\vspace{-0.2cm}
\caption{Variation of quasinormal modes w.r.t.\ $\lambda$ for GUP corrected de Sitter black hole in bumblebee gravity with $\alpha=0.1, \beta = 0.1, M=1, \mu=0.1, \Lambda_{eff} = 0.04$, $n=0$ and $l=2.$}
\label{fig_qnm07}
\end{figure}

\begin{figure}[!h]
   \centerline{
   \includegraphics[scale = 0.3]{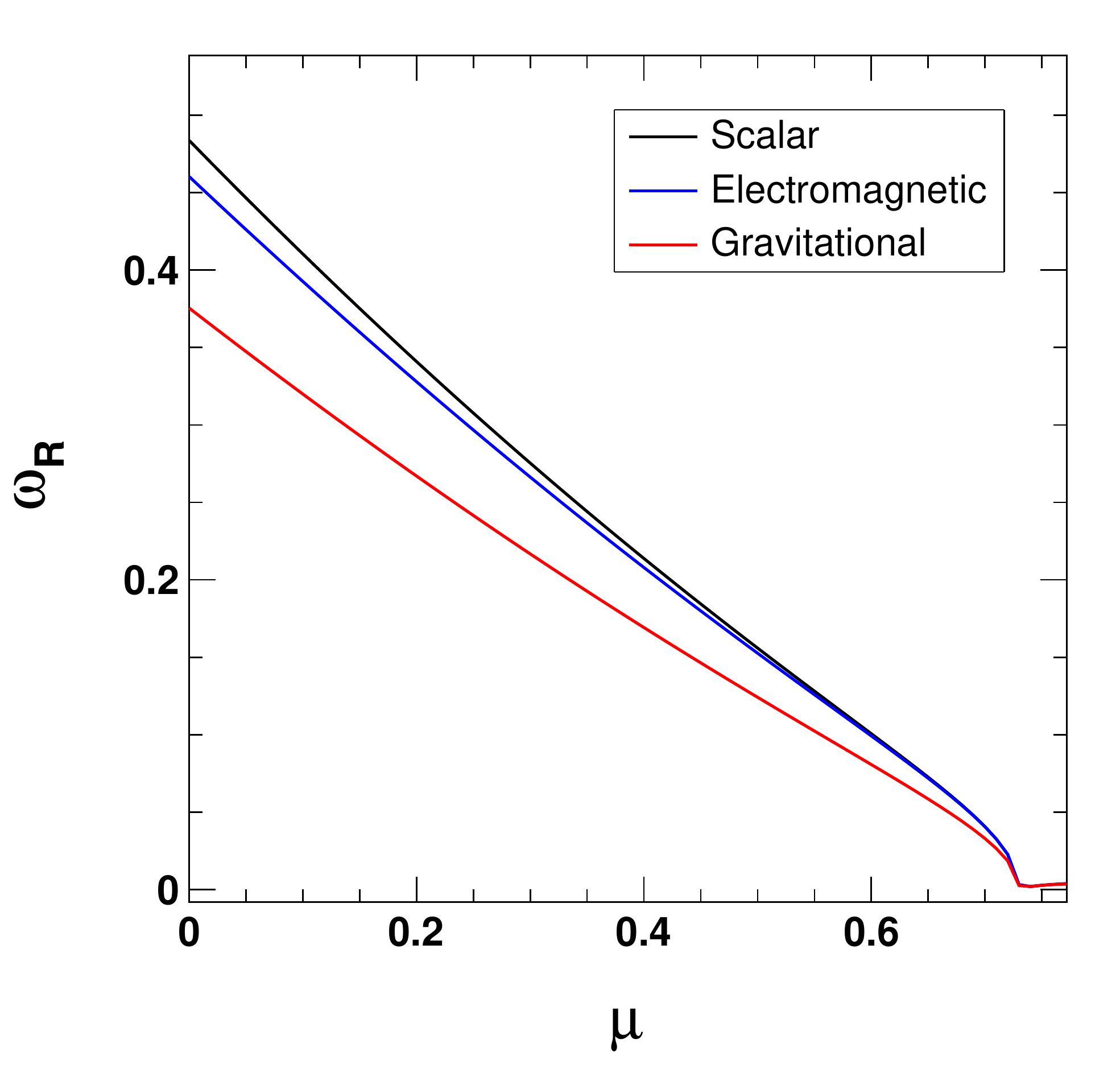}\hspace{1cm}
   \includegraphics[scale = 0.3]{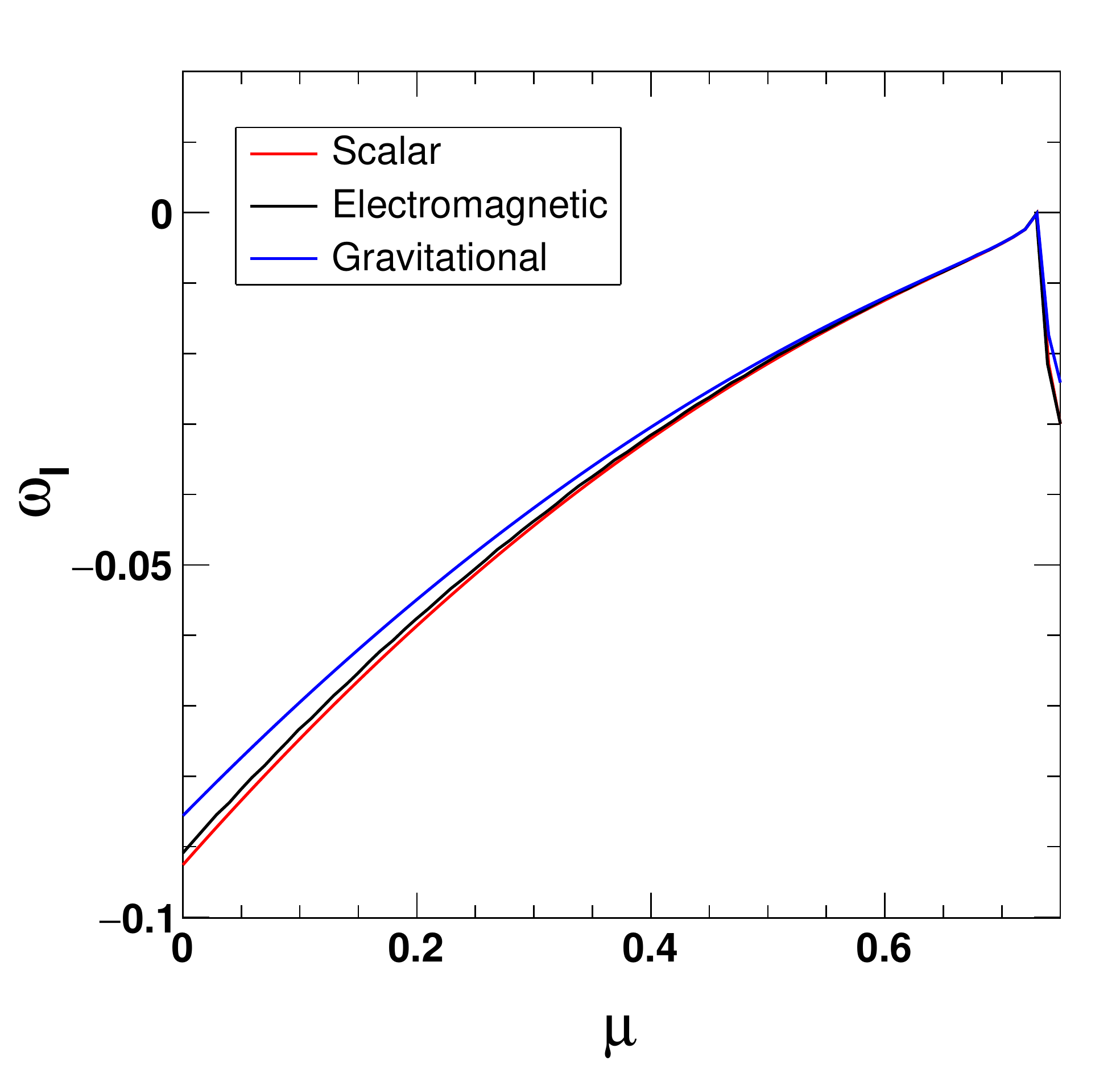}}
\vspace{-0.2cm}
\caption{Variation of quasinormal modes w.r.t.\ $\mu$ for GUP corrected de Sitter black hole in bumblebee gravity with $\alpha=0.1, \beta = 0.1, M=1, \lambda=0.1, \Lambda_{eff} = 0.002$, $n=0$ and $l=2.$}
\label{fig_qnm08}
\end{figure}

\begin{figure}[!h]
\centerline{
   \includegraphics[scale = 0.3]{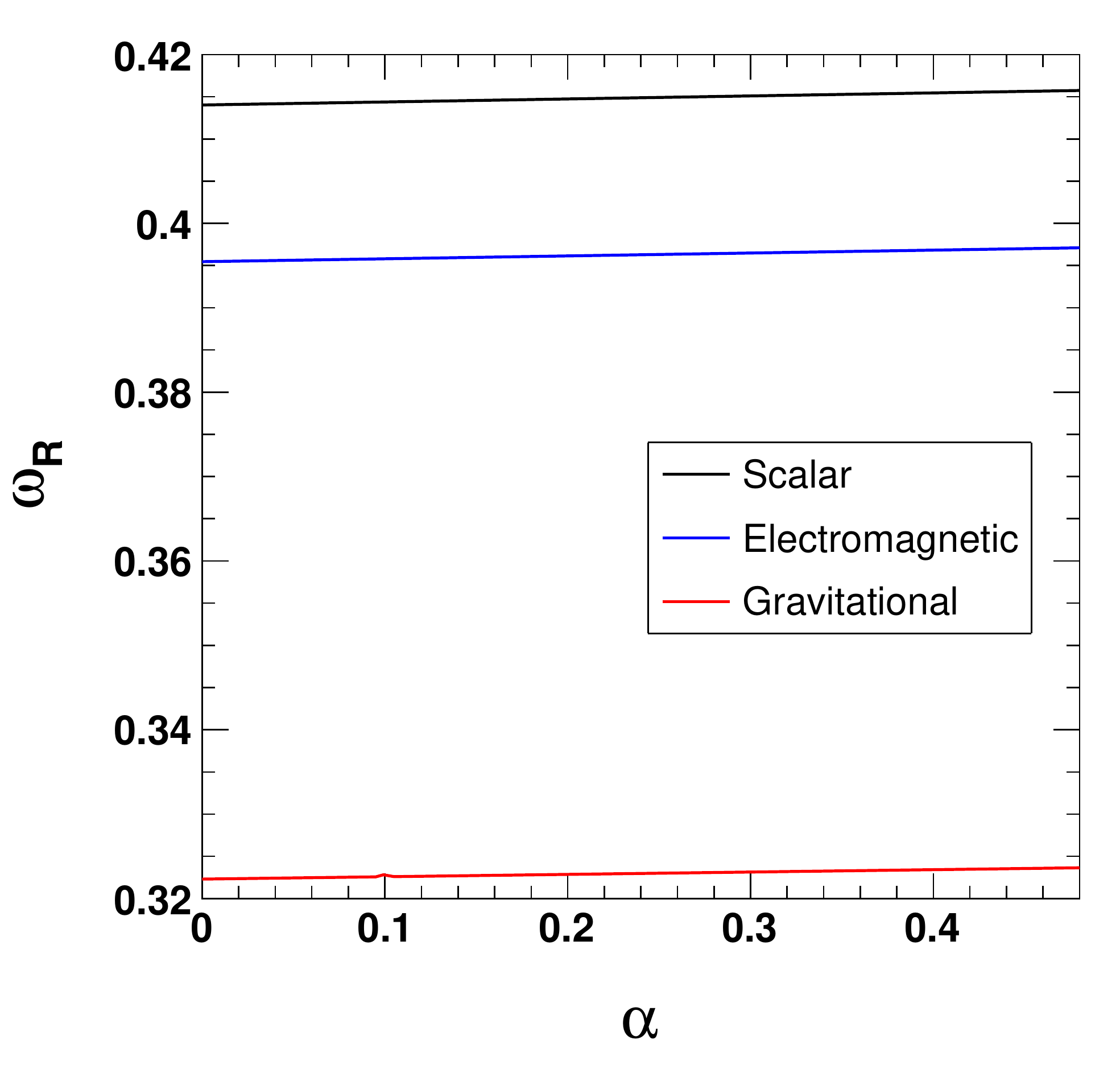}\hspace{1cm}
   \includegraphics[scale = 0.3]{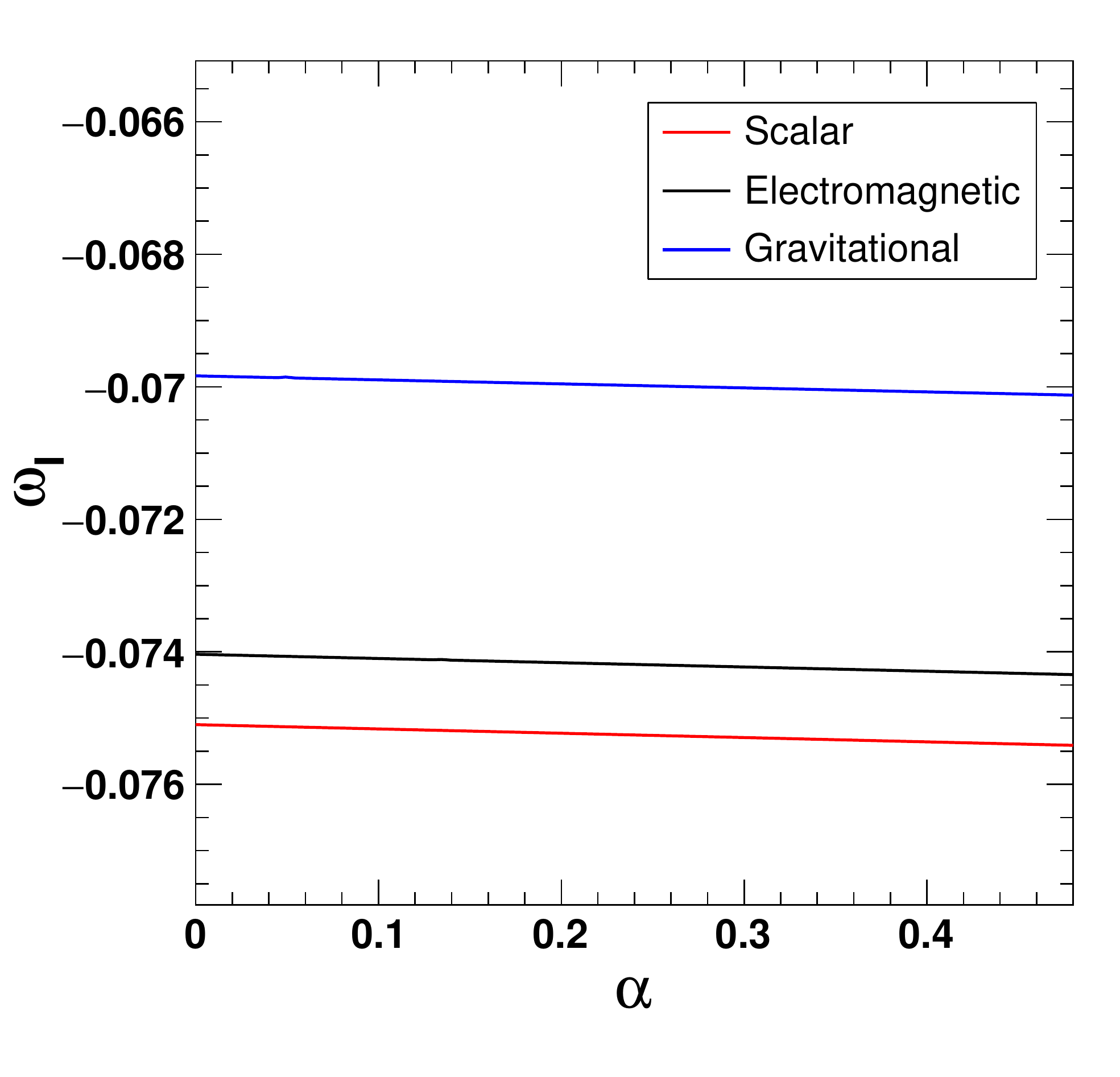}}
\caption{Variation of quasinormal modes w.r.t.\ $\alpha$ for GUP corrected anti-de Sitter black hole in bumblebee gravity with $\lambda=0.1, \beta = 0.1, M=1, \mu=0.1, \Lambda_{eff} = -0.001$, $n=0$ and $l=2.$}
\label{fig_qnm09}
\end{figure}

\begin{figure}[!h]
\centerline{
   \includegraphics[scale = 0.3]{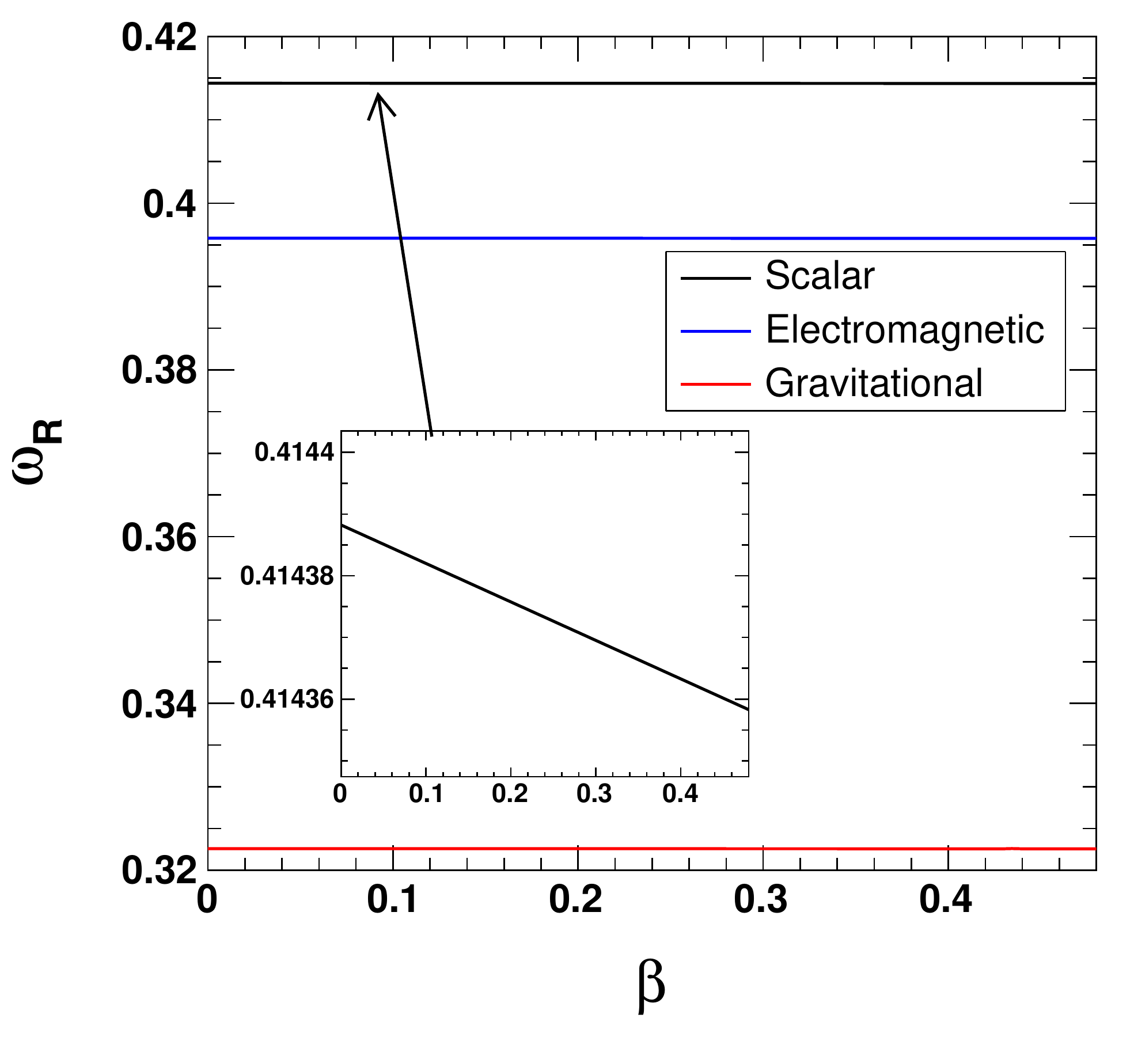}\hspace{1cm}
   \includegraphics[scale = 0.3]{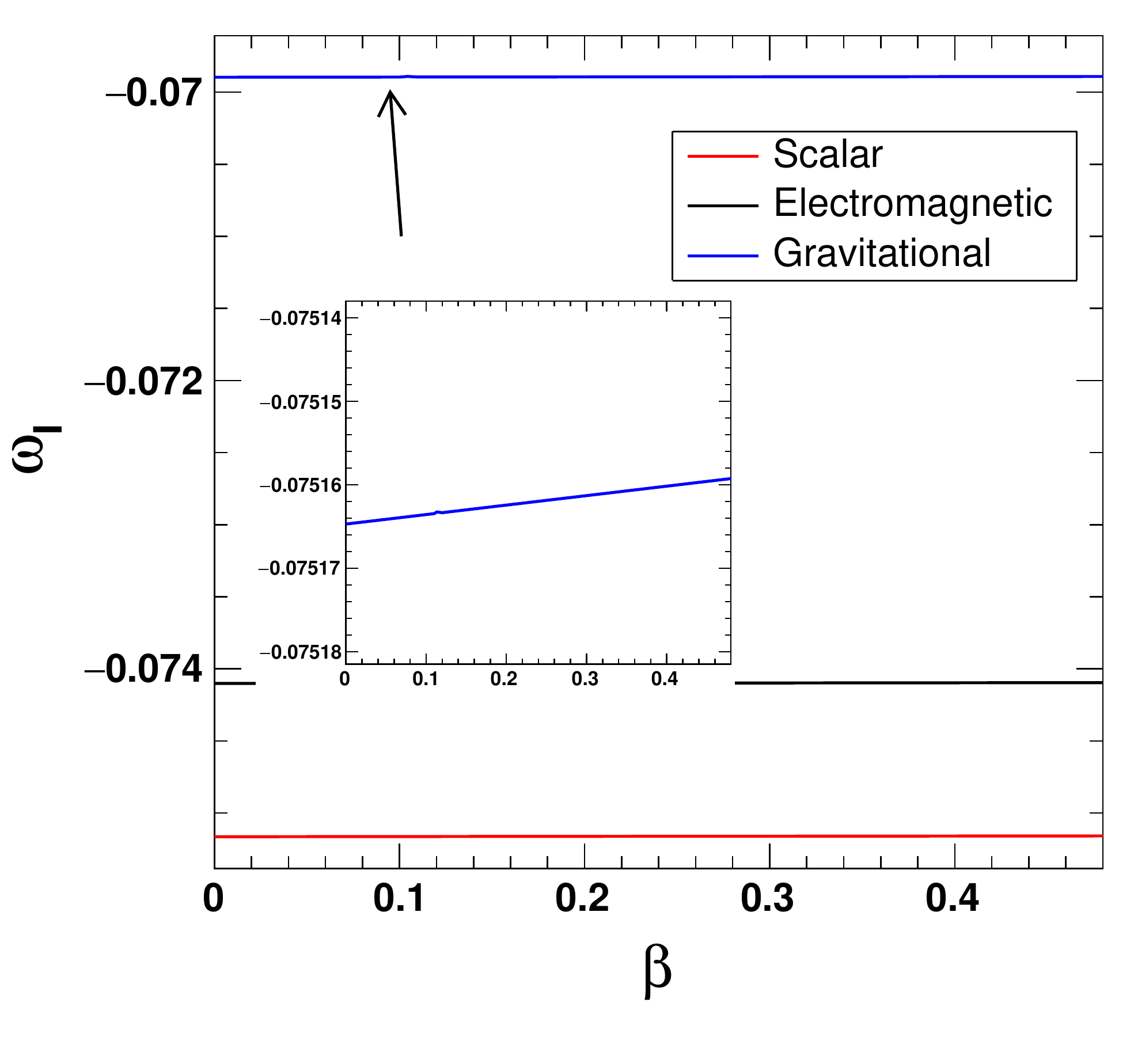}}
\vspace{-0.2cm}
\caption{Variation of quasinormal modes w.r.t.\ $\beta$ for GUP corrected anti-de Sitter black hole in bumblebee gravity with $\alpha=0.1, \lambda = 0.1, M=1, \mu=0.1, \Lambda_{eff} = -0.001$, $n=0$ and $l=2.$}
\label{fig_qnm10}
\end{figure}

\begin{figure}[!h]
   \centerline{
   \includegraphics[scale = 0.3]{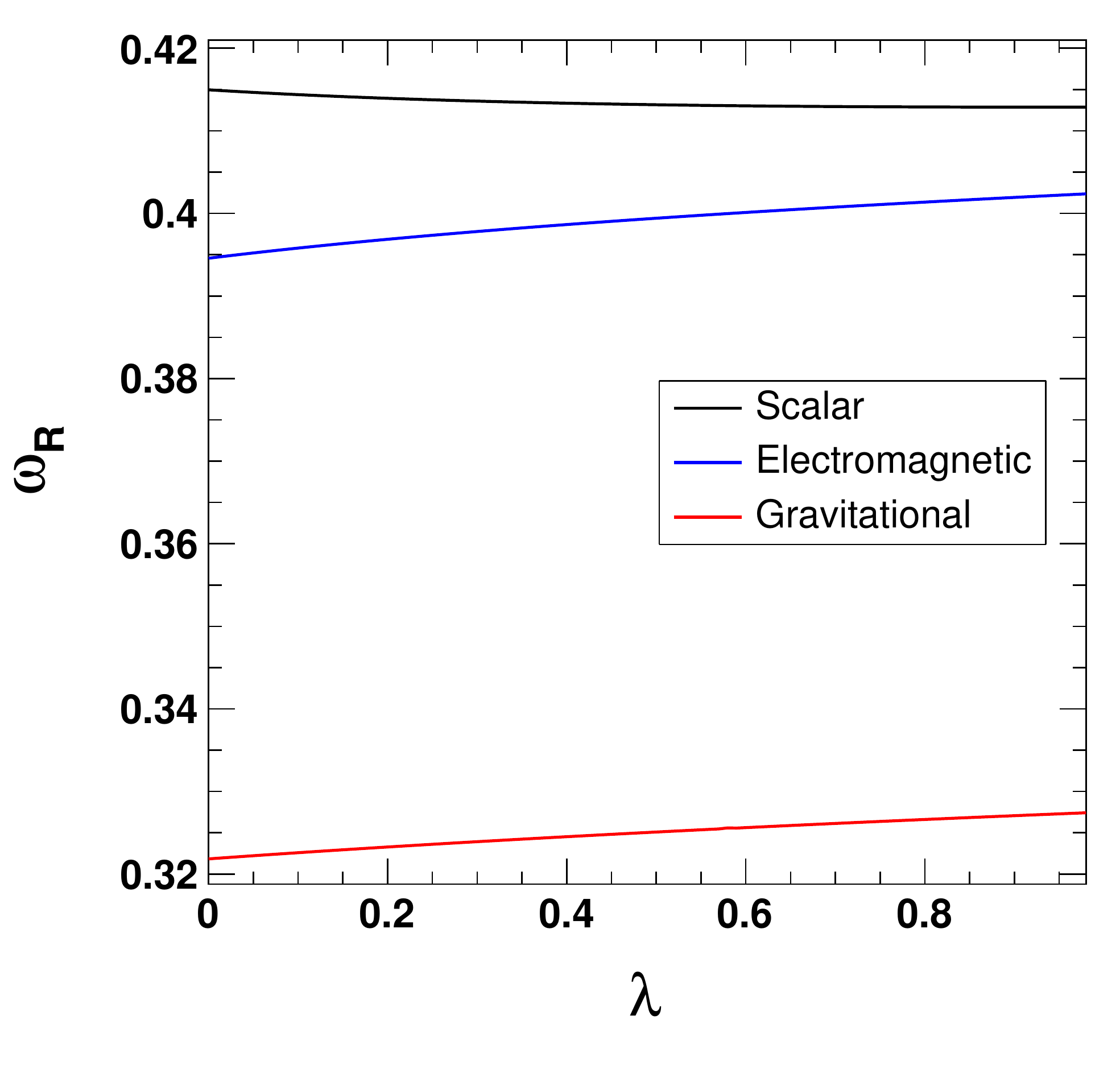}\hspace{1cm}
   \includegraphics[scale = 0.3]{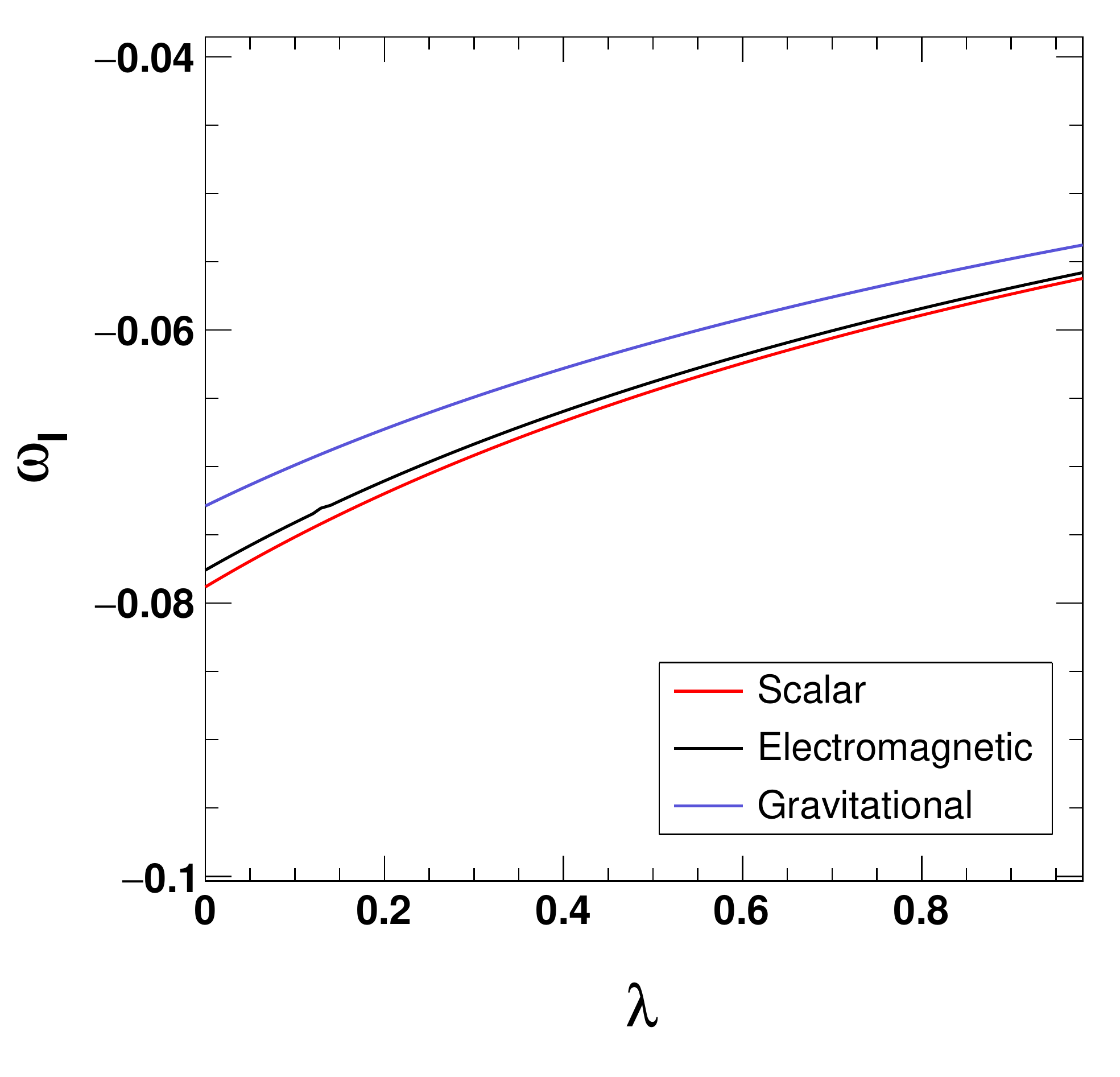}}
\vspace{-0.2cm}
\caption{Variation of quasinormal modes w.r.t.\ $\lambda$ for GUP corrected anti-de Sitter black hole in bumblebee gravity with $\alpha=0.1, \beta = 0.1, M=1, \mu=0.1, \Lambda_{eff} = -0.001$, $n=0$ and $l=2.$}
\label{fig_qnm11}
\end{figure}

\begin{figure}[!h]
   \centerline{
   \includegraphics[scale = 0.3]{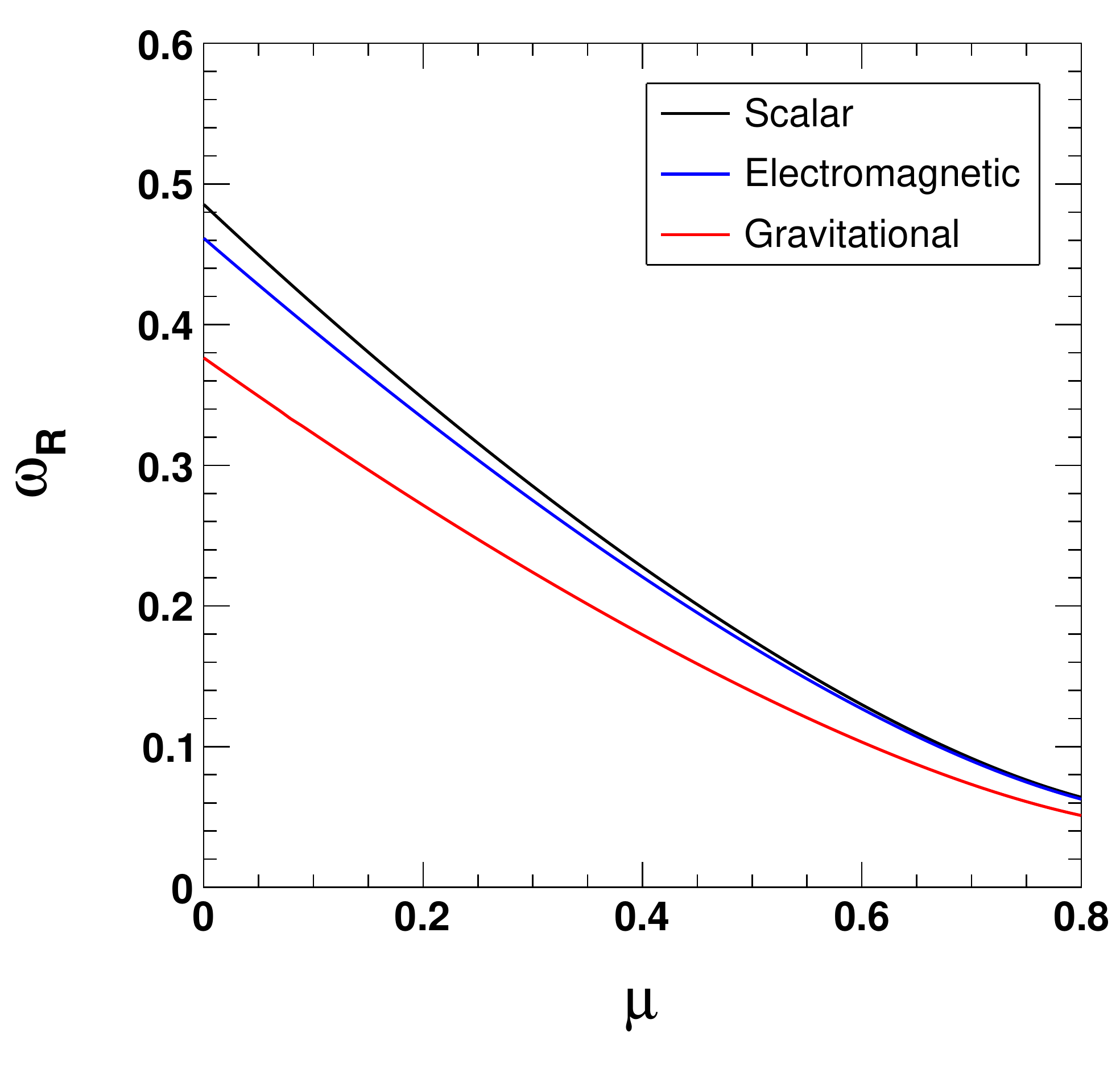}\hspace{1cm}
   \includegraphics[scale = 0.3]{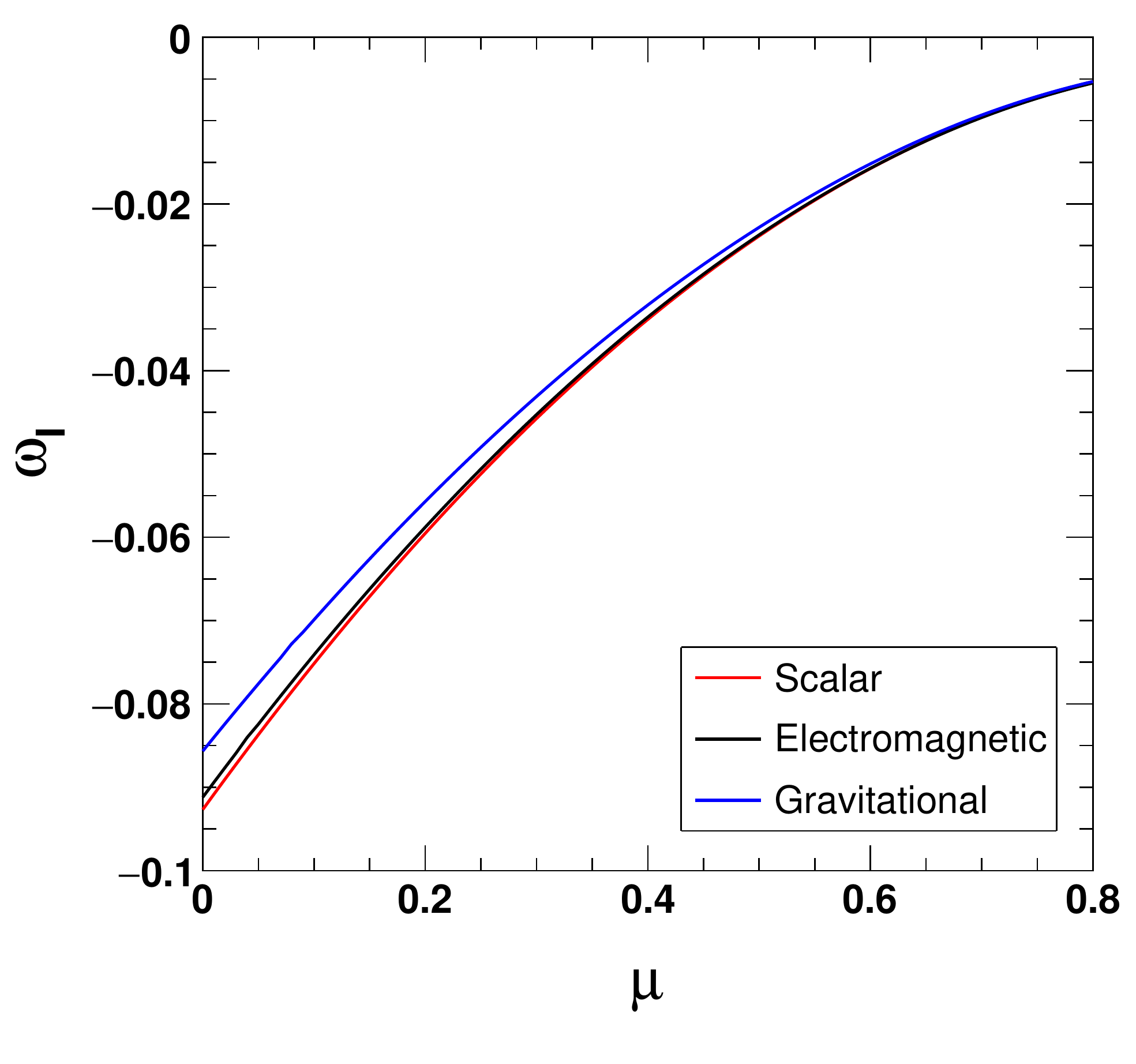}}
\vspace{-0.2cm}
\caption{Variation of quasinormal modes w.r.t.\ $\mu$ for GUP corrected anti-de Sitter black hole in bumblebee gravity with $\alpha=0.1, \beta = 0.1, M=1, \lambda=0.1, \Lambda_{eff} = -0.001$, $n=0$ and $l=2.$}
\label{fig_qnm12}
\end{figure}

\begin{figure}[!h]
\centerline{
   \includegraphics[scale = 0.35]{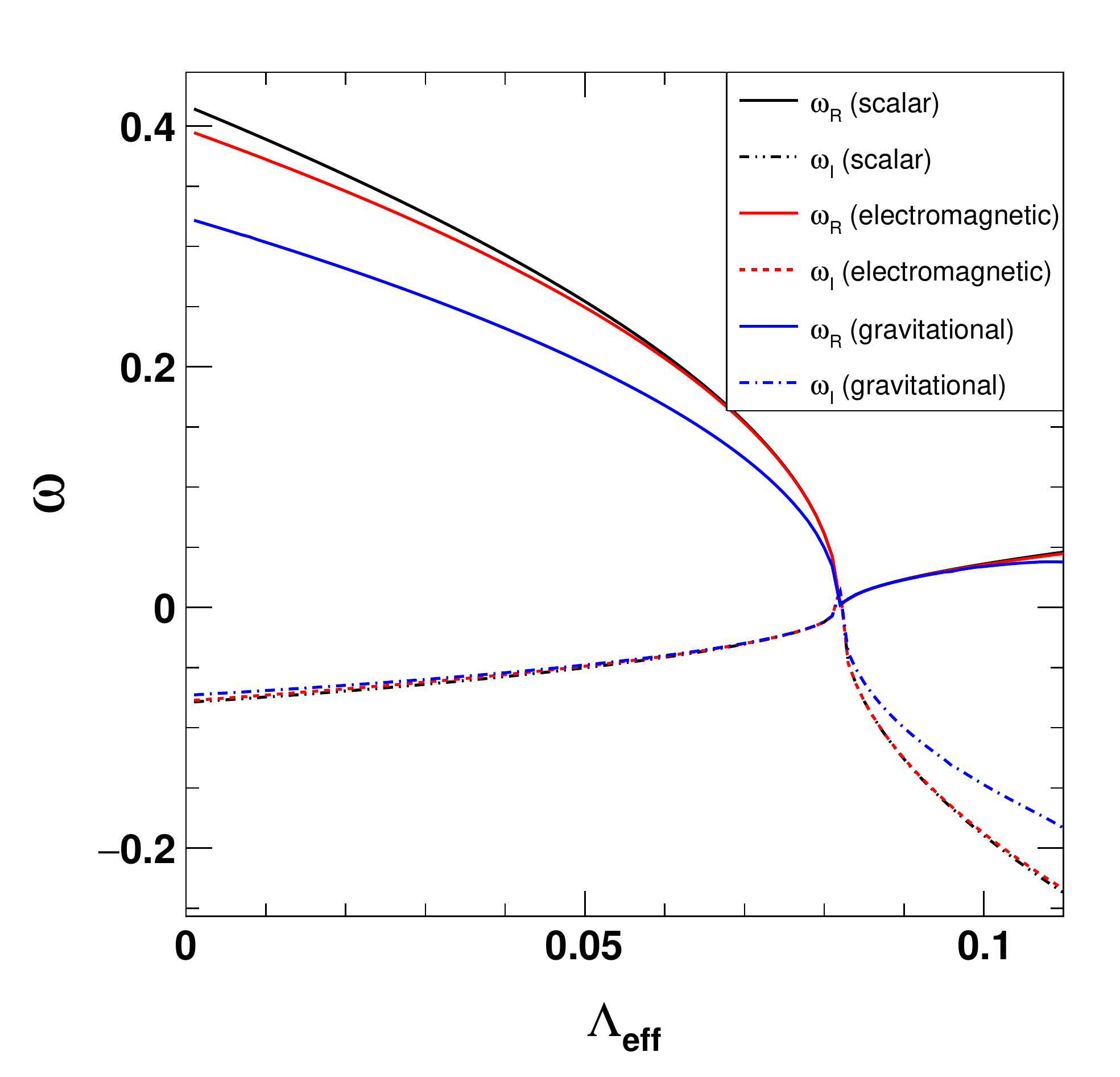}}
   \centerline{
   \includegraphics[scale = 0.3]{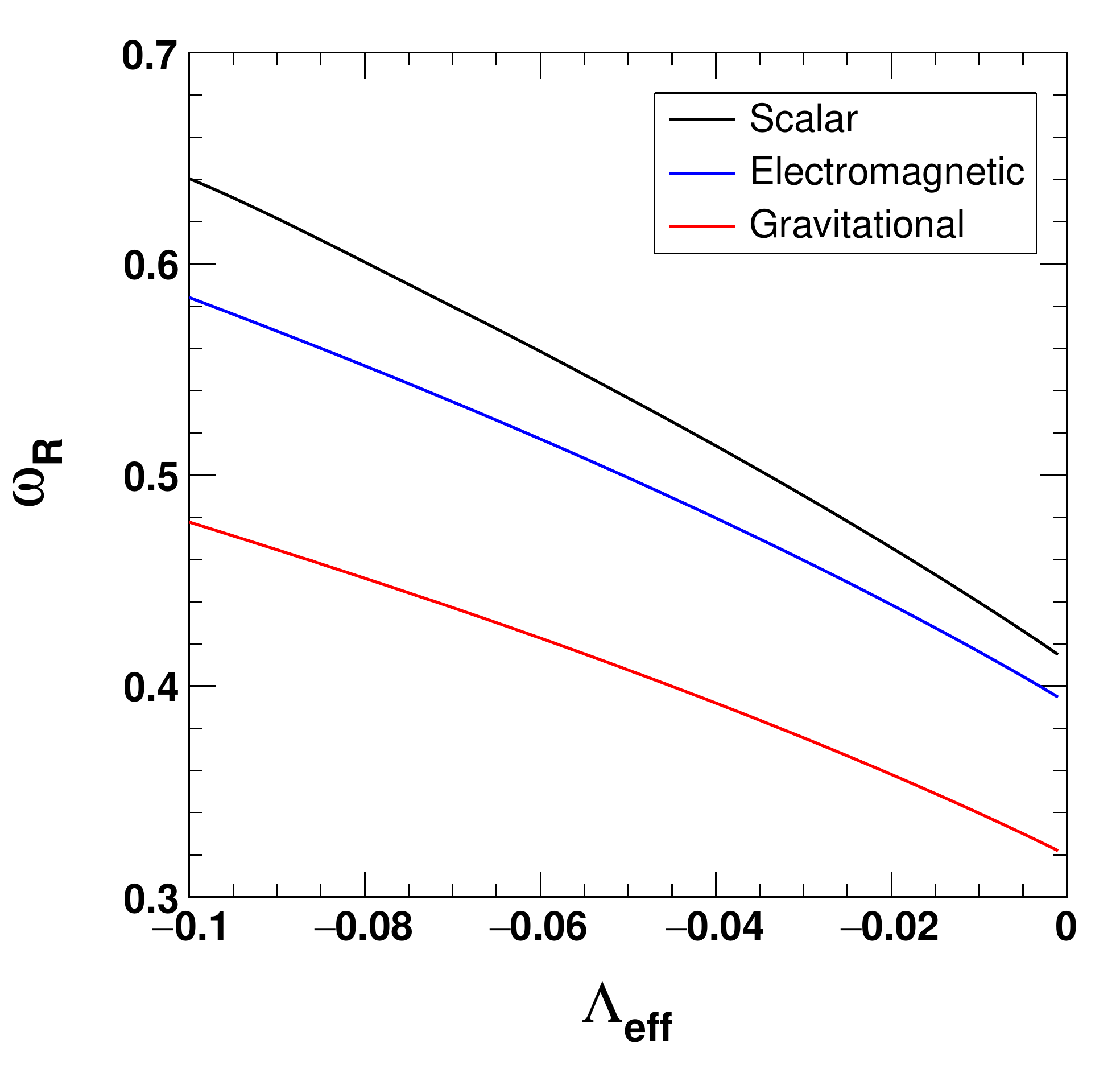}\hspace{1cm}
   \includegraphics[scale = 0.3]{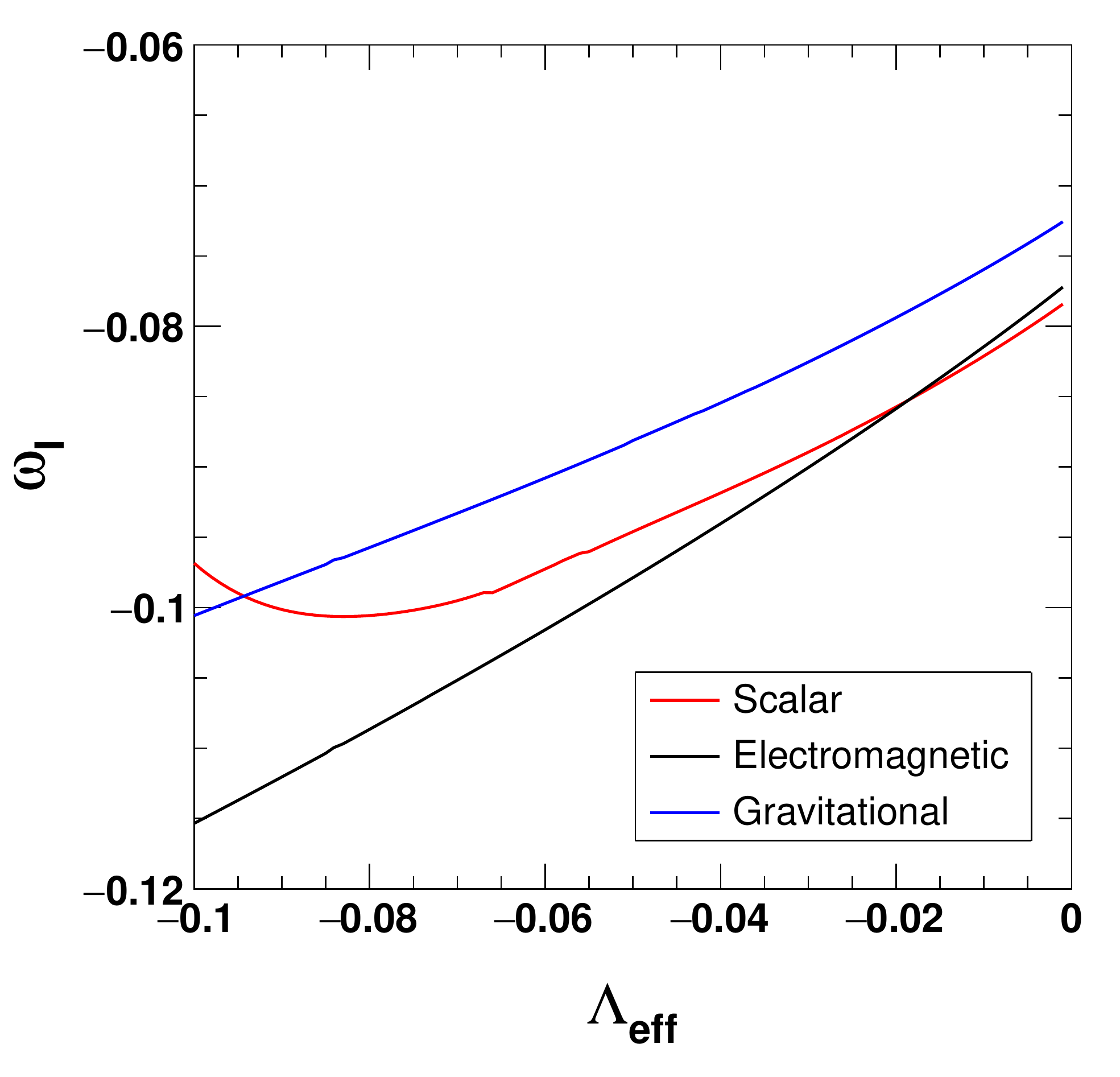}}
\vspace{-0.2cm}
\caption{Variation of quasinormal modes w.r.t.\ $\Lambda_{eff}$ for GUP corrected black hole in bumblebee gravity with $\alpha = 0.1, \beta = 0.1, M=1, \mu=0.1, n=0, l=2$ and  $\lambda=0.01.$}
\label{fig_qnm13}
\end{figure}
\subsection{WKB method with Pad\'e Approximation for Quasinormal modes}
In this work, we shall use higher order WKB method to calculate the quasinormal modes 
of the black holes defined in the previous sections.
The first order WKB method for calculating the quasinormal modes was first suggested 
by Schutz and Will in Ref.\ \cite{Schutz}. Later, the method was developed to higher 
orders \cite{Will_wkb, Konoplya_wkb, Maty_wkb}. In Ref.\ \cite{Maty_wkb} it was 
suggested that averaging of the Pad\'e approximations can be implemented to WKB method 
and later it was seen that the method improves the results of quasinormal modes with 
a higher accuracy \cite{Konoplya_wkb}. Here we shall use the Pad\'e averaged 6th order WKB 
approximation method.

We have shown the numerical values of quasinormal modes for scalar, electromagnetic 
and gravitational perturbations for different $l$ values in 
Tables \ref{tab01}, \ref{tab02} and \ref{tab03}. Each table shows the quasinormal 
modes obtained from $6$th order WKB approximation method, Pad\'e averaged $6$th order 
WKB approximation method and time domain analysis. To have a better idea on the 
errors, we have calculated rms error $\vartriangle_{rms}$ from the results given by 
Pad\'e WKB and also calculated $\Delta_6$. The error $\Delta_6$ is defined 
as \cite{Konoplya_wkb} 
\begin{equation}
\Delta_6 = \dfrac{\vline \; \omega_7 - \omega_5 \; \vline}{2},
\end{equation}
where $\omega_7$ and $\omega_5$ are the quasinormal modes obtained from the $7$th and $5$th order Pad\'e averaged WKB method.

In Fig.\ \ref{fig_qnm01}, we have plotted the real and imaginary quasinormal 
frequencies w.r.t.\ GUP parameter $\alpha$ with a vanishing cosmological constant i.e.\ for the black hole defined by the metric \eqref{flat_GUP_BH_sol}. It is seen that an 
increase in $\alpha$ increases the oscillation frequencies of GWs linearly for all 
the three perturbations viz., scalar, electromagnetic and gravitational 
perturbations. The decay rate also increases with increase in $\alpha$. It is seen 
that for the gravitational perturbation, the quasinormal frequencies and the decay 
rate are lowest while for the scalar perturbation, they are maximum. An increase in the 
second GUP parameter $\beta$, on the other hand, decreases the oscillation 
frequencies and the decay rate of GWs (see Fig.\ \ref{fig_qnm02}).  Thus the
effects of these two parameters on the quasinormal modes are opposite.

From Fig.\ \ref{fig_qnm03}, it is seen that the Lorentz symmetry breaking imposes 
different type of changes on the real part of quasinormal modes depending on the type 
of perturbation for the black hole \eqref{flat_GUP_BH_sol}. For the scalar perturbation, 
with increase in $\lambda$, the real quasinormal frequency decreases. On the hand, in the cases of gravitational perturbation and electromagnetic perturbation, the real 
quasinormal frequency increases very slowly with increase in $\lambda$. The 
decay rate of 
oscillations decrease with increase in the parameter $\lambda$ for all the three 
perturbations. So presence of Lorentz symmetry breaking allows the quasinormal 
frequencies to propagate further by decreasing the decay rate. The global monopole 
term $\mu$ shows a different behaviour on the quasinormal modes 
(see Fig.\ \ref{fig_qnm04}). With increase in $\mu$, quasinormal frequencies decrease 
for all the three perturbations and they finally merge when $\mu$ approaches $1$. 
Similar observation is made for the decay rate of the oscillations also. The decay 
rate decreases and approaches to zero for all the three perturbation cases. Note that 
these observations are done for the black hole metric \eqref{flat_GUP_BH_sol}.

In Fig.\ \ref{fig_qnm05}, we have plotted the real and imaginary quasinormal 
frequencies for the GUP corrected de Sitter black hole with global monopole with 
respect to the GUP parameter $\alpha$. In this case we have considered the values of 
effective cosmological constant $\Lambda_{eff}$ allowed by the constraint range 
provided by the equation \eqref{Lambda_bound}. It is observed that in case of the de 
Sitter black hole, the quasinormal frequencies for scalar and electromagnetic 
perturbations are close to each other. The real quasinormal frequencies as well as 
the imaginary parts increase with increase in the GUP parameter $\alpha$. However in 
comparison to the case for black hole \eqref{flat_GUP_BH_sol}, we observe more 
variation of the quasinormal frequencies with respect to $\alpha$. In case of the 
decay rate or the imaginary quasinormal modes, we see that for $\alpha=0$, the decay 
rates for all the perturbations are almost equal and with increase in $\alpha$, the 
decay rates changes differently resulting maximum decay rate for the case of scalar 
perturbations and minimum for the case of gravitational perturbations. On the other 
hand, in Fig.\ \ref{fig_qnm06} for the de Sitter black hole, both real and  imaginary 
quasinormal frequencies show a different variation pattern with respect 
to $\beta$ in comparison with Fig.\ \ref{fig_qnm02}. The real quasinormal frequencies 
for both scalar and electromagnetic perturbation are very close to each other. The 
decay rates differ very slightly for all the three perturbations and they decrease 
and tend to merge with an increase in the parameter $\beta$. The Lorentz violation 
also imposes a different variation pattern on the quasinormal frequencies in case of 
de Sitter black hole. In Fig.\ \ref{fig_qnm07}, one can see that the real quasinormal 
frequencies for both scalar and electromagnetic perturbations approach each other 
with an increase in the parameter $\lambda$ and towards the higher values 
of $\lambda$, they are almost indistinguishable. However, the real quasinormal 
frequencies for the case of gravitational perturbation are smaller and follows a 
similar trend. The decay rate for all the three perturbations, in case of the de 
Sitter black hole, decreases with increase in the Lorentz violation and becomes 
almost indistinguishable in higher values of $\lambda$. From Fig.\ \ref{fig_qnm08}, it 
is found that the variation of real and imaginary quasinormal frequencies for scalar, 
electromagnetic and gravitational perturbations with respect to $\mu$ also follow a 
similar trend but in both graphs, beyond $\mu=0.7$, the results for all the 
perturbations are identical and show a different behaviour.

In case of the anti-de Sitter black hole, from Fig.s \ref{fig_qnm09}, \ref{fig_qnm10} 
and \ref{fig_qnm11} one can see that the variation trend is similar to the previous 
cases for the black hole \eqref{flat_GUP_BH_sol}, but the differences of the 
quasinormal modes obtained from different perturbations are comparatively large. 
However, variation of quasinormal modes with respect to the parameter $\mu$ shows 
that the oscillation frequencies and decay rates for different perturbations may 
become identical or very close at higher values of $\mu$ as seen from Fig.\
\ref{fig_qnm12}. 

Finally, the effect of effective cosmological constant on the quasinormal modes is 
shown in Fig\ \ref{fig_qnm13}. It is seen that in the de Sitter regime, with increase 
in the effective de Sitter curvature, real quasinormal frequencies for scalar, 
electromagnetic and axial gravitational perturbations decrease and approach 
towards $0$ as $\Lambda_{eff} \rightarrow 0.08$ and beyond this, quasinormal 
frequencies start to increase again slowly. Similarly, the decay rate 
decreases for all the three perturbations upto $\Lambda_{eff}=0.08$ and beyond 
this, decay rate increases drastically. Important point to note here is that 
upto $\Lambda_{eff}=0.08$, the decay rates for all the three perturbations are 
very close to each other while, beyond this point the axial gravitational 
perturbation shows a smaller decay rate and the scalar and electromagnetic 
perturbations show identical decay rates of oscillations. On the other hand, 
for the anti-de Sitter background curvature, the real quasinormal modes are 
distinguishable and they decrease gradually with increase in the value of 
$\Lambda_{eff}$ towards $0$. In this regime, the decay rate decreases 
towards $\Lambda_{eff}=0.$ It is noticed that the decay rate for the 
electromagnetic perturbation is comparatively higher and scalar perturbation 
gives smaller decay rate for smaller values of $\Lambda_{eff}.$ However, when  
$\Lambda_{eff}$ approaches $0$, decay rate for scalar perturbation becomes
 maximum and for gravitational perturbation becomes minimum. So, it is seen 
that the decay rates as well as the quasinormal frequencies are highly 
$\Lambda_{eff}$ dependent and could be useful to comment on the nature of 
$\Lambda_{eff}$ once we have sufficient astrophysical observational 
data on quasinormal modes.

This study shows that the GUP signatures may be found from the quasinormal 
modes for de Sitter, anti-de Sitter or asymptotically flat black holes. 
However, in case of de Sitter black holes, it may be difficult to distinguish 
the quasinormal modes from the scalar perturbation and electromagnetic 
perturbation. Similarly, we see that Lorentz violation and hence the 
signatures of QG may be obtained from the quasinormal modes for all the three 
types of black holes. But, in this case also for the de Sitter black holes, it 
might be difficult to distinguish between scalar and electromagnetic 
quasinormal modes. On the other hand, it is found that the presence of global 
monopoles might make it difficult to distinguish among scalar, electromagnetic 
and gravitational quasinormal modes for both the de Sitter and anti-de Sitter
black holes.

For a clear picture, we compare our results with some previous results in 
Table \ref{compare_table}. The quasinormal modes in GUP corrected 
Schwarzschild black holes in GR have been studied in Ref. \cite{gupbh} for 
scalar perturbation. The quasinormal modes in bumblebee gravity have been 
studied for the first time in Ref. \cite{qnm_bumblebee}. However, in this 
study, no other ingredients apart from Lorentz violation have been used. We 
list some values of quasinormal modes from these two studies in Table 
\ref{compare_table} along with the corresponding quasinormal modes for the 
black hole defined by equation \eqref{flat_GUP_BH_sol}. We see that the 
percentage deviation of quasinormal modes for this new black hole from the 
GUP corrected black hole in GR is maximum $24.5\%$ in the Table 
\ref{compare_table} for $\alpha=0.06$ and $\beta=0.0$. With an increase in 
$\beta$, we notice a decrease in the deviation of quasinormal modes. An 
increase in $\beta$ may tend to nullify the effect of Lorentz violation as we 
have seen previously. Again, we see significant deviations of quasinormal 
modes from that of the general black hole in bumblebee gravity and it clearly 
shows the impacts of the GUP parameters and global monopoles. We have also 
listed the deviations of quasinormal modes from those in GR which show that 
the quasinormal modes deviate significantly in this new black hole solution 
with three ingredients viz., Lorentz violation, global monopole and GUP. Once 
we obtain significant experimental results in the near future, our study might 
help to differentiate such black hole solutions from those in GR. In other 
words, the study might help to have some experimental signatures on 
quasinormal modes for Lorentz violation, global monopole and GUP.

\begin{table}[ht]
\caption{Quasinormal modes of the AdS black hole with $n= 0, \lambda=0.1, \alpha=0.1, \beta = 0.01, M=1, \mu=0.1$ and $\Lambda_{eff} = -\,0.0001$ for the 
scalar perturbation.}
\begin{center}
\begin{small}
\begin{tabular}{|ccc|cc|} 
\hline
\;\;$l$   & WKB                    & \multicolumn{1}{l}{Pad\'e averaged WKB}         & $\vartriangle_{rms}$ & $\Delta_6$         \\ 
\hline
$l=1$ & $0.247217 - 0.0753594 i$   & \multicolumn{1}{l}{$0.247228 - 0.0753061 i$} & $2.12875\times10^{-6}$ &    $0.000013278$    \\
$l=2$ &   $0.411469 - 0.0747242 i$ & \multicolumn{1}{l}{$0.41147 - 0.0747213 i$}   & $2.90155\times10^{-7}$ &    $1.28146\times10^{-6}$ \\
$l=3$ &  $0.575856 - 0.0745578 i$ & \multicolumn{1}{l}{$0.575857 - 0.0745575 i$}   &   $4.82207\times10^{-8}$ & $2.00283\times10^{-7}$ \\
$l=4$ &  $0.740284 - 0.0744899 i$ & \multicolumn{1}{l}{$0.740284 - 0.0744898 i$}    &                                         $1.3031\times10^{-8}$ &  $1.75206\times10^{-8}$ \\ 
\hline
\;\;$l$   & Time domain    & $R^2$  &   & \multicolumn{1}{l}{}  \\ 
\cline{1-3}
$l=1$ & $0.246915373-0.0754907607 i$ & $0.99999254$   &  & \multicolumn{1}{l}{}  \\
$l=2$ & $0.418147972 - 0.0750250597 i$ & $0.9987652$  &       & \multicolumn{1}{l}{}  \\
$l=3$ &  $0.565259100-0.0764900414 i$   &    $0.9935930$    &      & \multicolumn{1}{l}{}  \\
$l=4$ &  $0.744106171-0.0747287532 i$ &  $0.9997818$    &    & \multicolumn{1}{l}{}  \\
\cline{1-3}
\end{tabular}
\end{small}
\end{center}
\label{tab01}
\end{table}

\begin{table}[ht]
\caption{Quasinormal modes of the AdS black hole with $n= 0, \lambda=0.1, \alpha=0.1, \beta = 0.01, M=1, \mu=0.1$ and $\Lambda_{eff} = -\,0.0001$ for the 
electromagnetic perturbation.}
\begin{center}
\begin{small}
\begin{tabular}{|ccc|cc|} 
\hline
\;\;$l$   & WKB                    & \multicolumn{1}{l}{Pad\'e averaged WKB}         & $\vartriangle_{rms}$ & $\Delta_6$         \\ 
\hline
$l=1$ & $0.215849 - 0.0721029 i$   & \multicolumn{1}{l}{$0.215873 - 0.0720365 i$} & $4.21545\times10^{-6}$ &    $4.28102\times10^{-6}$   \\
$l=2$ &   $0.393215 - 0.0736172 i$ & \multicolumn{1}{l}{$0.393216 - 0.0736145 i$}   & $2.67569\times10^{-7}$ &    $2.1195\times10^{-7}$ \\
$l=3$ &  $0.56292 - 0.0740012 i$ & \multicolumn{1}{l}{$0.56292 - 0.0740008 i$}   &   $4.68354\times10^{-8}$ & $4.6607\times10^{-8}$ \\
$l=4$ &  $0.730254 - 0.074155 i$ & \multicolumn{1}{l}{$0.730255 - 0.074155 i$}    &                                         $1.31401\times10^{-8}$ & $1.4897\times10^{-8}$ \\ 
\hline
\;\;$l$   & Time domain    & $R^2$  &   & \multicolumn{1}{l}{}  \\ 
\cline{1-3}
$l=1$ & $0.222487805 - 0.0738152670 i$ & $0.99897467$   &  & \multicolumn{1}{l}{}  \\
$l=2$ & $0.402254608-0.0742019865 i$ & $0.9979984$  &       & \multicolumn{1}{l}{}  \\
$l=3$ &  $0.555350578-0.0749674026 i$   &    $0.9977753$    &      & \multicolumn{1}{l}{}  \\
$l=4$ &  $0.735835101 - 0.0750020432 i$ &  $0.9993621$    &    & \multicolumn{1}{l}{}  \\
\cline{1-3}
\end{tabular}
\end{small}
\end{center}
\label{tab02}
\end{table}

\begin{table}[ht]
\caption{Quasinormal modes of the AdS black hole with $n= 0, \lambda=0.1, \alpha=0.1, \beta = 0.01, M=1, \mu=0.1$ and $\Lambda_{eff} = -\,0.0001$ for the 
gravitational perturbation.}
\begin{center}
\begin{small}
\begin{tabular}{|ccc|cc|} 
\hline
\;\;$l$   & WKB                    & \multicolumn{1}{l}{Pad\'e averaged WKB}         & $\vartriangle_{rms}$ & $\Delta_6$         \\ 
\hline
$l=2$ &   $0.32046 - 0.0694711 i$ & \multicolumn{1}{l}{$0.320458 - 0.0694916 i$}   & $5.73105\times10^{-7}$ & $9.38238\times10^{-6}$ \\
$l=3$ &  $0.513528 - 0.0721067 i$ & \multicolumn{1}{l}{$0.513528 - 0.0721067 i$}   &   $5.01761\times10^{-9}$ & $2.52104\times10^{-8}$\\
$l=4$ &  $0.692601 - 0.073066 i$ & \multicolumn{1}{l}{$0.692601 - 0.0730659 i$}    &                                         $3.41634\times10^{-9}$ & $1.02767\times10^{-8}$ \\ 
\hline
\;\;$l$   & Time domain    & $R^2$  &   & \multicolumn{1}{l}{}  \\ 
\cline{1-3}
$l=2$ & $0.320554335-0.0696256238 i$ & $0.99999977$  &       & \multicolumn{1}{l}{}  \\
$l=3$ &  $0.509424202-0.0721478170 i$   &    $0.9987712$    &      & \multicolumn{1}{l}{}  \\
$l=4$ &  $0.682412482 - 0.0730500202 i$ &  $0.9940553$    &    & \multicolumn{1}{l}{}  \\
\cline{1-3}
\end{tabular}
\end{small}
\end{center}
\label{tab03}
\end{table}

\begin{table}
\caption{Comparison of quasinormal modes of black hole \eqref{flat_GUP_BH_sol} 
with GUP corrected black hole considered in Ref.\ \cite{gupbh} and bumblebee 
black hole obtained in Ref.\ \cite{qnm_bumblebee} for scalar perturbation with 
$n=0$, $l=2$, $M=1$, $\lambda=0.1$ and $\mu=0.1$. Here GUPBH represents the GUP 
corrected black hole in GR obtained in Ref. \cite{gupbh}, BBH represents a
bumblebee black hole obtained in Ref.\ \cite{qnm_bumblebee}, $\triangle_{13}$, 
$\triangle_{23}$ and $\triangle_{GR}$ represent percentage deviations in 
quasinormal modes of black hole defined in \eqref{flat_GUP_BH_sol} from GUPBH, 
BBH and GR cases respectively.} 
\begin{center}
\begin{small}
\begin{tabular}{|c|c|c|c|c|c|c|} 
\hline
$(\alpha, \beta)$ & GUPBH & BBH & Black hole \eqref{flat_GUP_BH_sol}  & $\triangle_{13}$   & $\triangle_{23}$ & $\triangle_{GR}$ \\ 
\hline
    $(0.0, 0.0)$     &  $0.4836\, -0.0968 i$   &       $0.4813\, -0.0966 i$  &  $0.4110\, -0.0746 i$   &  $15.4\%$    &   $15.0\%$  &      $15.4\%$     \\ 
\hline
     $(0.0, 0.03)$    &  $0.4318\, -0.0864 i$   &  $0.4813\, -0.0966 i$   & $0.4095\, -0.0744 i$    &  $5.7\%$    &  $15.3\%$   &    $15.7\%$       \\ 
\hline
 $(0.06, 0.0)$    &   $0.5496\, -0.1100 i$  &     $0.4813\, -0.0966 i$  &  $0.4167\, -0.0757 i$   &  $24.5\%$    &   $13.8\%$  &     $14.2\%$         \\ 
\hline
    $(0.06, 0.02)$    & $0.5038\, -0.1008 i$    &      $0.4813\, -0.0966 i$  &  $0.4157\, -0.0755 i$    &   $17.8\%$   &   $14.0\%$  &   $14.4\%$        \\ 
\hline
\end{tabular}
\end{small}
\end{center}
\label{compare_table}
\end{table}

\section{Evolution of Scalar, Electromagnetic and Gravitational Perturbations on the Black hole geometries}\label{section6}
In this section, we study the evolution of the scalar, electromagnetic and 
gravitational perturbations using the time domain integration method described 
in Ref.\ \cite{gundlach}. Defining $\psi(r_*, t) = \psi(i \Delta r_*, j \Delta t) = \psi_{i,j} $, $V(r(r_*)) = V(r_*,t) = V_{i,j}$, we can express equation 
\eqref{scalar_KG} in the following form:
\begin{equation}
\dfrac{\psi_{i+1,j} - 2\psi_{i,j} + \psi_{i-1,j}}{\Delta r_*^2} - \dfrac{\psi_{i,j+1} - 2\psi_{i,j} + \psi_{i,j-1}}{\Delta t^2} - V_i\psi_{i,j} = 0.
\end{equation}
Now, using initial conditions $\psi(r_*,t) = \exp \left[ -\dfrac{(r_*-k_1)^2}{2\sigma^2}  \right]$ and $\psi(r_*,t)\vert_{t<0} = 0$ (here $k_1$ and $\sigma$ are 
median and width of the initial wave-packet), time evolution of the scalar field can 
be expressed as
\begin{equation}
\psi_{i,j+1} = -\,\psi_{i, j-1} + \left( \dfrac{\Delta t}{\Delta r_*} \right)^2 (\psi_{i+1, j + \psi_{i-1, j}}) + \left( 2-2\left( \dfrac{\Delta t}{\Delta r_*} \right)^2 - V_i \Delta t^2 \right) \psi_{i,j}.
\end{equation}
Here during the numerical procedure we have kept $\frac{\Delta t}{\Delta r_*} < 1$ 
in order to satisfy the Von Neumann stability condition.
Using the same procedure for electromagnetic perturbation and gravitational 
perturbation we have calculated the corresponding time profiles. 

\begin{figure}[htbp]
\centerline{
   \includegraphics[scale = 0.3]{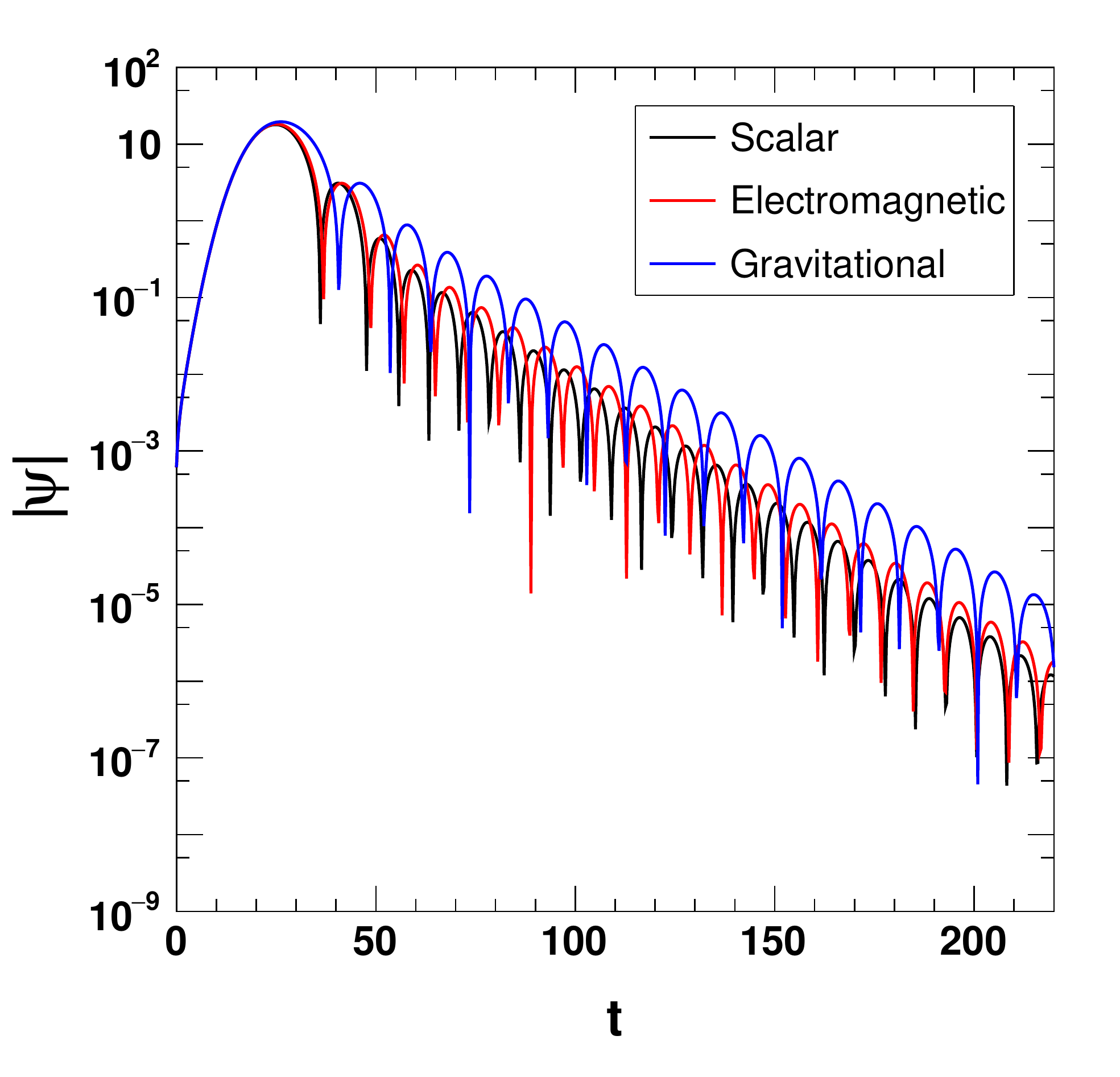}\hspace{0.8cm}
   \includegraphics[scale = 0.3]{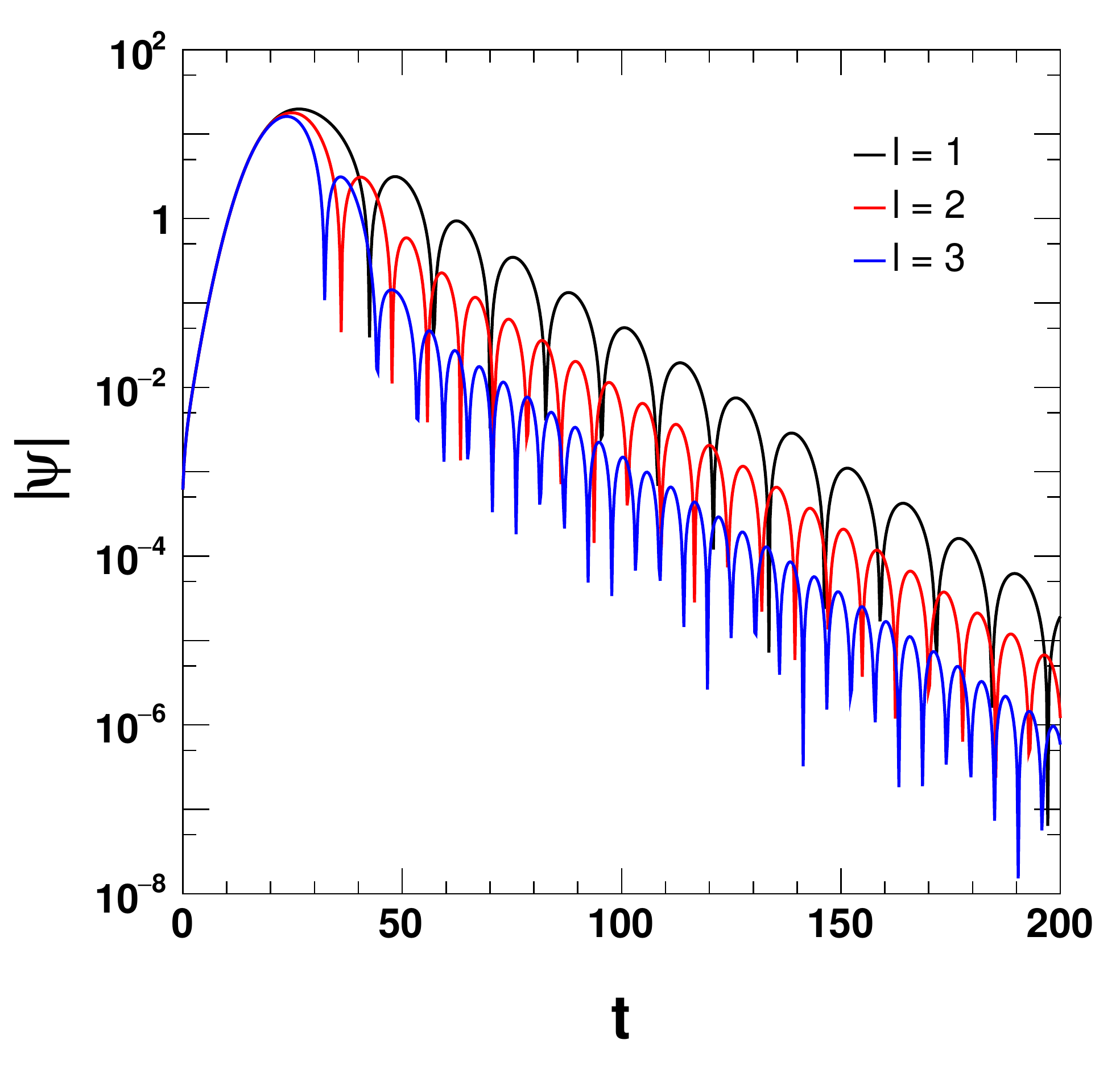}}
\vspace{-0.2cm}
\caption{Time domain profile with $n= 0, \lambda=0.1, \alpha=0.1, \beta = 0.01, M=1, \mu=0.1$ and $\Lambda_{eff} = -\,0.0001$. For the plot on right, we have 
used only the scalar perturbation with different $l$ values and for the plot 
on left we have used only $l=2$ for all three perturbations.}
\label{time01}
\end{figure}

\begin{figure}[htbp]
\centerline{
   \includegraphics[scale = 0.3]{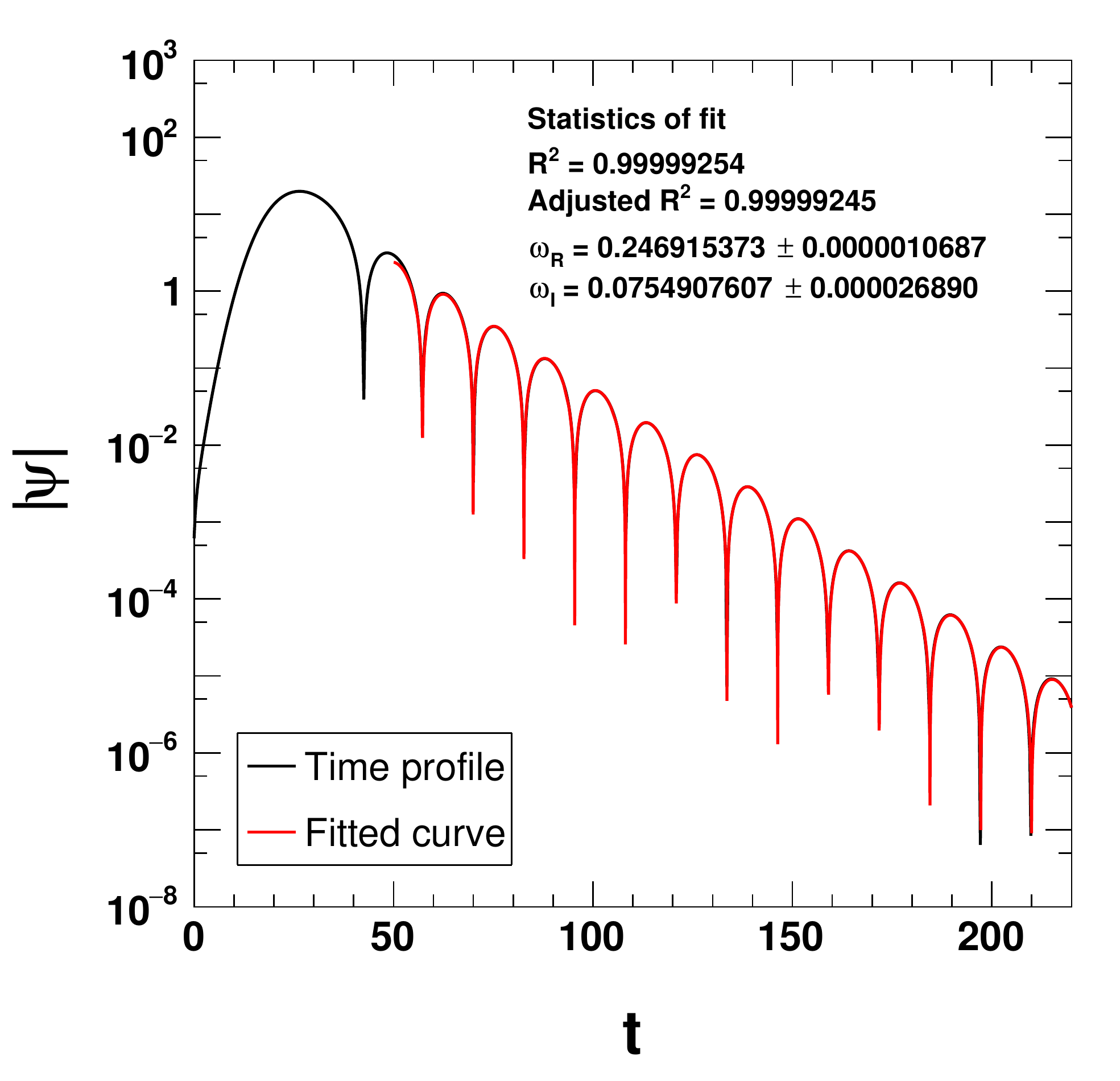}}
\vspace{-0.2cm}
\caption{Estimating quasinormal modes by fitting the time domain profile with $n= 0, l=1, \lambda=0.1, \alpha=0.1, \beta = 0.01, M=1, \mu=0.1$ and $\Lambda_{eff} = -\,0.0001$ for scalar perturbation.}
\label{time02}
\end{figure}

In Fig.\ \ref{time01}, in the plot on the left, we have shown the time 
profiles for scalar, electromagnetic and gravitational perturbations 
for $n= 0, l=2, \lambda=0.1, \alpha=0.1, \beta = 0.01, M=1, \mu=0.1$ and $\Lambda_{eff} = -0.0001$. 
It is seen that the oscillation frequency of gravitational perturbation is 
lower than the other two perturbations supporting the previously obtained 
results. In the plot on the right, we have shown the dependency of multipole 
moment $l$ on the quasinormal modes for scalar perturbation. The real 
quasinormal frequencies increase with increase in the multipole moment $l$. 
However, the multipole moment has a very small impact on the decay rate of the 
oscillations. These results agree well with the Table \ref{tab01} where we 
have calculated the quasinormal modes using Pad\'e averaged $6$th order WKB 
approximation method. Further, we have used the time domain profiles to 
calculate the quasinormal modes using the Levenberg Marquardt algorithm 
\cite{levenberg1944, marq1963, berti2007}. In Fig.\ \ref{time02}, we have 
shown the estimation of quasinormal modes by fitting the time domain profile 
with $n=0, l=1, \lambda = 0.1, \alpha =0.1, \beta = 0.01, M=1, \mu=0.1$ and 
$\Lambda_{eff} = -\,0.0001$ for the scalar perturbation. In Tables \ref{tab01}, \ref{tab02} and \ref{tab03}, we have shown the quasinormal modes obtained from the time domain analysis along with those obtained from 6th order WKB and 6th order Pad\'e averaged WKB methods. We see that the quasinormal frequencies
obtained from the time domain analysis show a good agreement with those obtained from the
WKB approximation method.

\section{Detection Possibilities of Quasinormal modes}\label{section6.5}
 In this section we discuss the possibility of detection of quasinormal 
modes from such black holes in brief. Following Ref.\ \cite{Ferrari2008}, we 
assume the mass of the black hole $M_{\bigstar} = \xi M_{\odot}$, where 
$\xi$ is a positive constant, $M_{\odot} = 1.48\times 10^5 cm$ and 
$M_{\bigstar}$ is equal to $M_{AdS(gup)}$, or $M_{dS(gup)}$, or $M_{gup}$ 
depending upon the value of the cosmological constant. In physical units, the 
quasinormal frequency and the decay time are expressed as \cite{Ferrari2008}:
\begin{align}
f &= \dfrac{c \; M_{\bigstar}\; \omega_R}{2 \pi \xi M_{\odot}} \;\;\;kHz,\\
&= \dfrac{32.26 \times M_{\bigstar}\; \omega_R}{\xi}  \;\;\;kHz,
\end{align}
and
\begin{align}
\tau &= \dfrac{\xi M_{\odot} }{M_{\bigstar} \;\omega_i \; c}  \;\;\;s,\\
&= \dfrac{\xi\times0.4937\times10^{-5}}{M_{\bigstar}\; \omega_i}  \;\;\;s.
\end{align}
These expressions can allow us to check the possibility of detection of GW 
signals coming from an perturbed black hole by GW detectors provided we know 
the sensitive frequency range of the detectors. For the ground based 
interferometers, like LIGO-Virgo GW detectors the sensitive 
frequency range is $f \in [12 \; Hz, 1.2 \; kHz]$. Using this range for the 
quasinormal modes from Tables \ref{tab01}, \ref{tab02} and \ref{tab03} for 
$l=2$ calculated using Pad\'e averaged WKB method, we have for the scalar 
perturbation,
\begin{equation}
11.05 M_{\odot} \lesssim M_{AdS(gup)} \lesssim 1105.14 M_{\odot},
\end{equation}
for the electromagnetic perturbation,
\begin{equation}
10.56 M_{\odot} \lesssim M_{AdS(gup)} \lesssim 1056.11 M_{\odot},
\end{equation}
and for the gravitational perturbation,
\begin{equation}
8.61 M_{\odot} \lesssim M_{AdS(gup)} \lesssim 860.69 M_{\odot}.
\end{equation}
Again, the LISA sensitivity range is from $0.1$ mHz to $1$ Hz 
\cite{Yamamoto22}. So, in case of LISA, for the scalar perturbation
\begin{equation}
13261.6 M_{\odot} \lesssim M_{AdS(gup)} \lesssim 1.33\times 10^8 M_{\odot},
\end{equation}
for the electromagnetic perturbation,
\begin{equation}
12673.3 M_{\odot} \lesssim M_{AdS(gup)} \lesssim 1.27\times 10^8 M_{\odot},
\end{equation}
and finally for the gravitational perturbation,
\begin{equation}
10328.3 M_{\odot} \lesssim M_{AdS(gup)} \lesssim 1.03\times10^8 M_{\odot}.
\end{equation}
If we consider the quasinormal modes from the oscillations of the black hole 
at the center of our Galaxy i.e.\ Sagittarius $A^\ast$, $M = (3.7\pm0.2)\times 10^6 M_{\odot}$ \cite{ghez_sgr}. 
It suggests that for all the three perturbations i.e.\ scalar, electromagnetic 
and gravitational perturbations, LIGO-Virgo can't detect the quasinormal 
frequencies from Sagittarius $A^\ast$. However, we see that the quasinormal 
frequencies from Sagittarius $A^\ast$ fall within the detection range of LISA. 
Hence, in the near future it might be possible to comment on the Lorentz 
violation, global monopole and GUP parameters, and would be possible 
constrain them using the quasinormal modes from a black hole.

\section{Sparsity of Hawking emission}\label{section7}
For a black hole at temperature $T_{bh}$ emitting Hawking radiation with 
frequency $\omega$ in the momentum interval $d^3\mathbf{k}$, the energy 
emitted per unit time or the total power of Hawking radiation is given 
by \cite{miao2017, gray2016}
\begin{equation}
\frac{dE(\omega)}{dt}\equiv P_{tot} = \sum_l \mathcal{T}_l (\omega)\frac{\omega}{e^{\omega/T_{bh}}-1} \hat{k} \cdot  \hat{n}~  \frac{d^3 k ~dA}{(2\pi)^3},
\end{equation}
where $\hat{n}$ is the unit vector normal to the surface element $dA$ and 
$\mathcal{T}_l(\omega)$ is the greybody factor. For massless particles 
$\left|\mathbf{k}\right|=\omega$ and hence the total power of Hawking 
radiation can be rewritten as
\begin{equation}\label{eq_Ptot}
P_{tot}=\sum_l \int_{0}^{\infty} P_l\left(\omega\right) d\omega,
\end{equation}
where
\begin{equation}\label{eq_Pl}
P_l\left(\omega\right)=\frac{A}{8\pi^2} \mathcal{T}_l(\omega)\frac{\omega^3}{e^{\omega/T_{bh}}-1} 
\end{equation}
is the power emitted per unit frequency in the $l^{th}$ mode. Here the area 
$A$ is a multiple of the horizon area. Although for Schwarzschild black hole, 
$A$ is taken to be $(27/4 )$ times the horizon area, we shall consider 
$A = A_h$ i.e.\ the black hole event horizon area as this consideration will 
not impact the qualitative result of the study \cite{miao2017}.

The excitation of massless uncharged scalar fields around the bumblebee black 
hole is governed by the Klein-Gordon equation defined in equation \eqref{scalar_KG}. A portion of the radiation being emitted from the black hole is reflected back 
by the effective potential of the perturbation while the remaining portion is transmitted out. The greybody
factor gives a measurement to the transmission probability of the outgoing Hawking quanta to 
reach the infinity without being back-scattered by this effective 
potential.

\subsection{Bounds on the greybody factor}
\begin{figure}[!pht]
\centerline{
   \includegraphics[scale = 0.28]{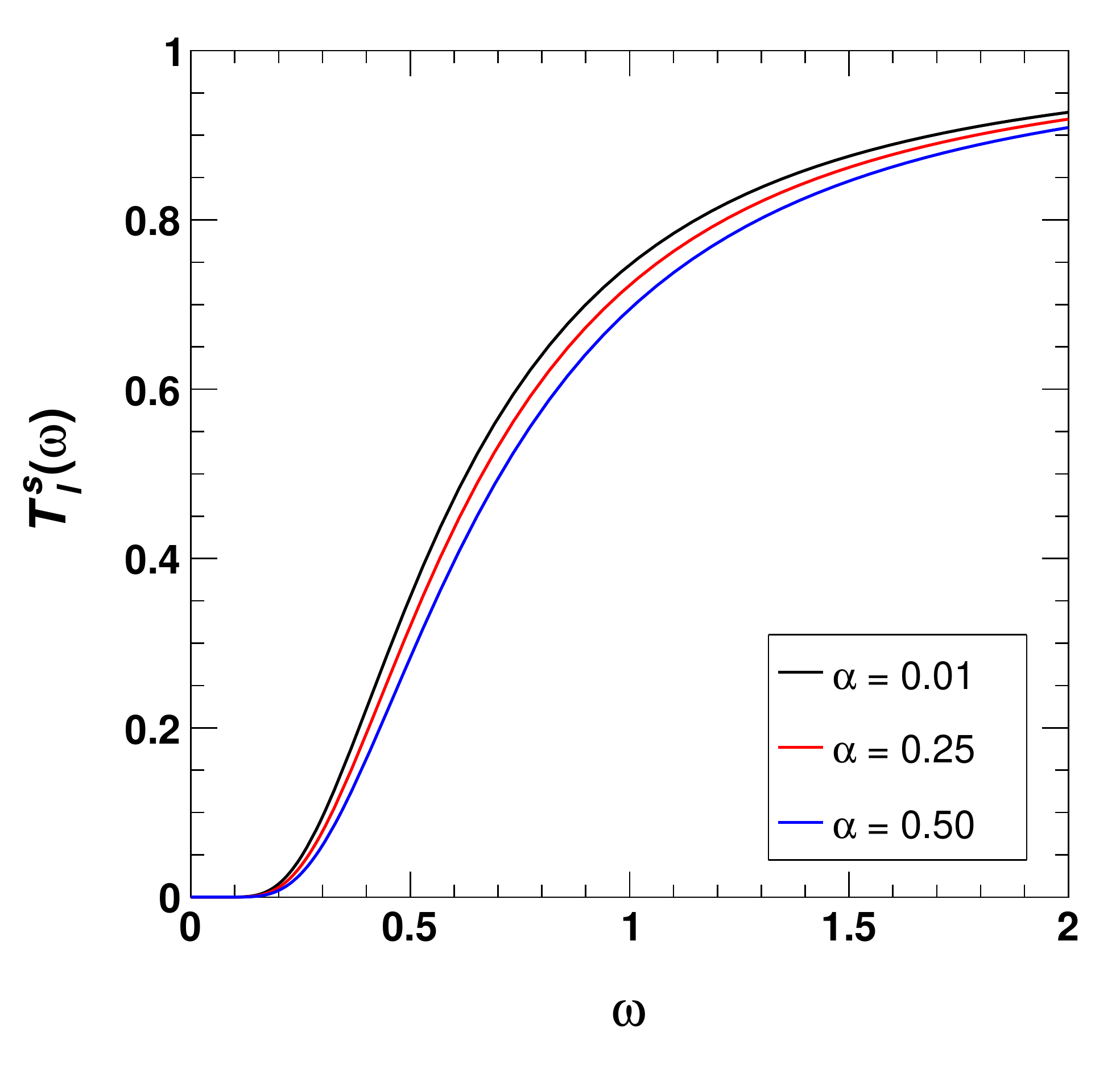}\hspace{1cm}
   \includegraphics[scale = 0.28]{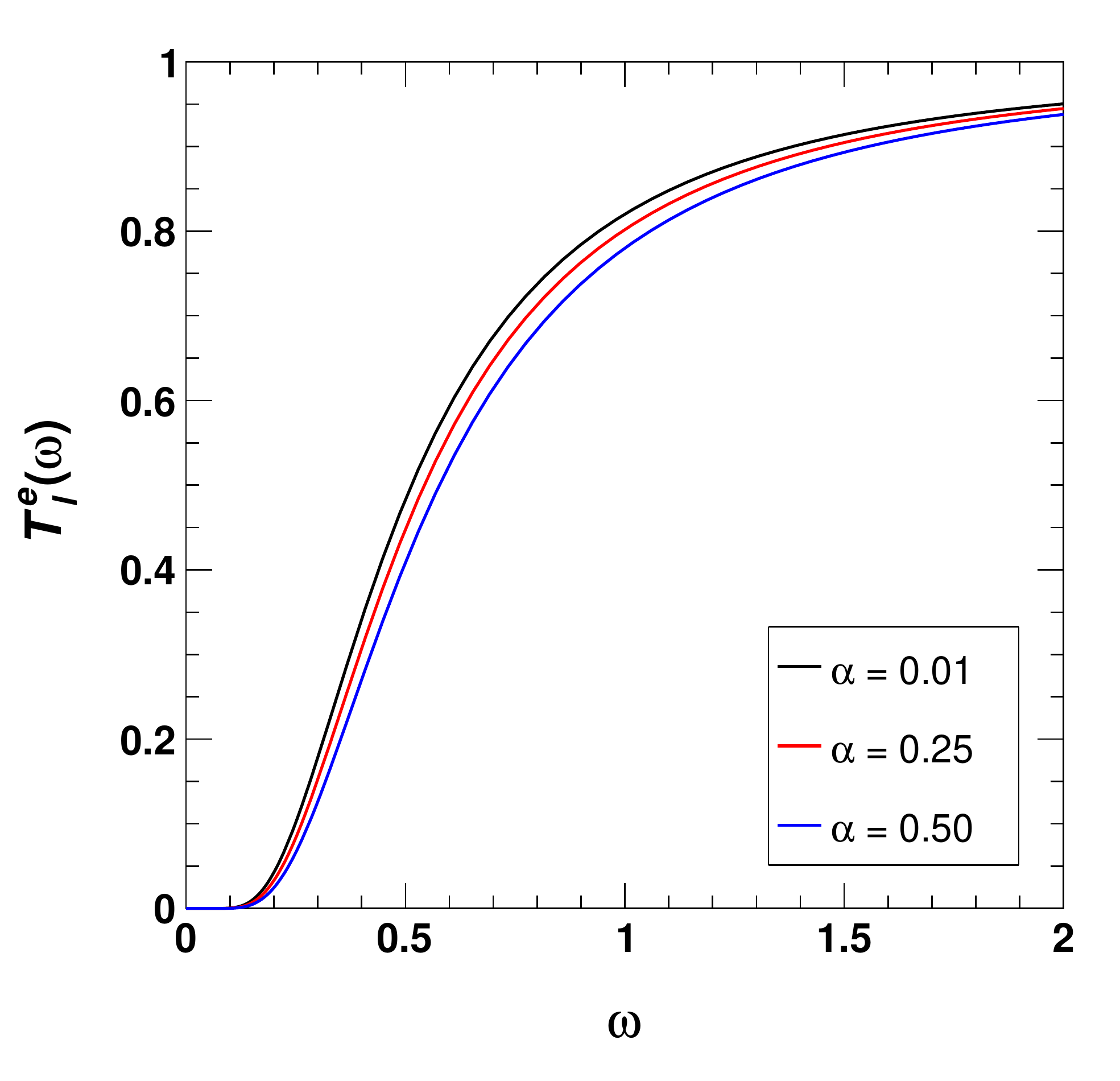}}
   \centerline{
   \includegraphics[scale = 0.28]{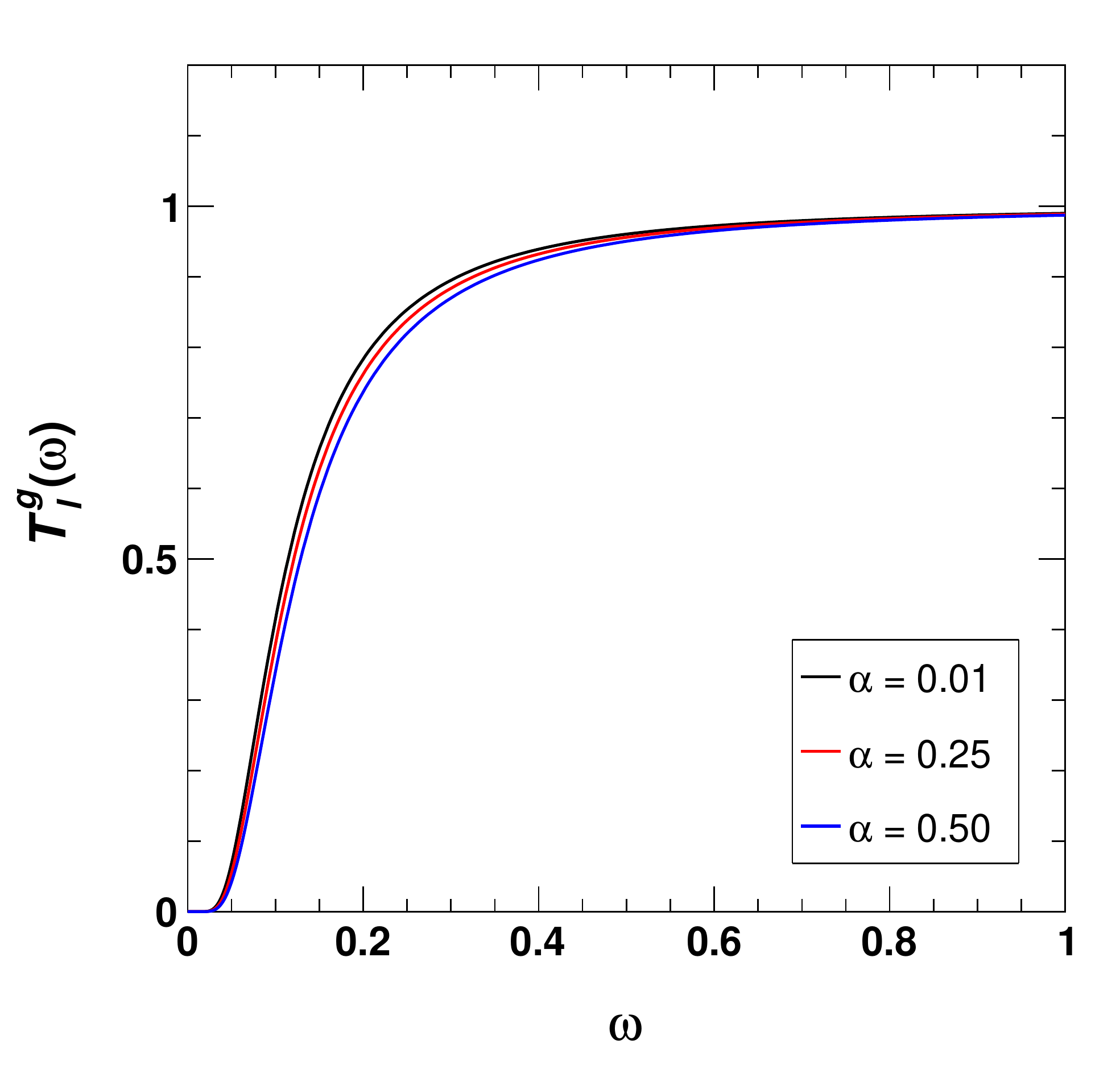}}
\vspace{-0.2cm}
\caption{Greybody factors of the black hole with $\lambda=0.01, \beta = 0.01, M=1, \mu=0.1$ and $\Lambda_{eff} = 0.0$ for different values of $\alpha$.}
\label{grey01}
\end{figure}

\begin{figure}[!phb]
\centerline{
   \includegraphics[scale = 0.28]{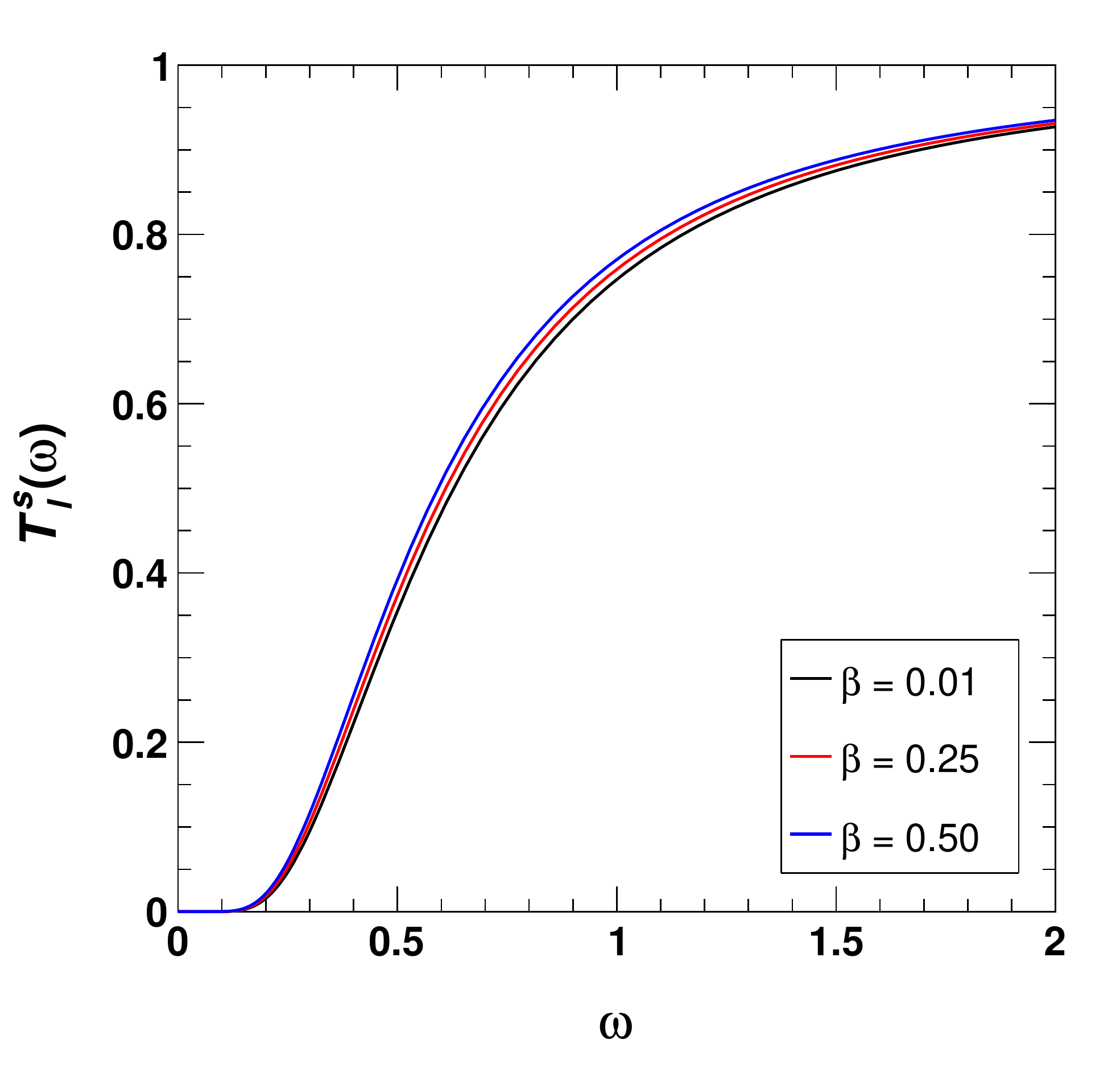}\hspace{1cm}
   \includegraphics[scale = 0.28]{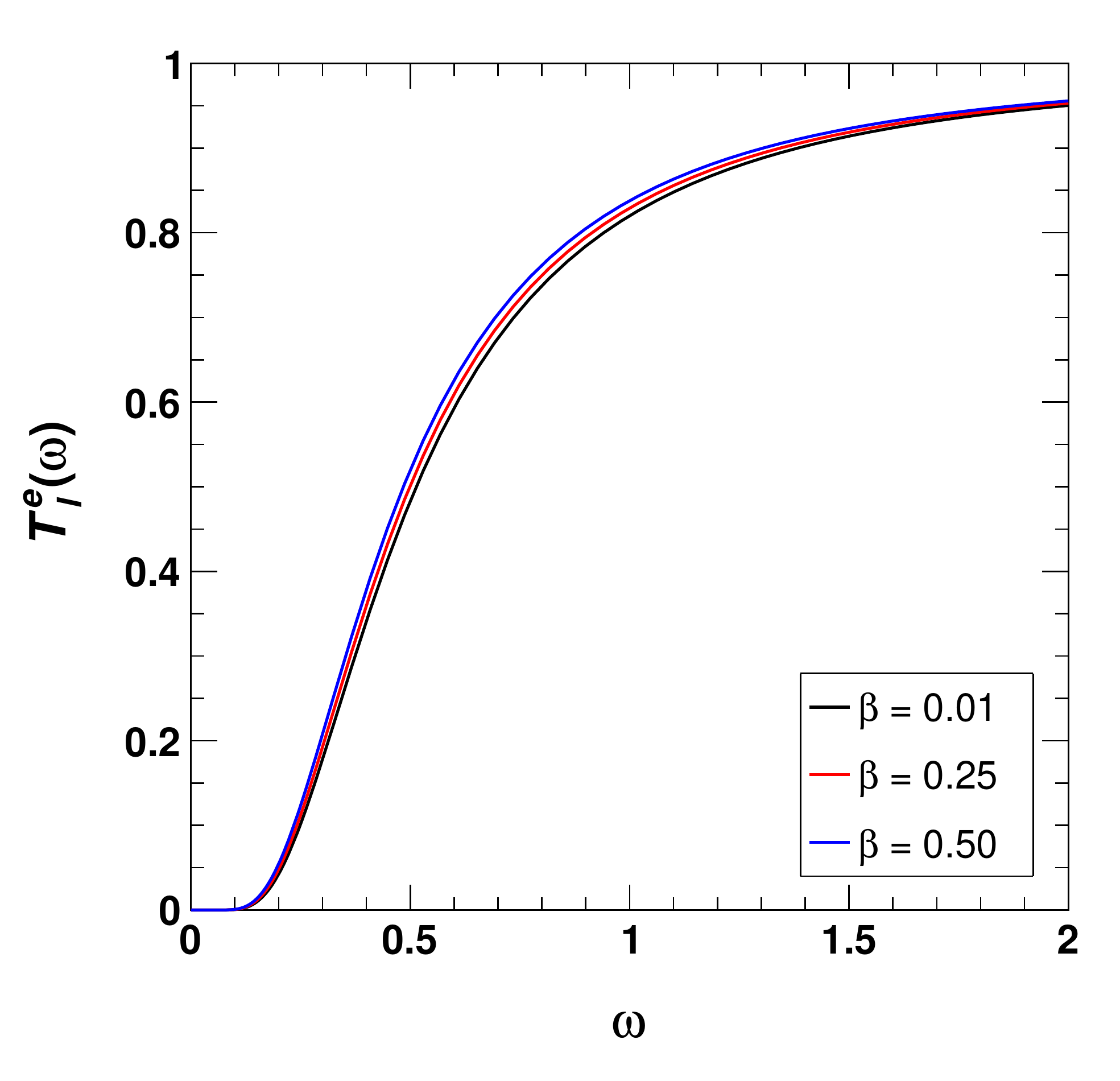}}
   \centerline{
   \includegraphics[scale = 0.28]{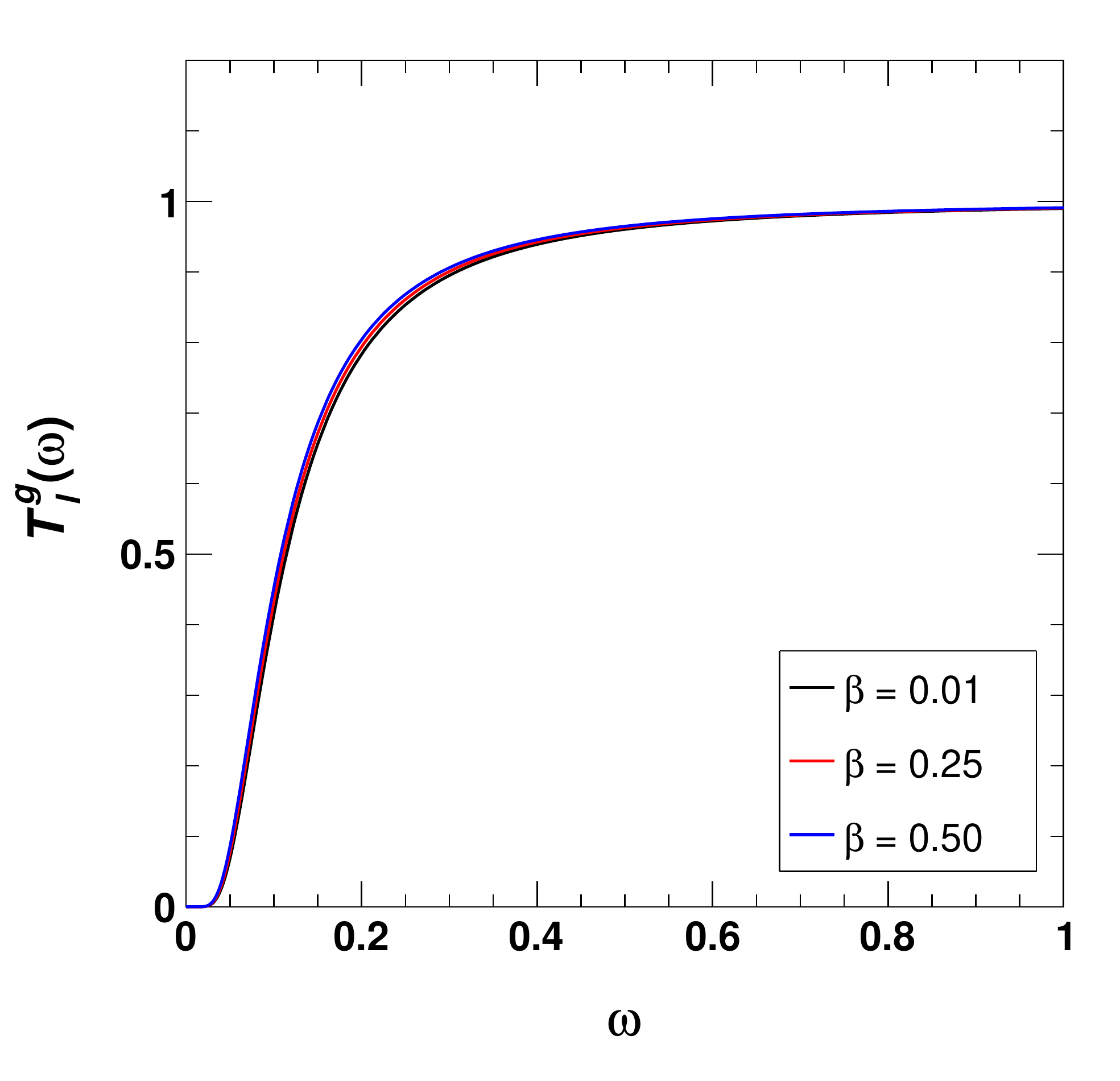}}
\vspace{-0.2cm}
\caption{Greybody factors of the black hole with $\lambda=0.01, \alpha = 0.01, M=1, \mu=0.1$ and $\Lambda_{eff} = 0.0$ for different values of $\beta$.}
\label{grey02}
\end{figure}

\begin{figure}[!pht]
\centerline{
   \includegraphics[scale = 0.28]{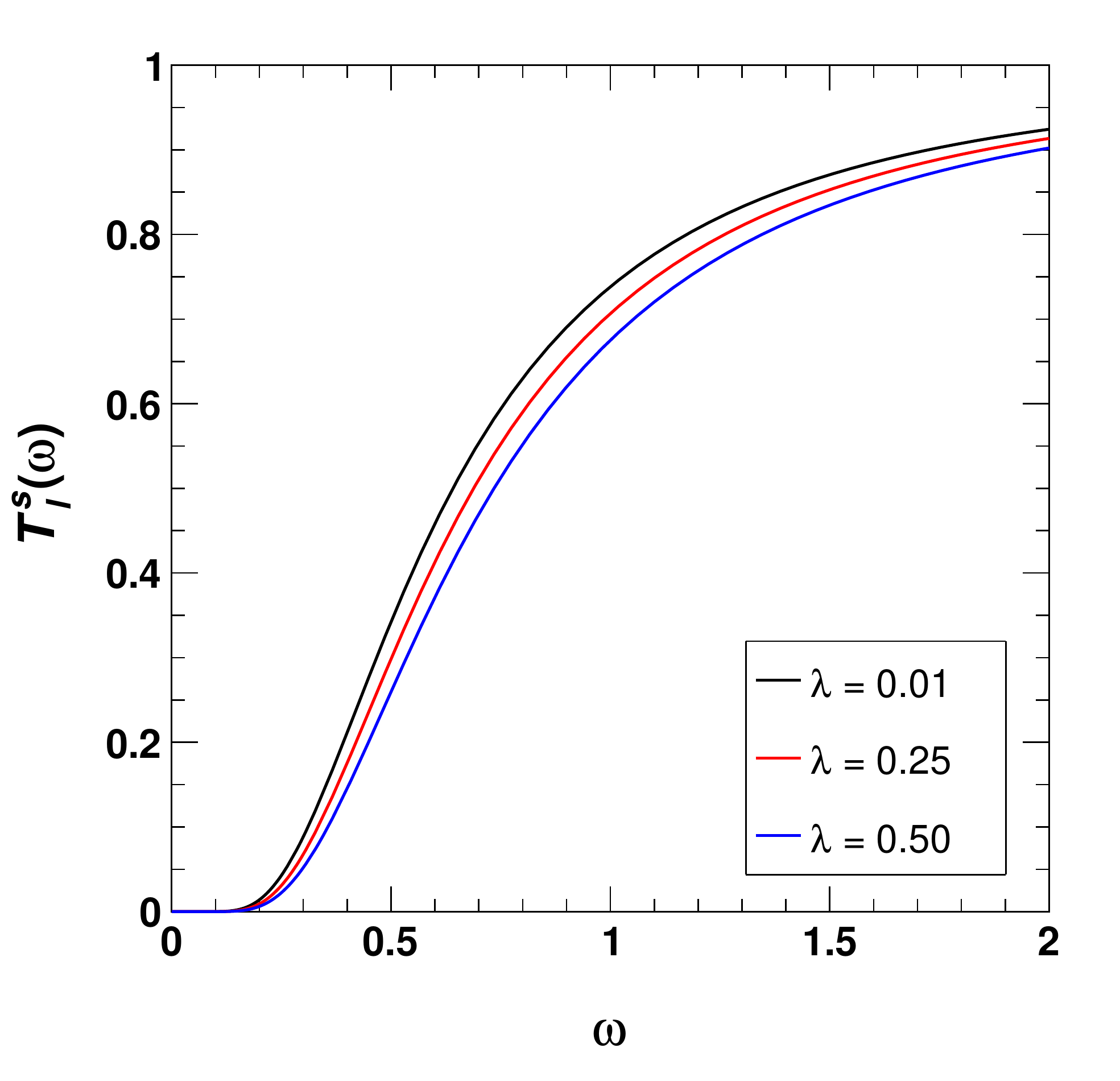}\hspace{1cm}
   \includegraphics[scale = 0.28]{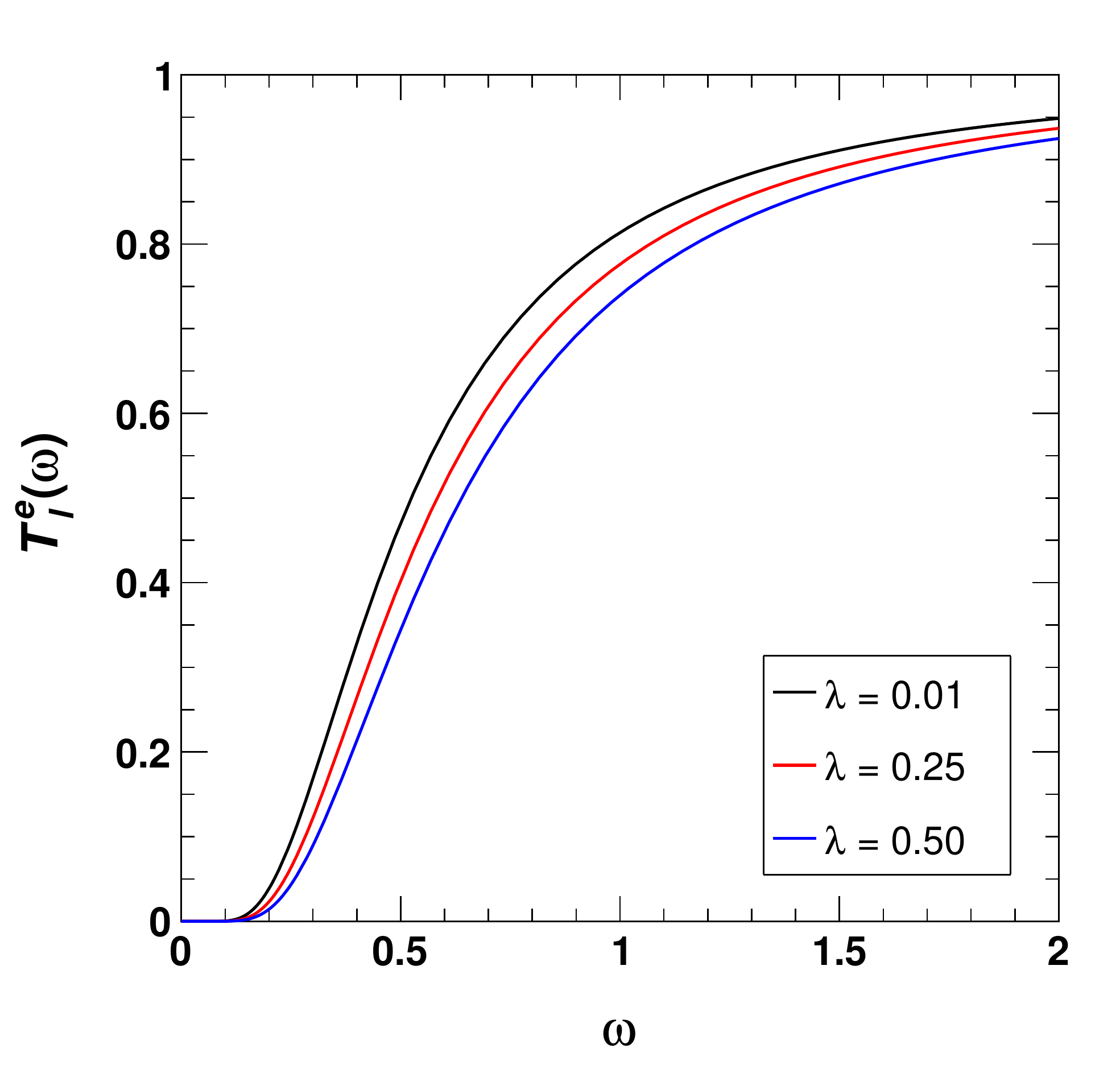}}
   \centerline{
   \includegraphics[scale = 0.28]{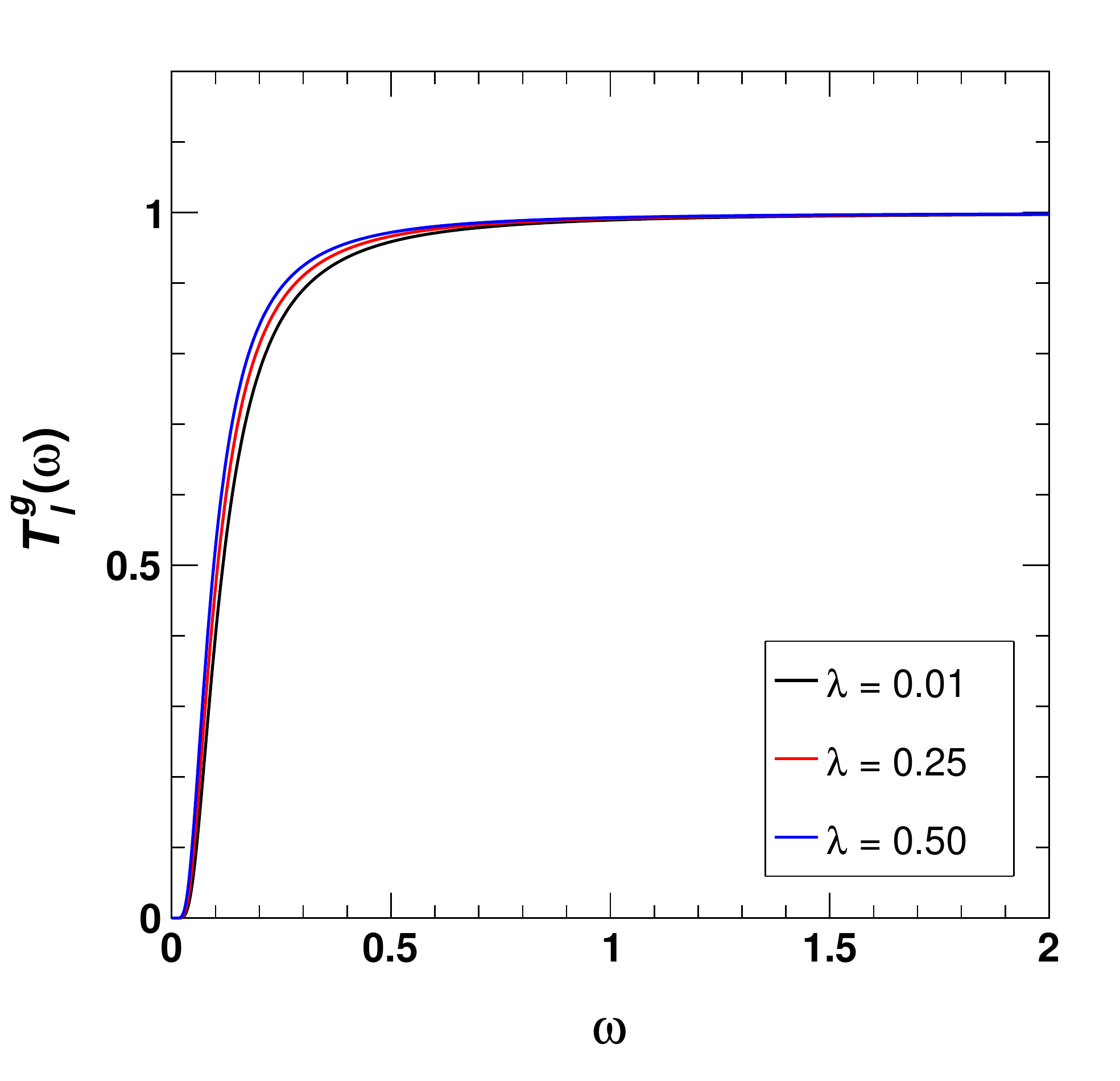}}
\vspace{-0.2cm}
\caption{Greybody factors of the black hole with $\alpha=0.1, \beta = 0.01, M=1, \mu=0.1$ and $\Lambda_{eff} = 0.0$ for different values of $\lambda$.}
\label{grey03}
\end{figure}

\begin{figure}[!phb]
\centerline{
   \includegraphics[scale = 0.28]{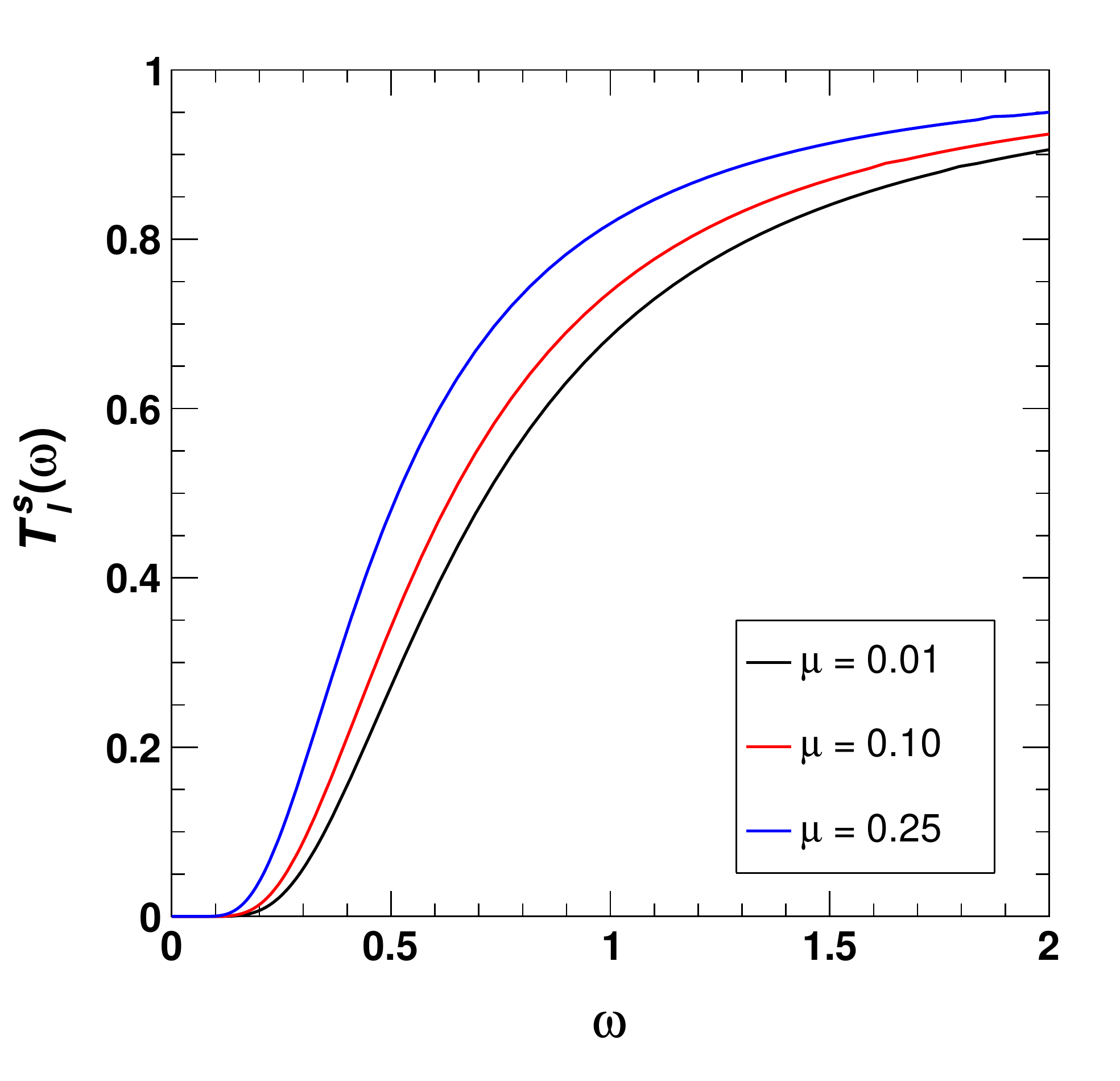}\hspace{1cm}
   \includegraphics[scale = 0.28]{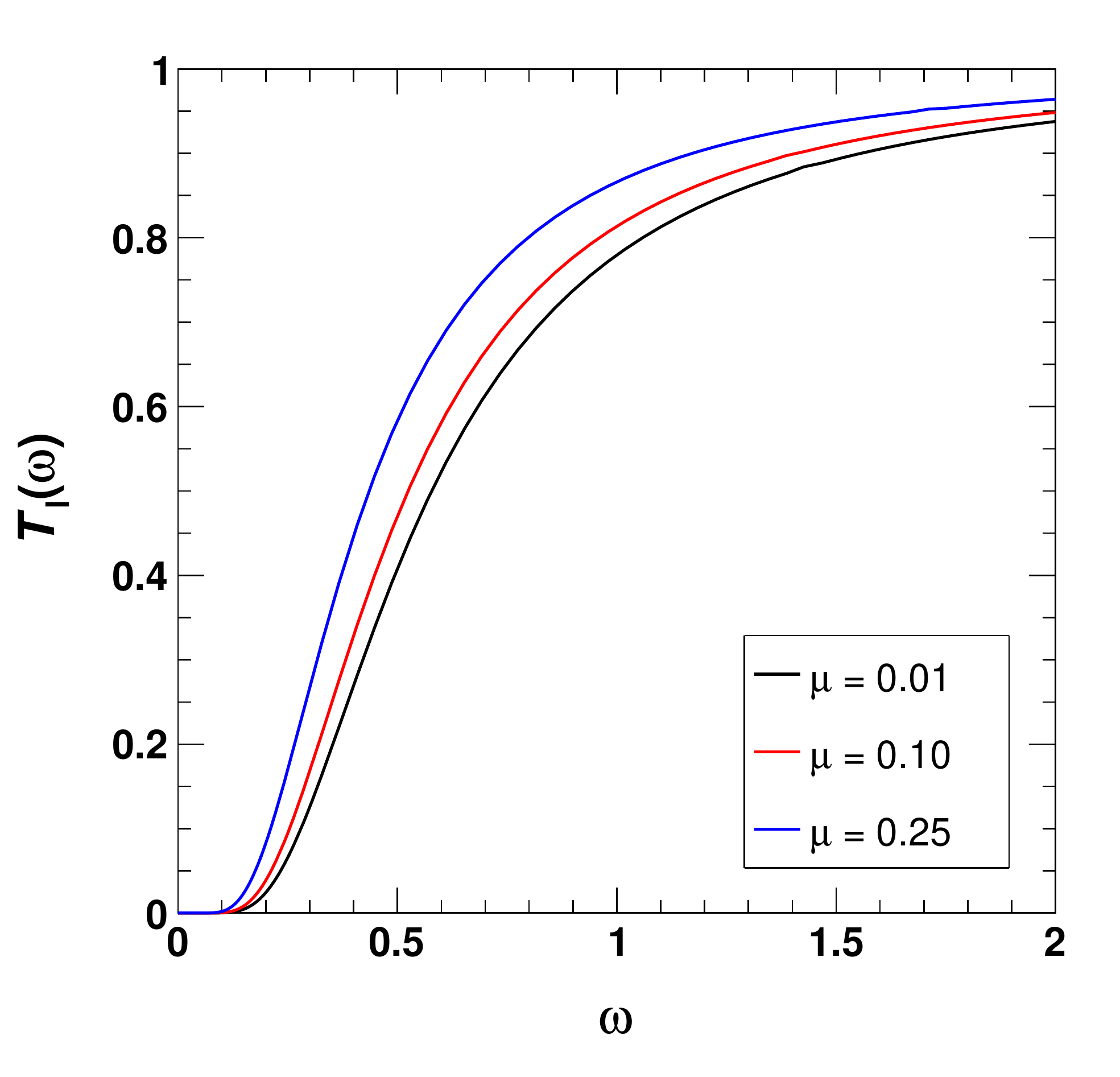}}
   \centerline{
   \includegraphics[scale = 0.28]{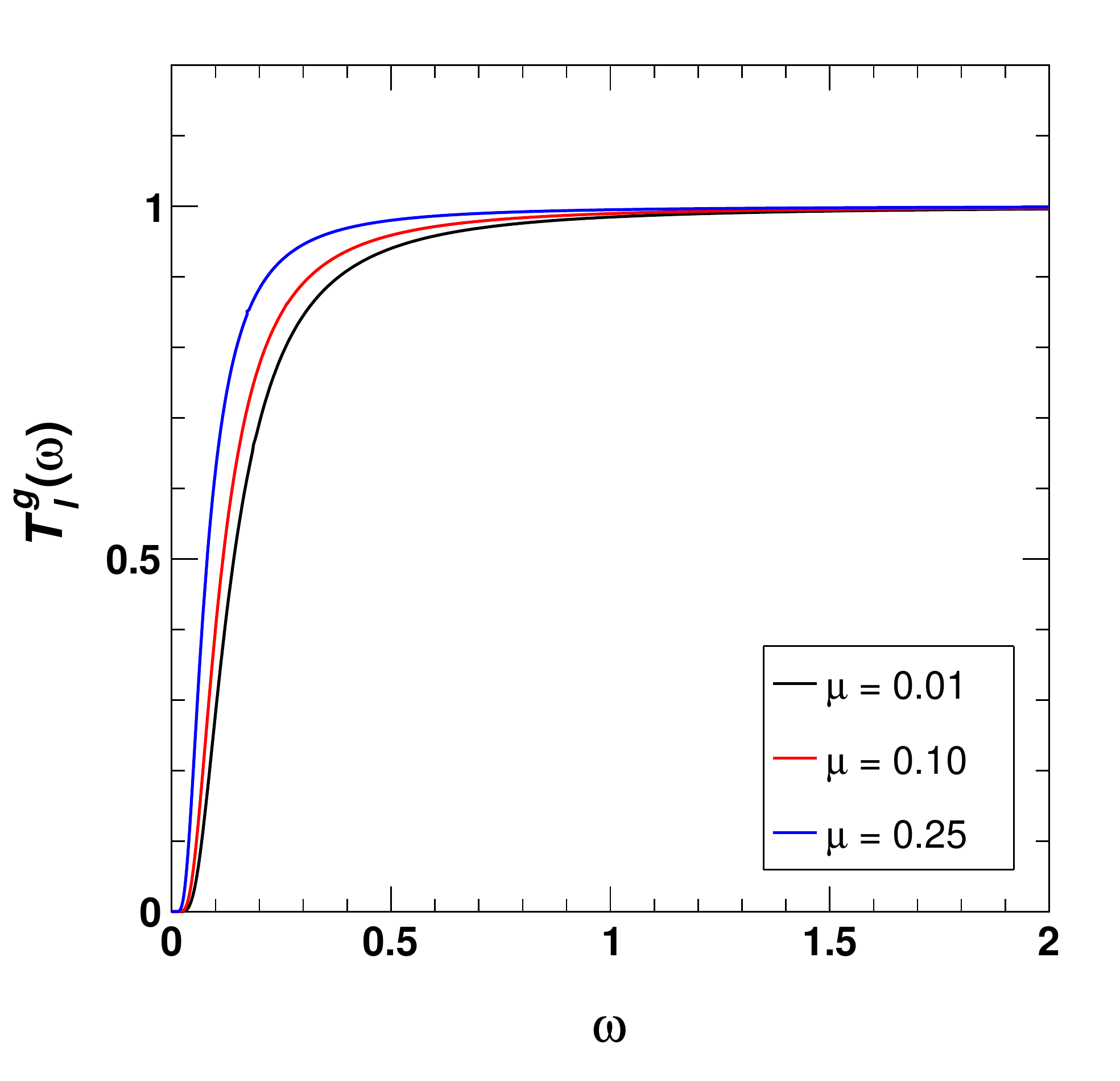}}
\vspace{-0.2cm}
\caption{Greybody factors of the black hole with $\alpha=0.1, \beta = 0.01, M=1, \lambda=0.01$ and $\Lambda_{eff} = 0.0$ for different values of $\mu$.}
\label{grey04}
\end{figure}

Although there are many methods to obtain the greybody factor of a black hole, 
in this work we follow Refs.\ \cite{visser1999, boon2008, boon2008_2, boon2014, chowdhury2020} to obtain the greybody factor of the black hole defined by 
the metric \eqref{flat_GUP_BH_sol}. The general bound on the greybody factor \cite{visser1999} is given by
\begin{equation}\label{eq_bound}
\mathcal{T}_l(\omega)\geq sech^2\left\{ \int_{-\infty}^{\infty} \vartheta dr_* \right \},
\end{equation}
where \begin{equation}\label{eq_bound1}
\vartheta=\frac{\sqrt{\left[h'(r)\right]^2+\left[ \omega^2-V_{eff}-h(r)^2 \right]^2}}{2h(r)}.
\end{equation}
The arbitrary function $h(r)$ has to be positive definite everywhere and 
satisfy the boundary condition, $h(\infty)=h(r_+)=\omega$ for the 
bound~\eqref{eq_bound} to hold. However, as a particularly simple choice of 
$h(r)$, we consider for the present case that 
\begin{equation}\label{eq_h_unchrgd}
h(r)=\omega.
\end{equation}
Substituting equation \eqref{eq_h_unchrgd} in equation \eqref{eq_bound1} and 
using the definition of $r_*$, we get
\begin{equation}\label{eq_bound2}
\int_{-\infty}^{\infty} \vartheta dr_*=\int_{r_+}^{\infty} \frac{V_{eff}}{2 \omega f(r)}  dr.
\end{equation}
Equation \eqref{eq_bound} in conjunction with equations \eqref{Vs}, 
\eqref{Ve} \eqref{Vg} and \eqref{eq_bound2} yields relatively simple 
expressions for the lower bound of the greybody factor as
\begin{equation}\label{eq_tl_s}
\mathcal{T}^{s}_l(\omega)\geq sech^2\left\{\frac{(1-\mu) M \left(2 (\lambda +1)l(l+1)+1-\mu \right)}{\omega \sqrt{(1+\lambda )} \left(\beta(1-  \mu ^2) +8 M^2+2 \alpha  (\mu -1) M\right)} \right\},
\end{equation}
\begin{equation}\label{eq_tl_em}
\mathcal{T}^{e}_l(\omega)\geq sech^2\left\{\frac{2 \sqrt{\lambda +1} \;l (l+1) (1-\mu) M}{\omega  \left(\beta(1-  \mu ^2)+8 M^2+2 \alpha  (\mu -1) M\right)} \right\},
\end{equation}
\begin{equation}\label{eq_tl_g}
\mathcal{T}^{g}_l(\omega)\geq sech^2\left\{\frac{(1-\mu) M (-4 \lambda +2 (\lambda +1) l (l+1)-\mu -3)}{\sqrt{(\lambda +1)} \;\omega  \left(\beta(1-  \mu ^2) +8 M^2+2 \alpha  (\mu -1) M\right)} \right\},
\end{equation}
where $\mathcal{T}^{s}_l(\omega)$, $\mathcal{T}^{e}_l(\omega)$ and 
$\mathcal{T}^{g}_l(\omega)$ denote the greybody factors for the scalar 
perturbation, electromagnetic perturbation and gravitational perturbation 
respectively. The expressions show that the GUP correction 
factors $\alpha$ and $\beta$ impact the greybody factor of a black hole along
with the Lorentz violation term $\lambda$ and the global monopole term $\mu$.
It should be noted that these expressions are derived for the vanishing 
effective cosmological constant only. To have a better qualitative idea we 
have plotted the greybody factors for different sets of parameters of the 
black hole. In Fig.\ \ref{grey01}, we have plotted the greybody factors with 
respect to frequency $\omega$ for different values of the first GUP parameter 
$\alpha$ for the scalar, electromagnetic and 
gravitational perturbations. It is seen that in case of scalar perturbations, 
the greybody factors decrease comparatively rapidly with the increasing value
of $\alpha$ than the other two cases. Thus, for all the three perturbations, 
we have a clear conclusion that an increase in the first GUP parameter 
$\alpha$, decreases the greybody factor of the black hole. It implies that an 
increase in $\alpha$ decreases the probability of Hawking radiation to reach 
the spatial infinity. On the other hand, the increase in the second GUP 
parameter $\beta$ increases the greybody factor of the black hole in case of 
all the three perturbations (see Fig.\ \ref{grey02}). However, in both cases, 
the greybody factor for the gravitational perturbation has less dependency on 
the parameters $\alpha$ and $\beta$ than the other two perturbations. That is 
GUP correction has less impact on the gravitational perturbation in this 
respect. Similarly, the dependencies of the greybody factors 
on the parameters $\lambda$ and $\mu$ for the all three perturbations are shown
in Fig.\ \ref{grey03} and Fig.\ \ref{grey04} respectively. We see that the 
pattern of dependency on $\lambda$ is similar to the dependency on the 
parameter $\alpha$, whereas the dependency on $\mu$ is similar to the case of
$\beta$. However, the impacts of $\lambda$ and $\beta$ are highest for the 
cases of electromagnetic perturbation and scalar perturbation respectively. 
Among all four parameters, the parameter $\mu$, i.e.\ the global monopole seems
to have the dominant influence on the greybody factors of all perturbations. 
Moreover, in all cases the greybody factor for the gravitational perturbation
approaches very quickly towards its peak value $1$ with respect to the
frequency $\omega$ in comparison to the other two perturbations and hence
the probability of the hawking radiation to reach the spatial infinity is
high in this perturbation.                    

Using equations \eqref{eq_tl_s}, \eqref{eq_tl_em} and \eqref{eq_tl_g} in 
equation \eqref{eq_Ptot}, one can obtain the total power spectra of the black 
hole for scalar, electromagnetic and gravitational perturbations respectively. 
With the help of the black hole power spectra, it is possible to have a 
quantitative idea on the sparsity of the black hole, which is discussed in the 
following subsection.

\subsection{Sparsity of the Hawking radiation}
\begin{figure}[!h]
\centerline{
   \includegraphics[scale = 0.28]{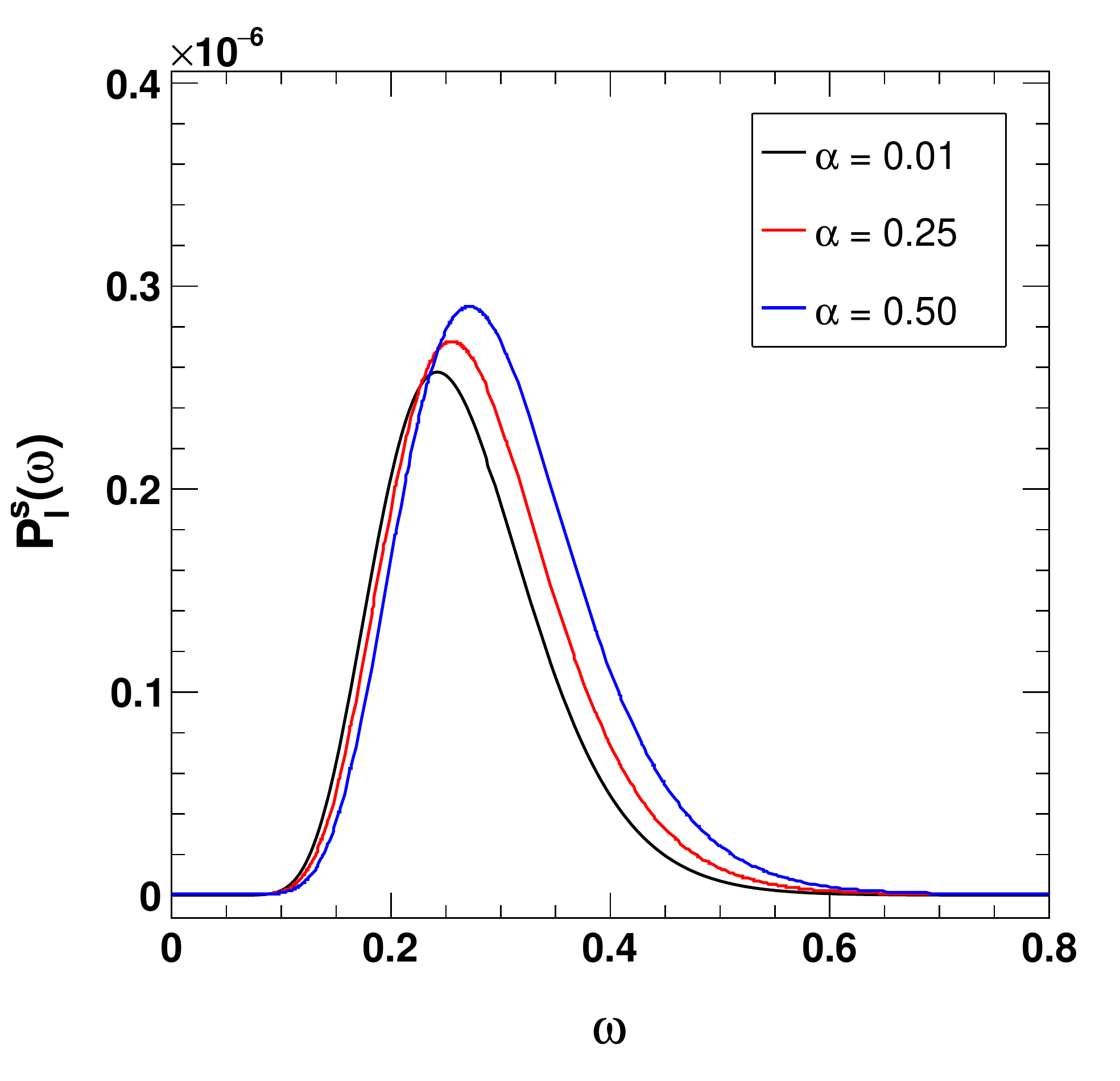}\hspace{1cm}
   \includegraphics[scale = 0.28]{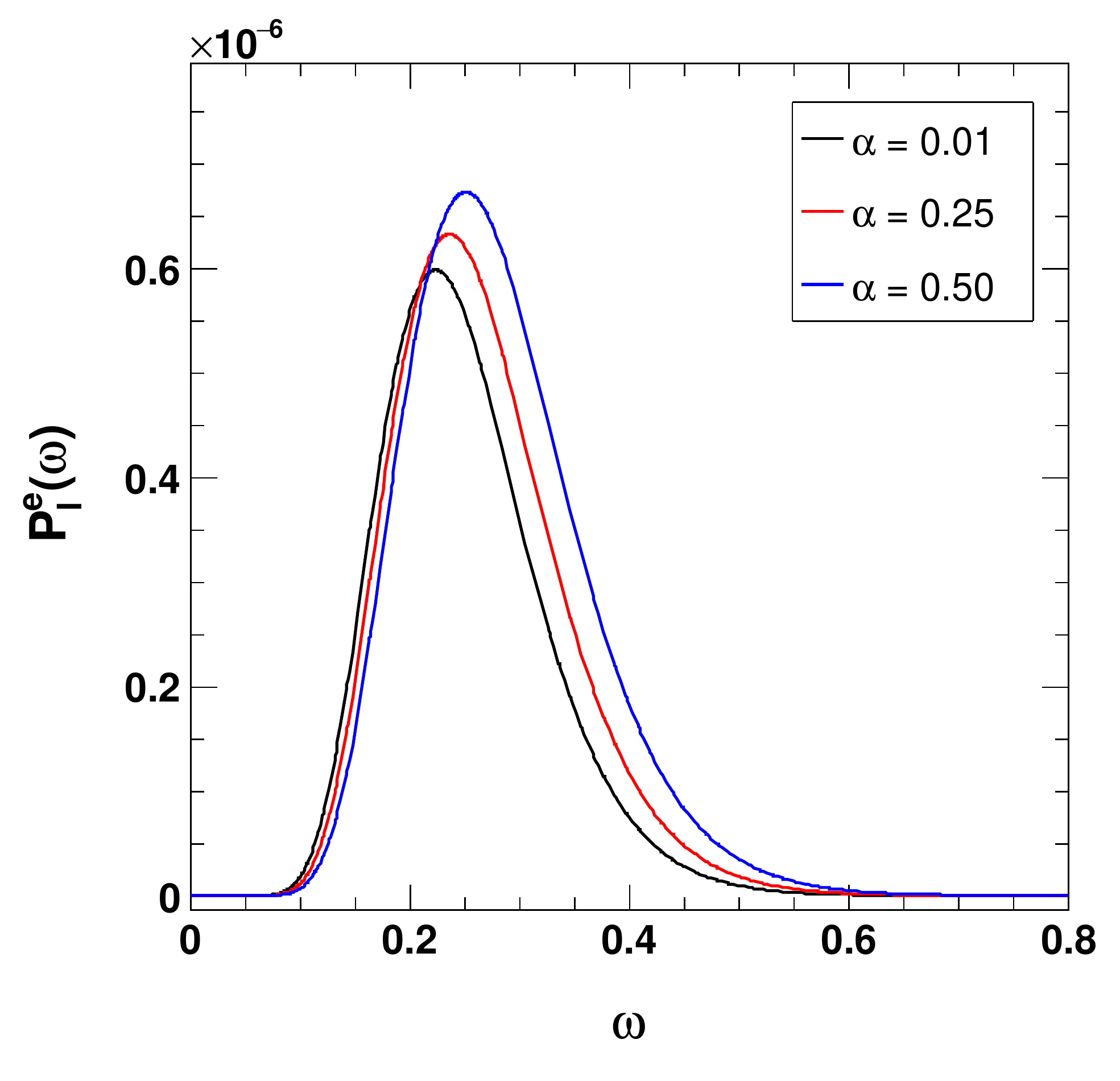}}
   \centerline{
   \includegraphics[scale = 0.28]{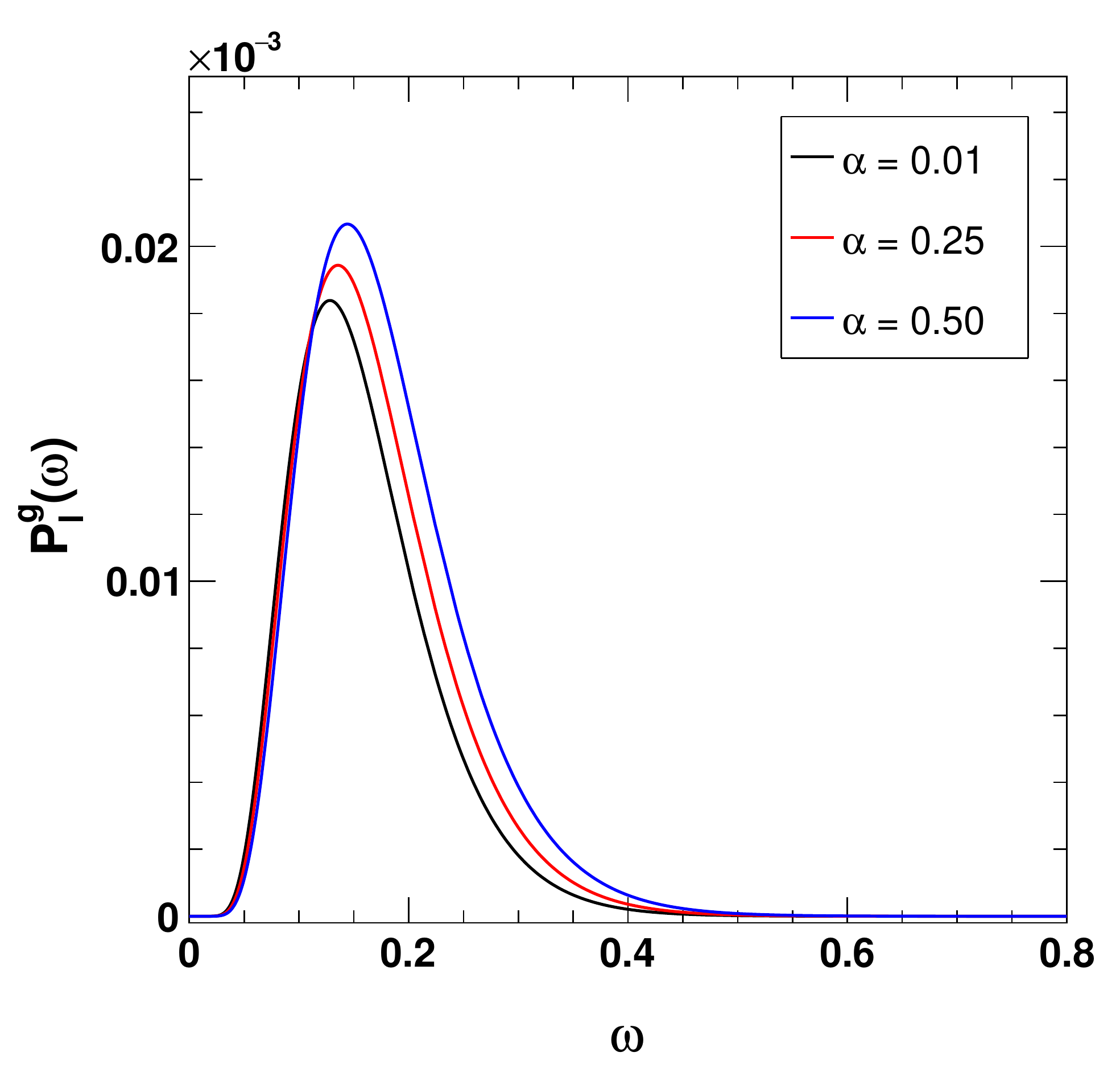}}
\vspace{-0.2cm}
\caption{Power spectrum of the black hole with $\lambda=0.01, \beta = 0.01, M=1, \mu=0.1$ and $\Lambda_{eff} = 0.0$ for different values of $\alpha$.}
\label{P01}
\end{figure}

\begin{figure}[!h]
\centerline{
   \includegraphics[scale = 0.28]{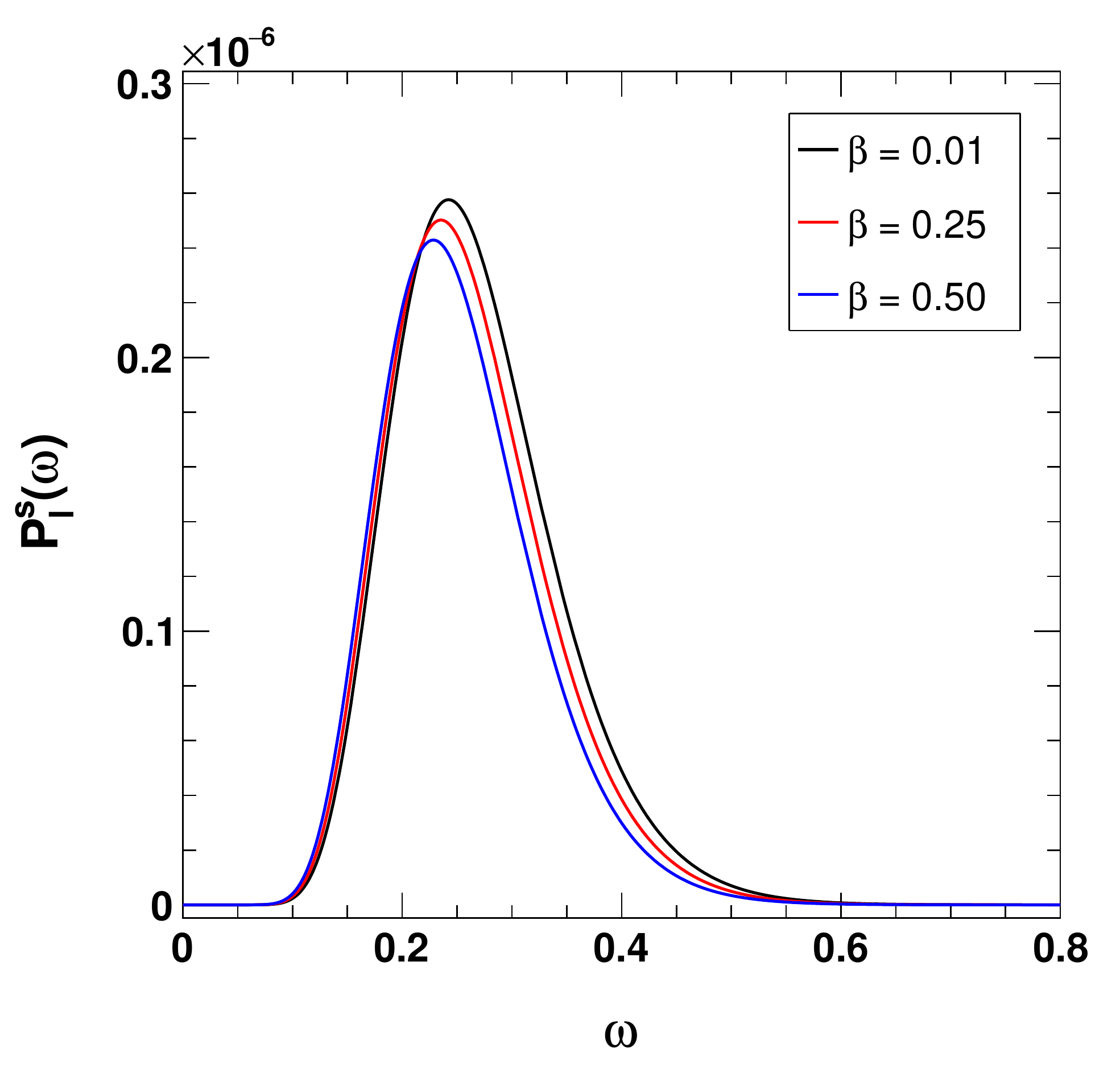}\hspace{1cm}
   \includegraphics[scale = 0.28]{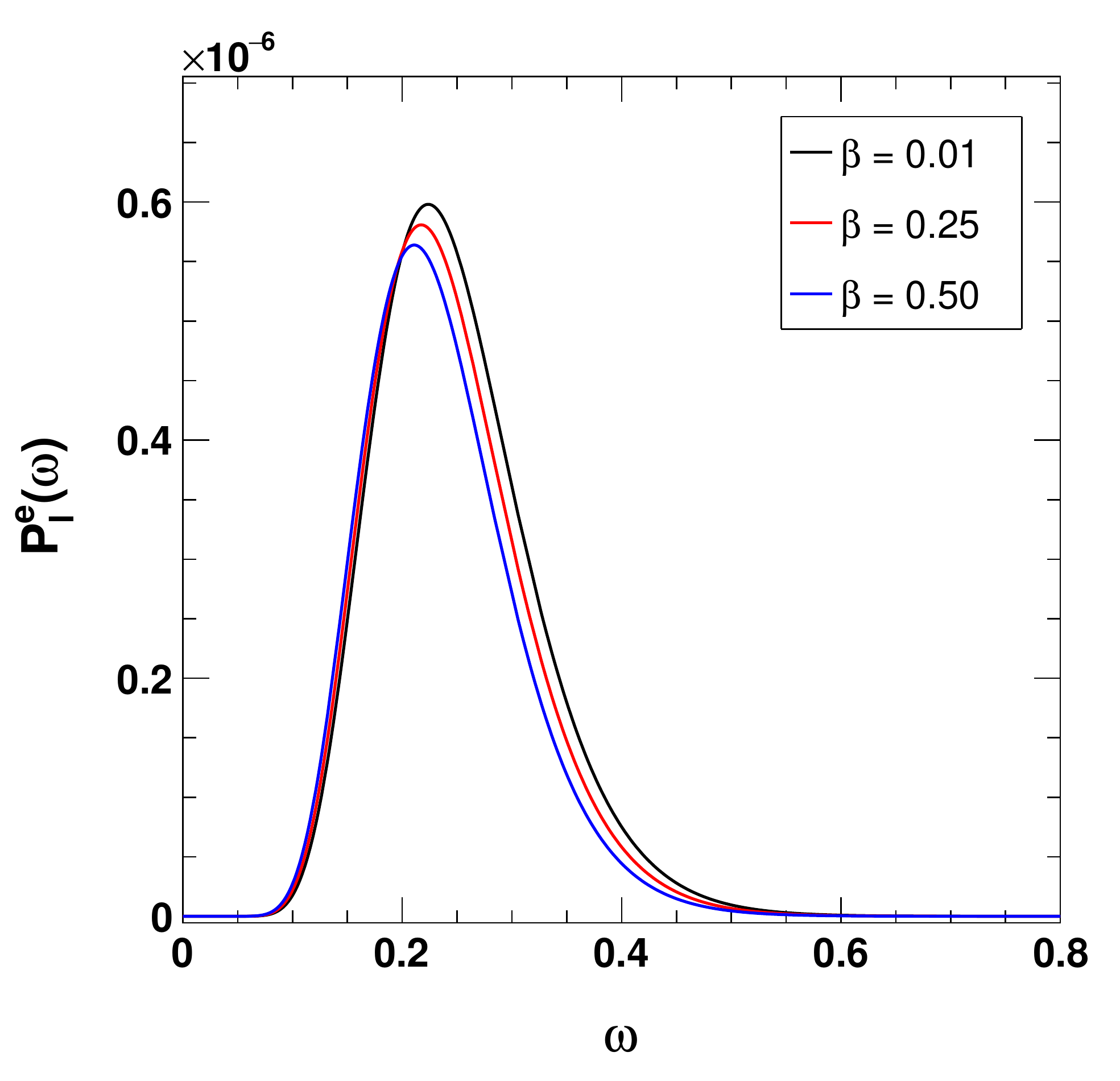}}
   \centerline{
   \includegraphics[scale = 0.28]{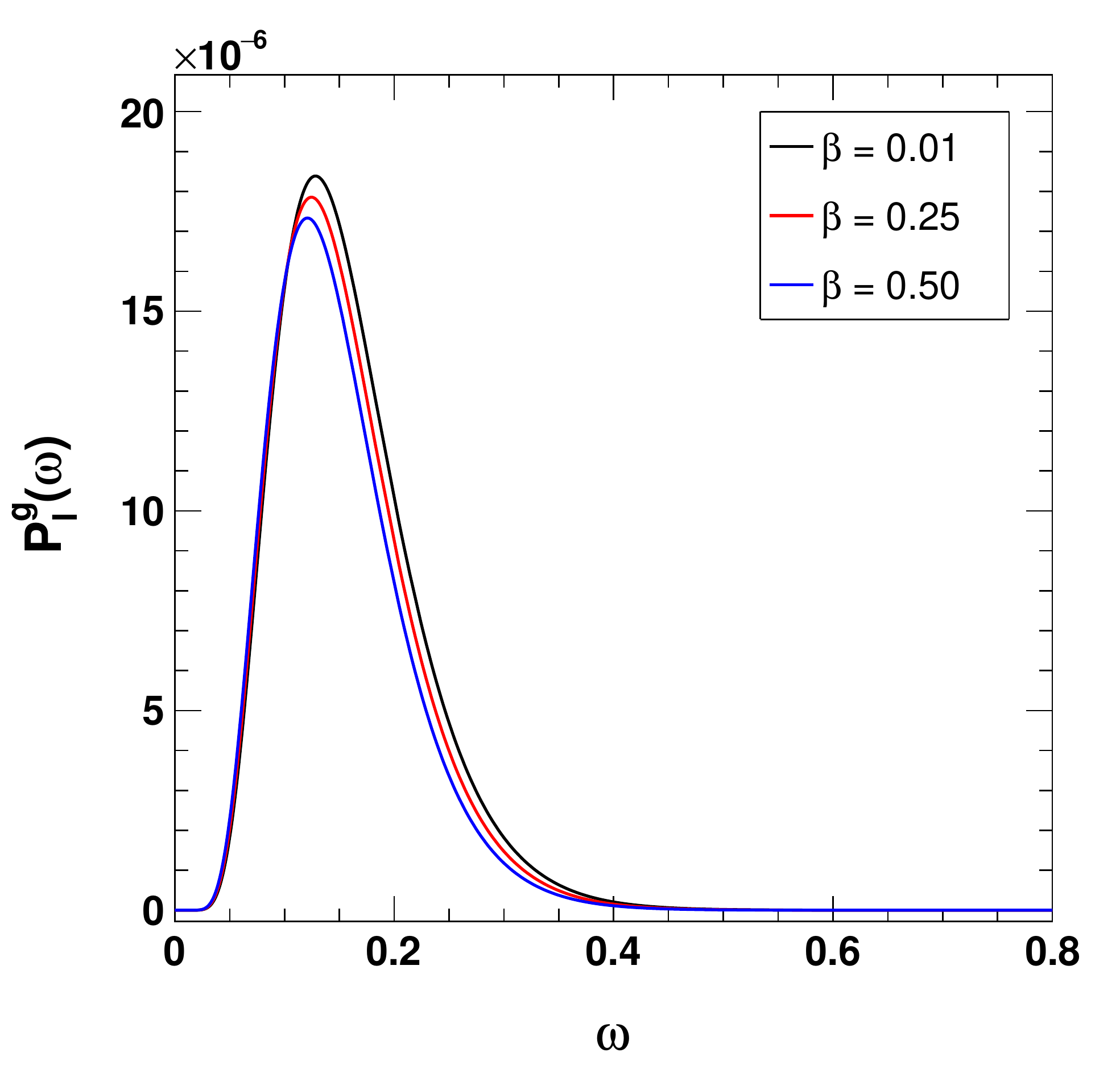}}
\vspace{-0.2cm}
\caption{Power spectrum of the black hole with $\lambda=0.01, \alpha = 0.01, M=1, \mu=0.1$ and $\Lambda_{eff} = 0.0$ for different values of $\beta$.}
\label{P02}
\end{figure}

\begin{figure}[!h]
\centerline{
   \includegraphics[scale = 0.28]{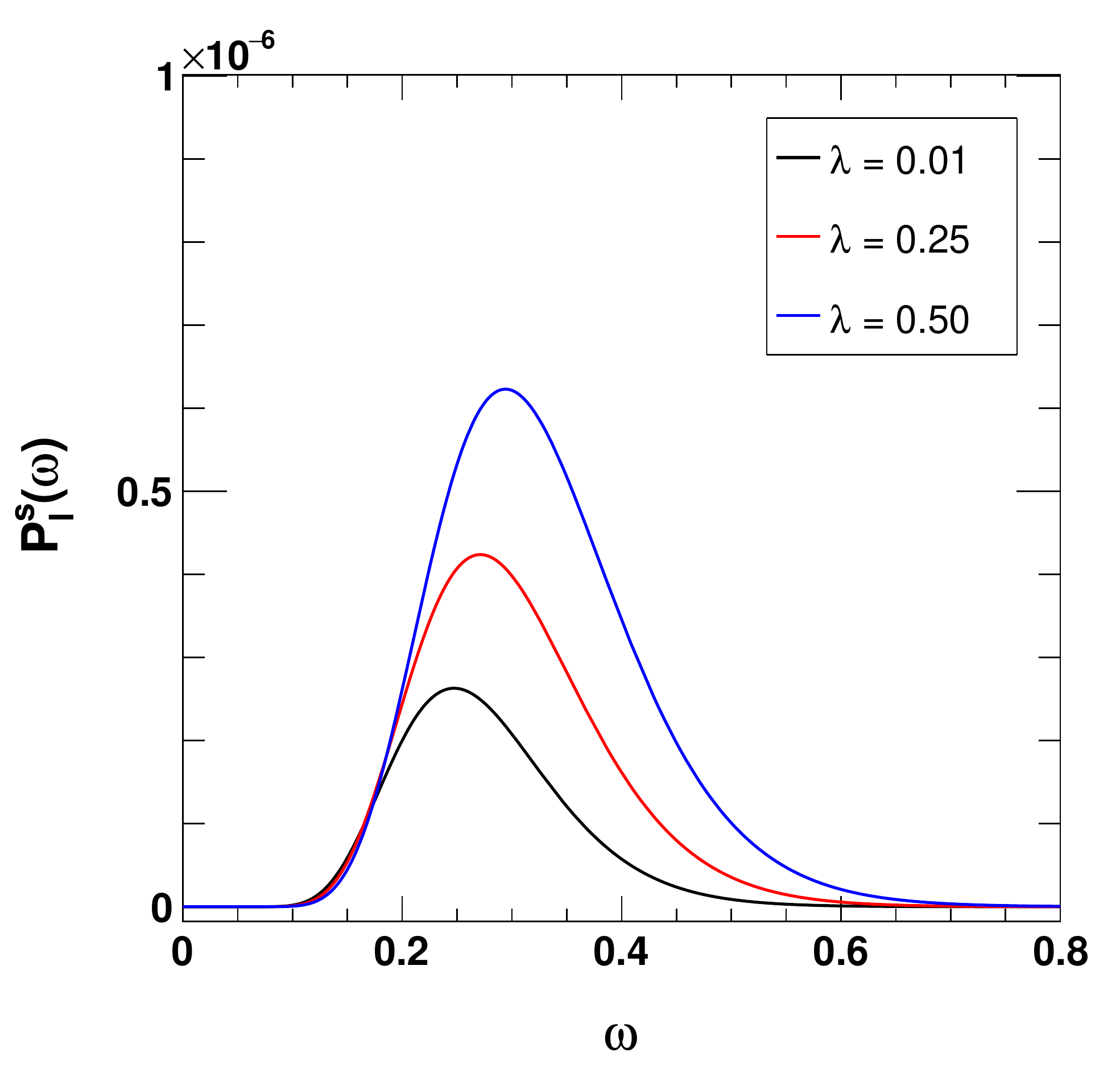}\hspace{1cm}
   \includegraphics[scale = 0.28]{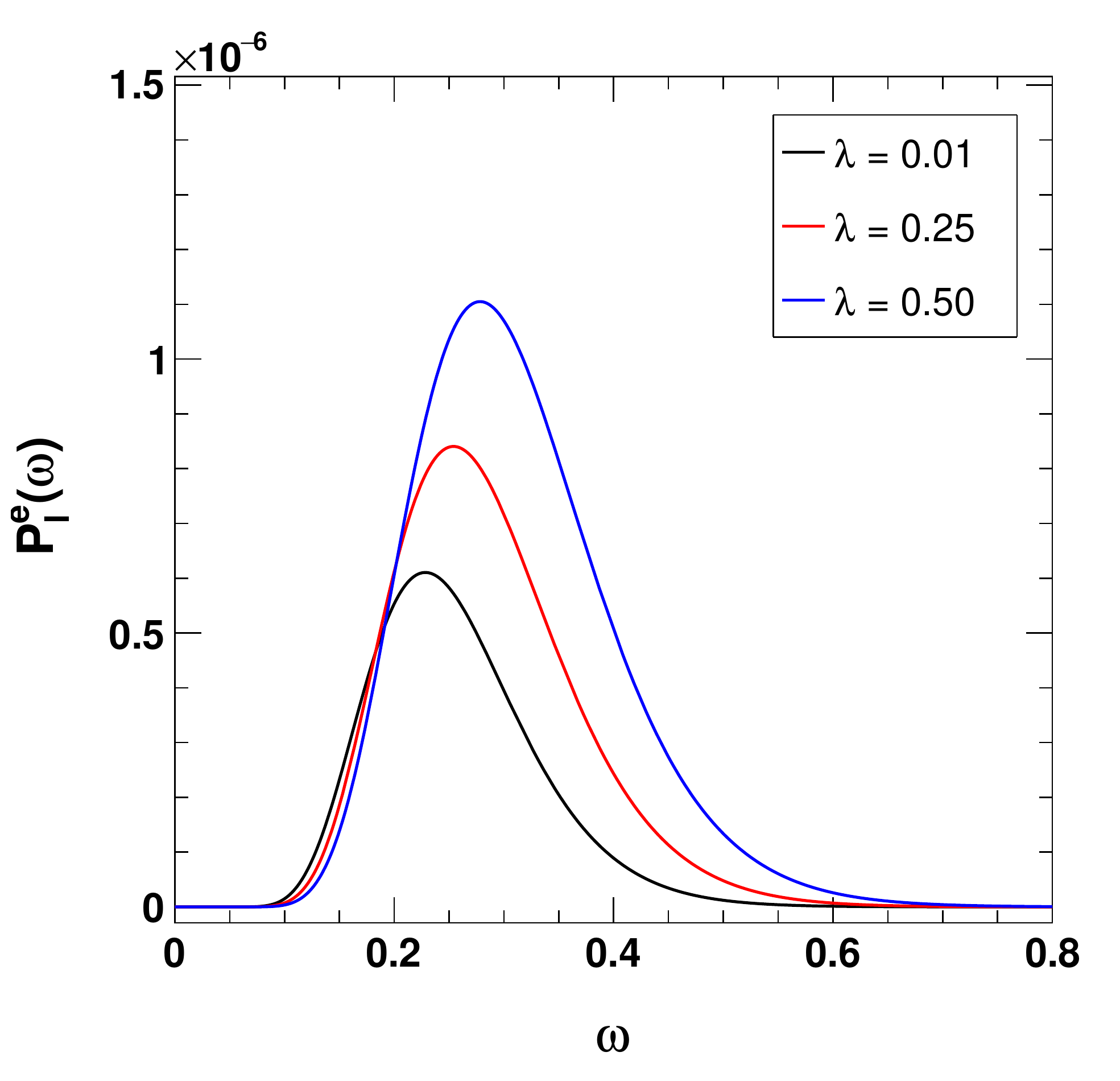}}
   \centerline{
   \includegraphics[scale = 0.28]{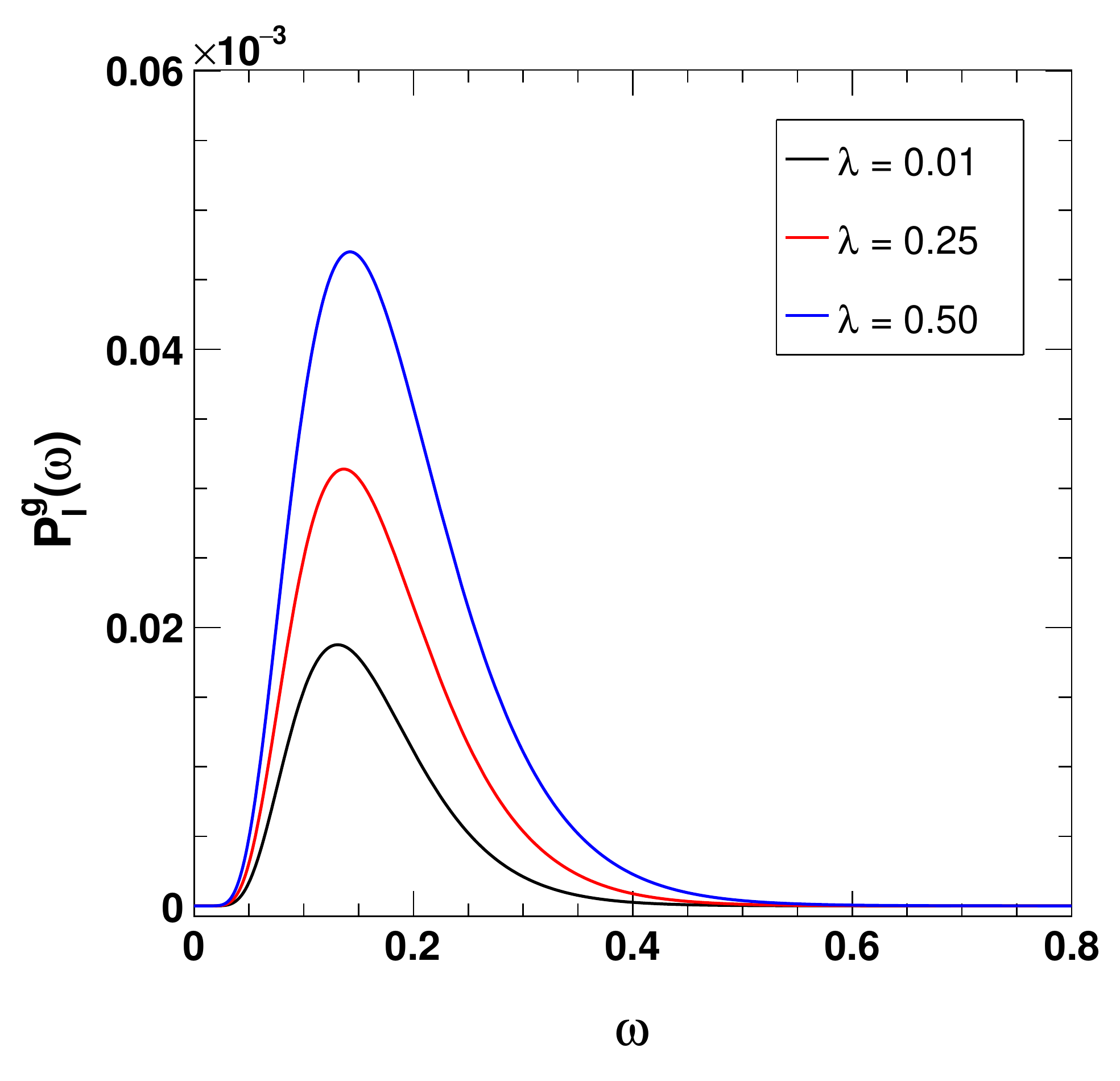}}
\vspace{-0.2cm}
\caption{Power spectrum of the black hole with $\alpha=0.1, \beta = 0.01, M=1, \mu=0.1$ and $\Lambda_{eff} = 0.0$ for different values of $\lambda$.}
\label{P03}
\end{figure}

The $P_l(\omega)$ distribution with respect to frequency $\omega$ provides the 
necessary insights to study the sparsity of the black hole. In Fig.\ \ref{P01},
we have shown the Hawking radiation power spectrum of the black hole for 
different values of GUP parameter $\alpha$. It is observed that with increase 
in the parameter value, the peak of the distribution shifts towards the higher 
frequency and the total Hawking radiation emitted is increased for all the 
three cases of perturbation. 
However, it is observed that for the gravitational perturbation, total Hawking 
radiation emitted is higher. But, for the other GUP parameter $\beta$, we 
observe an opposite scenario (see Fig.\ \ref{P02}). The Lorentz violation can 
also increase the total Hawking radiation emitted. From Fig.\ \ref{P03}, it is 
seen that an increase in the parameter $\lambda$ increases the total emission 
and the peak of the distribution shifts towards the higher frequency range. 
However, the shift is very small in case of the gravitational perturbation. 
Finally, we consider the variation of the power spectrum for different values 
of the monopole term $\mu$ and see that an increase in the parameter $\mu$ 
decreases the total radiation emitted and shifts the peak of the distribution 
towards the lower frequency ranges. One can see that the changes are 
significant for the gravitational perturbations also. From the previous 
results it is seen that in this case, the parameter $\mu$ has a higher 
influence over the power spectrum of the black hole. 

\begin{figure}[!h]
\centerline{
   \includegraphics[scale = 0.28]{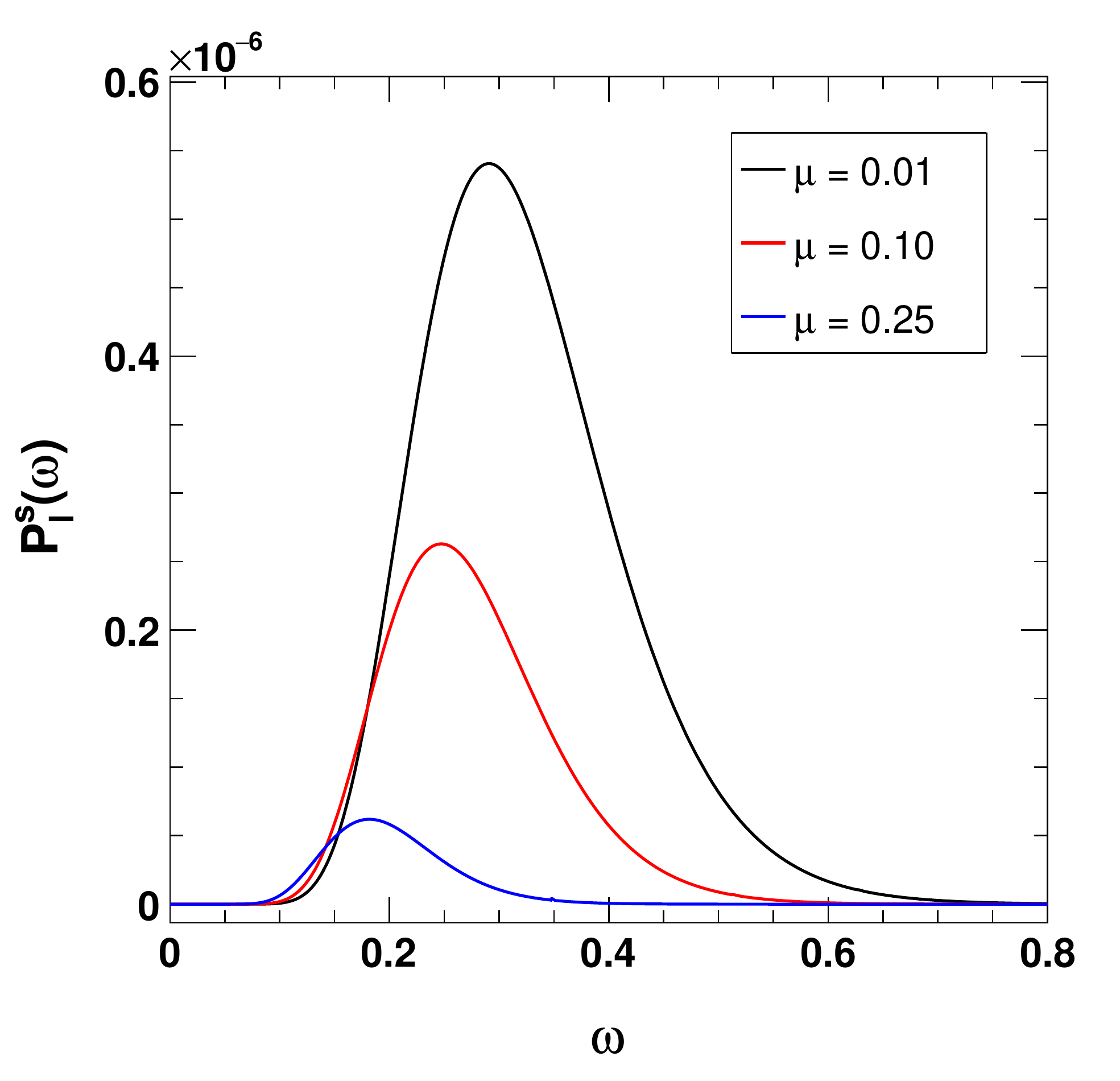}\hspace{1cm}
   \includegraphics[scale = 0.28]{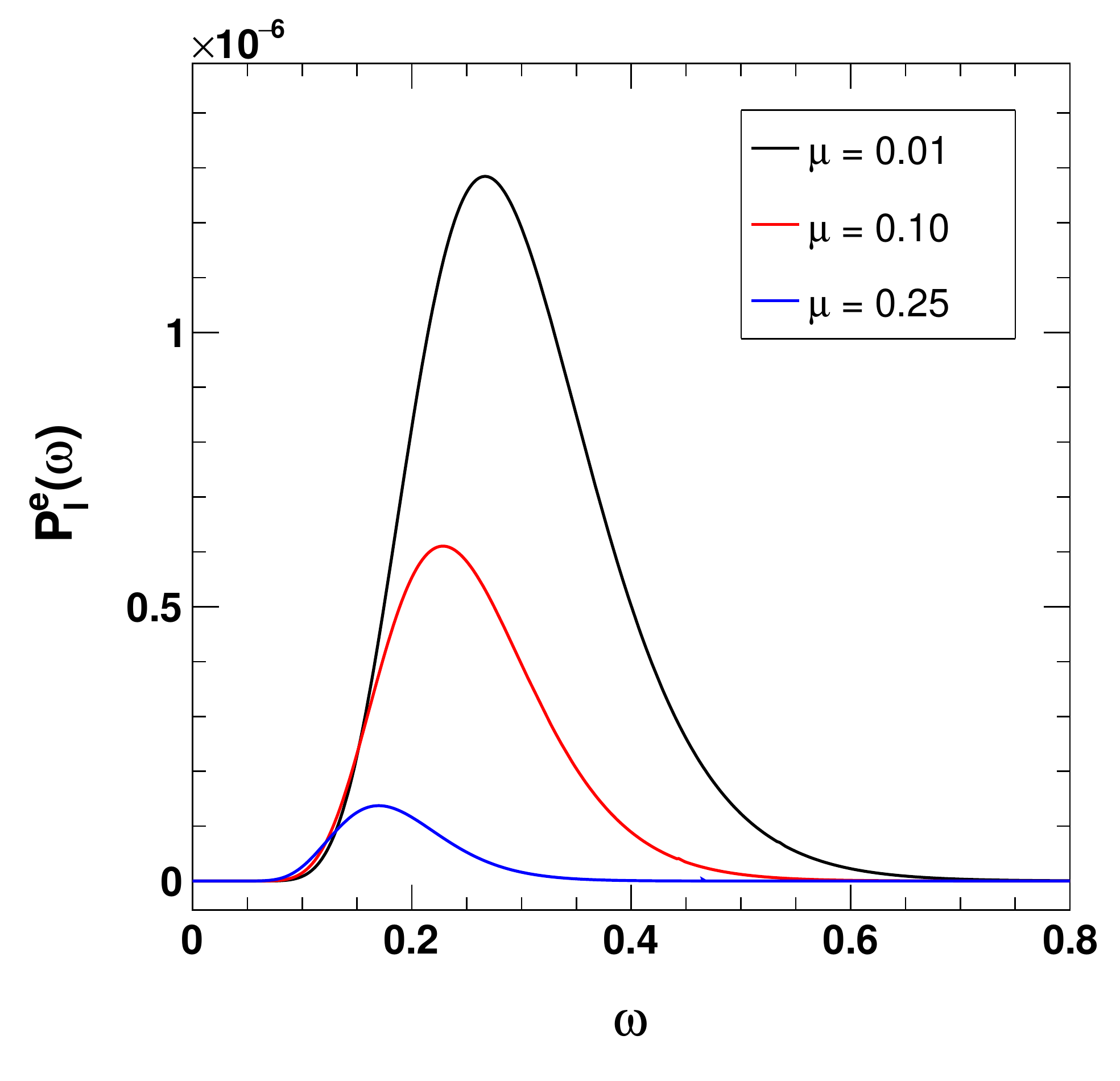}}
   \centerline{
   \includegraphics[scale = 0.28]{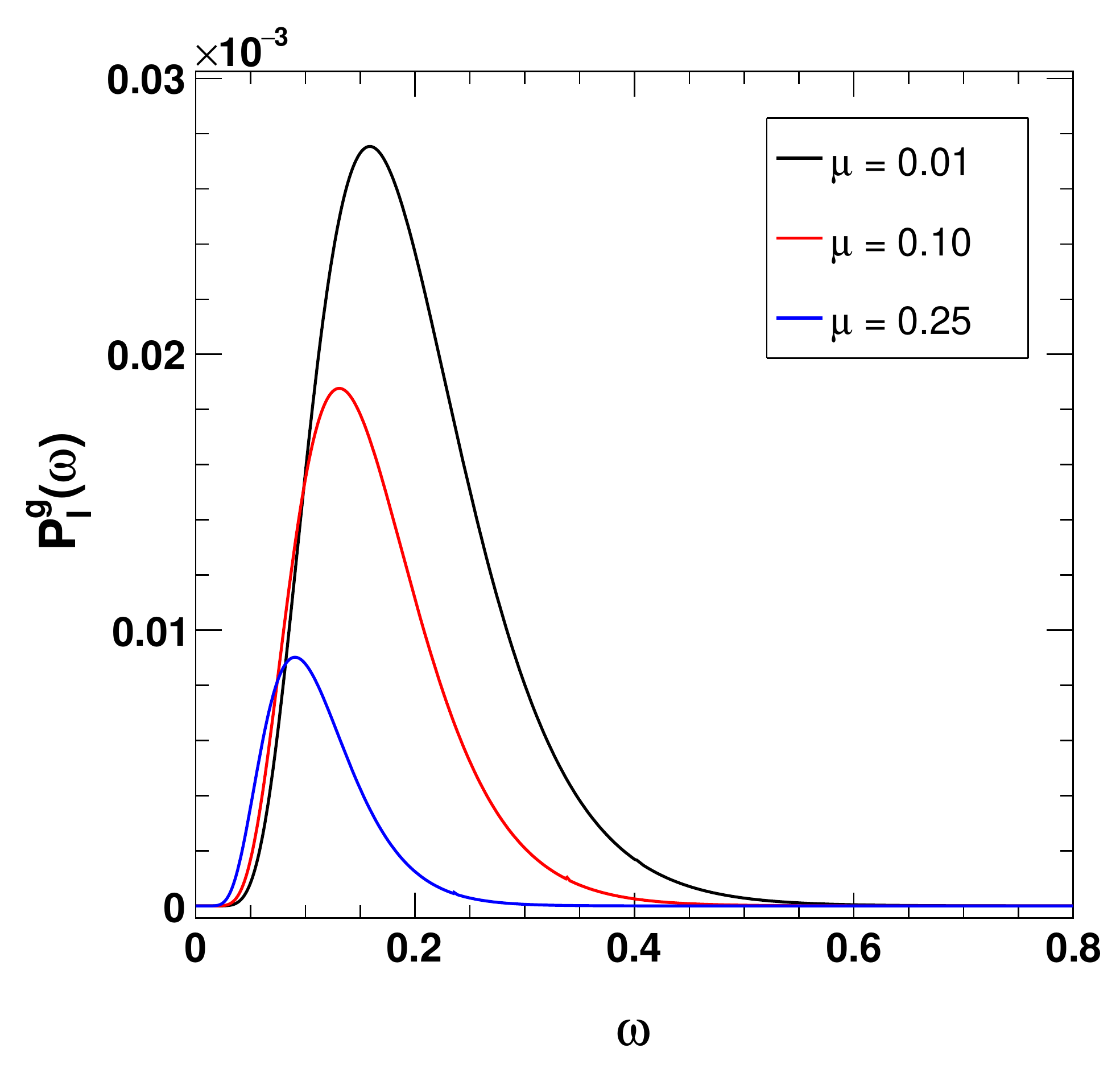}}
\vspace{-0.2cm}
\caption{Power spectrum of the black hole with $\alpha=0.1, \beta = 0.01, M=1, \lambda=0.01$ and $\Lambda_{eff} = 0.0$ for different values of $\mu$.}
\label{P04}
\end{figure}

To have a better idea on the radiation emitted by the black holes, we
consider the sparsity of the black holes which will give a dimensionless
quantitative measure of the Hawking radiation from these black holes.
So, for a quantitative idea on the sparsity of Hawking radiation, we define 
a dimensionless parameter \cite{gray2016, hod2016, hod2015, miao2017,
chowdhury2020} as
\begin{equation}\label{eq_eta}
\eta=\frac{\tau_{gap}}{\tau_{emission}}.
\end{equation}
The parameter $\tau_{gap}$ in the above expression is the average time interval 
 between the emission of two successive Hawking radiation quanta, defined as
\begin{equation}\label{eq_tgap}
\tau_{gap}=\frac{\omega_{peak}}{P_{tot}},
\end{equation}
where $\omega_{peak} $ is the frequency corresponding to the peak of the black 
hole power spectrum, which can be easily determined from the power spectrum 
distribution curves we have studied. $\tau_{emission}$  is the characteristic 
time for the emission of individual Hawking quantum, defined by
\begin{equation}\label{eq_temm_unchrgd}
\tau_{emission} \geq \tau_{localisation}=\frac{2 \pi}{\omega_{peak}},
\end{equation}
where $\tau_{localisation}$, the localisation time-scale is the characteristic 
time taken by the emitted wave field  with frequency $\omega_{peak}$ to 
complete one cycle of oscillation. So, clearly $\eta\gg 1$ implies extremely 
sparse Hawking cascade i.e.\ the time interval between successive Hawking 
quanta emission is much larger than the time required for the emission of 
individual Hawking quantum and $\eta\ll 1$ suggests that the Hawking radiation 
flow is almost continuous.

\begin{table}
 \caption{Numerical values of $\omega_{max}, P_{max}$ and the dimensionless parameter $\eta_{max}=\tau_{gap}/\tau_{localisation}$ for the $l=1$ mode with $\alpha=0.1, \beta=0.01, M=1$ and $\mu=0.1$. $\omega_{max}$ and $P_{max}$ are expressed in Planck's unit i.e.\ $G=c=\hbar=k_B=l_p=1.$}
\begin{center}
\begin{tabular}{|c|c|c|c|ccc} 
\hline
   & \multicolumn{3}{c|}{Scalar perturbation} & \multicolumn{3}{c|}{Electromagnetic perturbation}                                                 \\ 
\hline
$\lambda$ & \;\;$0.01$ &\;\; $0.25$ &\;\; $0.50$                  & \multicolumn{1}{c|}{\;\;$0.01$} & \multicolumn{1}{c|}{\;\;$0.25$} & \multicolumn{1}{c|}{\;\;$0.50$}  \\ 
\hline
$\omega_{max}$ & $0.25$ & $0.27$ & $0.29$                  & \multicolumn{1}{c|}{$0.23$ } & \multicolumn{1}{c|}{$0.25$} & \multicolumn{1}{c|}{$0.28$}  \\ 
\hline
$P_{max}$ & $2.63\times10^{-7}$ & $4.24\times10^{-7}$ & $6.23\times10^{-7}$     & \multicolumn{1}{c|}{$6.10\times10^{-7}$} & \multicolumn{1}{c|}{$8.40\times10^{-7}$} & \multicolumn{1}{c|}{$1.10\times10^{-6}$}  \\ 
\hline
 $\eta_{max}$ & $205578.67$ & $138789.94$ & $101799.47$    & \multicolumn{1}{c|}{$78153.27$} & \multicolumn{1}{c|}{$63147.84$} & \multicolumn{1}{c|}{$52623.20$}  \\ 
\hline
   & \multicolumn{3}{c|}{Gravitational perturbation}  &                        &                        &                         \\ 
\cline{1-4}
$\lambda$ & $0.01$ & $0.25$ & $0.50$                   &                        &                        &                         \\ 
\cline{1-4}
$\omega_{max}$ & $0.130$ & $0.136$ & $0.142$                  &                        &                        &                         \\ 
\cline{1-4}
$P_{max}$ & $1.88\times10^{-5}$ & $3.14\times10^{-5}$ & $4.70\times10^{-5}$                 &                        &                        &                         \\ 
\cline{1-4}
 $\eta_{max}$ & $996.86$ & $586.50$ & $389.16$                   &                        &                        &                         \\
\cline{1-4}
\end{tabular}
\end{center}
\label{table_sparsity_01}
\end{table}

\begin{table}
 \caption{Numerical values of $\omega_{max}, P_{max}$ and the dimensionless parameter $\eta_{max}=\tau_{gap}/\tau_{localisation}$ for the $l=1$ mode with $\lambda=0.01, \beta=0.01, M=1$ and $\mu=0.1$. $\omega_{max}$ and $P_{max}$ are expressed in Planck's unit i.e. $G=c=\hbar=k_B=l_p=1.$}
\begin{center}
\begin{tabular}{|c|c|c|c|ccc} 
\hline
   & \multicolumn{3}{c|}{Scalar perturbation} & \multicolumn{3}{c|}{Electromagnetic perturbation}                                                 \\ 
\hline
$\alpha$ & \;\;$0.01$ &\;\; $0.10$ &\;\; $0.25$                  & \multicolumn{1}{c|}{\;\;$0.01$} & \multicolumn{1}{c|}{\;\;$0.10$} & \multicolumn{1}{c|}{\;\;$0.25$}  \\ 
\hline
$\omega_{max}$ & $0.242$  & $0.247$ & $0.256$                  & \multicolumn{1}{c|}{$0.224$ } & \multicolumn{1}{c|}{$0.228$} & \multicolumn{1}{c|}{$0.237$}  \\ 
\hline
$P_{max}$ & $2.58\times10^{-7}$ & $2.63\times10^{-7}$ & $2.72\times10^{-7}$ & \multicolumn{1}{c|}{$5.98\times10^{-7}$} & \multicolumn{1}{c|}{$6.10\times10^{-7}$} & \multicolumn{1}{c|}{$6.32\times10^{-7}$}  \\ 
\hline
 $\eta_{max}$ & $205578.6703$ & $205578.6727$ & $205578.6757$  & \multicolumn{1}{c|}{$78153.2686$} & \multicolumn{1}{c|}{$78153.2660$} & \multicolumn{1}{c|}{$78153.2662$}  \\ 
\hline
   & \multicolumn{3}{c|}{Gravitational perturbation}  &                        &                        &                         \\ 
\cline{1-4}
$\alpha$ & $0.01$ & $0.10$ & $0.25$                   &                        &                        &                         \\ 
\cline{1-4}
$\omega_{max}$ & $0.128$ & $0.131$ & $0.136$                  &                        &                        &                         \\ 
\cline{1-4}
$P_{max}$ & $1.84\times10^{-5}$ & $1.88\times10^{-5}$ & $1.94\times10^{-5}$   &                        &                        &                         \\ 
\cline{1-4}
 $\eta_{max}$ & $996.861124$ & $996.861207$ & $996.861209$                  &                        &                        &                         \\
\cline{1-4}
\end{tabular}
\end{center}
\label{table_sparsity_02}
\end{table}

\begin{table}
 \caption{Numerical values of $\omega_{max}, P_{max}$ and the dimensionless parameter $\eta_{max}=\tau_{gap}/\tau_{localisation}$ for the $l=1$ mode with $\lambda=0.01, \alpha=0.01, M=1$ and $\mu=0.1$. $\omega_{max}$ and $P_{max}$ are expressed in Planck's unit i.e. $G=c=\hbar=k_B=l_p=1.$}
\begin{center}
\begin{tabular}{|c|c|c|c|ccc} 
\hline
   & \multicolumn{3}{c|}{Scalar perturbation} & \multicolumn{3}{c|}{Electromagnetic perturbation}                                                 \\ 
\hline
$\beta$ & \;\;$0.01$ &\;\; $0.10$ &\;\; $0.25$                  & \multicolumn{1}{c|}{\;\;$0.01$} & \multicolumn{1}{c|}{\;\;$0.10$} & \multicolumn{1}{c|}{\;\;$0.25$}  \\ 
\hline
$\omega_{max}$ & $0.242$  & $0.240$ & $0.235$   & \multicolumn{1}{c|}{$0.224$} & \multicolumn{1}{c|}{$0.221$} & \multicolumn{1}{c|}{$0.217$}  \\ 
\hline
$P_{max}$ & $2.58\times10^{-7}$ & $2.55\times10^{-7}$ & $2.50\times10^{-7}$ & \multicolumn{1}{c|}{$5.98\times10^{-7}$} & \multicolumn{1}{c|}{$5.91\times10^{-7}$} & \multicolumn{1}{c|}{$5.81\times10^{-7}$}  \\ 
\hline
 $\eta_{max}$ & $205578.6703$ & $205578.6721$ & $205578.6721$  & \multicolumn{1}{c|}{$78153.2686$ } & \multicolumn{1}{c|}{$78153.2661$} & \multicolumn{1}{c|}{$78153.2660$}  \\ 
\hline
   & \multicolumn{3}{c|}{Gravitational perturbation}  &                        &                        &                         \\ 
\cline{1-4}
$\beta$ & $0.01$ & $0.10$ & $0.25$                   &                        &                        &                         \\ 
\cline{1-4}
$\omega_{max}$ & $0.128$ & $0.127$ & $0.125$    &                        &                        &                         \\ 
\cline{1-4}
$P_{max}$ & $1.84\times10^{-5}$ & $1.82\times10^{-5}$ & $1.78\times10^{-5}$  &                        &                        &                         \\ 
\cline{1-4}
 $\eta_{max}$ & $996.86112$ & $996.86115$ & $996.86117$      &                        &                        &                         \\
\cline{1-4}
\end{tabular}
\end{center}
\label{table_sparsity_03}
\end{table}

\begin{table}
 \caption{Numerical values of the dimensionless parameter $\eta_{max}=\tau_{gap}/\tau_{localisation}$ for the $l=1$ mode with $\alpha=0.1, \beta=0.01, M=1$ and $\lambda=0.01$. $\omega_{max}$ and $P_{max}$ are expressed in Planck's unit i.e. $G=c=\hbar=k_B=l_p=1.$}
\begin{center}
\begin{tabular}{|c|c|c|c|ccc} 
\hline
   & \multicolumn{3}{c|}{Scalar perturbation} & \multicolumn{3}{c|}{Electromagnetic perturbation}                                                 \\ 
\hline
$\mu$ & \;\;$0.01$ &\;\; $0.10$ &\;\; $0.25$                  & \multicolumn{1}{c|}{\;\;$0.01$} & \multicolumn{1}{c|}{\;\;$0.10$} & \multicolumn{1}{c|}{\;\;$0.25$}  \\ 
\hline
$\omega_{max}$ & $0.29$  & $0.25$ & $0.18$ & \multicolumn{1}{c|}{$0.27$} & \multicolumn{1}{c|}{$0.23$} & \multicolumn{1}{c|}{$0.17$ }  \\ 
\hline
$P_{max}$ & $5.41\times10^{-7}$ & $2.63\times10^{-7}$ & $6.19\times10^{-8}$ & \multicolumn{1}{c|}{$1.28\times10^{-6}$} & \multicolumn{1}{c|}{$6.10\times10^{-7}$} & \multicolumn{1}{c|}{$1.37\times10^{-7}$}  \\ 
\hline
 $\eta_{max}$ & $115592.33$ & $205578.67$ & $663721.86$  & \multicolumn{1}{c|}{$42461.52$} & \multicolumn{1}{c|}{$78153.27$} & \multicolumn{1}{c|}{$269116.10$}  \\ 
\hline
   & \multicolumn{3}{c|}{Gravitational perturbation}  &                        &                        &                         \\ 
\cline{1-4}
$\mu$ & $0.01$ & $0.10$ & $0.25$                   &                        &                        &                         \\ 
\cline{1-4}
$\omega_{max}$ & $0.16$ & $0.13$ & $0.09$  &                        &                        &                         \\ 
\cline{1-4}
$P_{max}$ & $2.75\times10^{-5}$ & $1.88\times10^{-5}$ & $0.90\times10^{-5}$  &                        &                        &                         \\ 
\cline{1-4}
 $\eta_{max}$ & $823.85$ & $996.86$ & $1435.48$   &                        &                        &                         \\
\cline{1-4}
\end{tabular}
\end{center}
\label{table_sparsity_04}
\end{table}

Now, to have a quantitative idea, we have shown the calculated sparsity of the 
black holes in Tables \ref{table_sparsity_01}, \ref{table_sparsity_02}, \ref{table_sparsity_03} and \ref{table_sparsity_04} for 
different black hole parameters and perturbations. In Table \ref{table_sparsity_01}, 
we have shown the $\omega_{max}$ i.e.\ the frequency corresponding to the 
peak $P_{max}$ in the power spectrum and the sparsity $\eta_{max}$ for scalar, 
electromagnetic and gravitational perturbations for different values of the 
Lorentz violation parameter $\lambda$. We see that with increase in the 
parameter $\lambda$, the sparsity decreases for all the three perturbations. 
In case of the scalar perturbations, the sparsity of the black hole is very 
high and in case of the gravitational perturbations, sparsity is less. This 
shows that in case of gravitational perturbations, the Hawking radiation is 
less sparse and the time between the emission of two successive Hawking quanta 
is comparatively less than that for the electromagnetic perturbations and 
scalar perturbations. The results also show that the variations are not very 
small and hence the Hawking radiation may be a useful way to obtain the 
signature of Lorentz violating and the possibility of quantum gravity 
in the near future.

In case of the GUP parameters (see Tables \ref{table_sparsity_02} 
and \ref{table_sparsity_03}), although the variations are monotonic and stable, they are very small and practically not possible to obtain the signatures of 
GUP corrections from the sparsity of the black holes. While in case of 
Table \ref{table_sparsity_04}, where we have considered the variations of 
sparsity of the black hole with the parameter $\mu$, it is seen that an 
increase in the value of $\mu$ increases the sparsity of the Hawking radiation 
cascade of the black hole drastically. The variations are more distinctive in 
case of the scalar perturbations and less distinctive in case of gravitational 
perturbations. These results show that the existence of a global monopole has 
a significant impacts over the sparsity of the black holes and a higher value 
of $\mu$ might make the black holes very sparse. This will increase the time 
gap between the two successive Hawking radiation quanta.

\subsection{Area Spectrum from Adiabatic Invariance}
Finally, following Refs.\ \cite{Majhi2011, qqjiang2012} we derive the area spectrum of the 
black hole which is useful for the further study of sparsity of the black 
hole. We apply the Wick rotation in the Lorentzian time and thus transforming 
time $t$ to $-i\tau$, we write the Euclideanized form of the metric in the 
following way: 
\begin{equation}
\label{A01}
ds^{2}=-\,|g_{tt}|\,d\tau^{2}-g_{rr}\, dr^{2}-r^{2}(d\theta^{2}+\sin^{2}\theta\, d\phi ^{2}),
\end{equation}
where  $ |g_{tt}| = \left(1-\mu-\frac{2M_{gup}}{r}\right)$, $g_{rr} = \left(1+\lambda\right)\left(1-\mu-\frac{2M_{gup}}{r}\right)^{-1}$
and $\tau$ means the Euclidean time. Considering the only dynamic freedom of 
adiabatic invariants to be the radial coordinate $r$, the adiabatic invariant can be 
simply expressed~\cite{Shahjalal2019} as
\begin{equation}
\label{A02}
\mathcal{J}=\oint p_{r}\,dq_{r}=\oint\int_{0}^{p_{r}}dp'_{r}dr,
\end{equation}
where $p_{r}$ is the conjugate momentum to the coordinate $q_{r}$. Now, considering 
only the outgoing path and ignoring whether the particle has mass or 
not \cite{umetsu2010}, from Hamilton's canonical equation,
\begin{equation}
\label{A03}
\dot{r}=\frac{\mathrm{d} r}{\mathrm{d} \tau }=\frac{\mathrm{d} H'}{\mathrm{d} p'_{r}},
\end{equation}
where $H'=M'$ and $M'=M_{gup}-\omega'$, $M'$ is the mass of the black 
hole from which a particle with energy $E'=\omega '$ tunnels through its 
horizon. Using equation \eqref{A03} into equation \eqref{A02}, we may write
\begin{equation}
\label{A04}
\mathcal{J}=\oint p_{r}\,dq_{r}=\oint\int_{0}^{H}\frac{dH'}{\dot{r}}dr=\oint\int_{0}^{M}dM'd\tau.
\end{equation}
This equation \eqref{A04} can be further written as
\begin{equation}
\label{A05}
\mathcal{J}=\int_{0}^{M}\frac{dM'}{T'_{\rm{BH}}},
\end{equation}
where we have used the assumption that the particles moving in the black hole 
background spacetime  have the same periodicity as the background spacetime \cite{gibbons1978}. At last, using the definitions of the black hole mentioned above, we 
found that
\begin{equation}
\label{A06}
\mathcal{J}=\pi r^{2}_{\rm H} (1+\lambda)^{-1/2}=\frac{A (1+\lambda)^{-1/2}}{4}.
\end{equation}
From the Bohr-Sommerfeld quantization rule, $\mathcal{J}=2\pi n$, $n=0, 1, 2, 3, \dots$, so the quantized area of the black hole is now given by
\begin{equation}
\label{A07}
A_{n}=8\pi n (1+\lambda)^{1/2}.
\end{equation}
Hence the area spectrum of the black hole is different from that of 
Schwarzschild one by a multiplicative factor of $(1+\lambda)^{1/2}$ and is 
given by
\begin{equation}
\label{A08}
\Delta A=A_{n}-A_{n-1}=8\pi (1+\lambda)^{1/2}.
\end{equation}
The equations \eqref{eq_Pl} with the above equation \eqref{A08} 
show that the power spectrum of the black hole is also quantized and is 
differed from the quantization rule of Schwarzschild black hole by a factor of 
$(1+\lambda)^{1/2}$. Hence, it is seen that the Lorentz violation can have 
impacts over the area spectrum of the black hole and consequently on the 
Hawking radiation emitted by the black hole.

\section{Conclusion}\label{section8}
 In this work, we have used three elements viz., bumblebee 
field, global monopole and GUP corrections to study the properties of 
the black hole in this configuration. We have previously mentioned 
that the bumblebee field can effectively result in Lorentz violation and 
in Lorentz violating theories topological defects like global monopoles 
may arise. Being isolated, global monopoles may be present in the 
universe till now and can influence the black hole properties 
including quasinormal modes as we have seen. The GUP corrections 
provide the quantum effects to the black hole and previous studies 
suggest that they are connected with Lorentz violation 
\cite{LGUP1, LGUP2}. Hence, inclusion of all these three ingredients 
may provide a realistic platform to study the properties of the black 
hole. It should be noted that such a configuration with all three 
ideas has not been studied before. So we believe this study will 
contribute significantly in the studies of the impacts by Lorentz 
violation, GUP and global monopole on the quasinormal modes. 

This study shows that the GUP correction can affect the quasinormal modes of the black 
holes. An increase in the GUP parameter $\alpha$ increases the quasinormal 
frequencies and the decay rates. While the other GUP parameter $\beta$ has an 
opposite impact on the quasinormal modes. It is seen that the impacts of the GUP 
parameters are more significant in the de Sitter black hole and less 
significant in the anti-de Sitter black hole.

The Lorentz violation also affects the quasinormal modes. We observe that 
for $\Lambda_{eff} = 0$ and for anti-de Sitter case, the scalar quasinormal frequency 
decreases with an increase in the violation parameter value but the 
electromagnetic and gravitational quasinormal modes increase with an increase in 
the value of the violation parameter. 
However, for the de Sitter black hole, quasinormal frequencies decrease for all the 
perturbation schemes considered in this work. The decay rate in all three cases of 
perturbations decreases with increase in violation parameter. On 
the other hand, an increase in the global monopole parameter results 
in decrease of the quasinormal frequency as well as the decay rate. 
The impact of the monopole term on quasinormal modes is higher in 
comparison to the GUP parameters and in de Sitter regime, the impact 
pattern of the second GUP deformation parameter $\beta$ resembles with it for small values of the parameters.

 The quasinormal modes in the de Sitter regime are quite different 
from those we have obtained for anti-de Sitter and vanishing 
cosmological constant cases. This variation can be definitely useful 
in inferring a bound on the cosmological constant provided we have 
significant experimental results from the next generation 
gravitational wave detectors in the near future.
One may note that there is a theoretical constraint on the de Sitter 
spacetime given by equation \eqref{Lambda_bound} which is controlled 
by the global monopole and the Lorentz violation term. For a viable de 
Sitter solutions, global monopole term should be less than $1$ and the 
Lorentz violation term should be greater than $-1$. An increase in 
both the parameters by respecting these limits reduces the ranges of 
viable de Sitter solutions. So, experimental constraints on global 
monopoles and the Lorentz violation parameter could be helpful in 
introducing bounds on the cosmological constant.

 Moreover, we see that this study might help to differentiate the impacts 
of GUP and global monopoles on the black hole. Although GUP and global 
monopole impacts may be apparently indistinguishable for very small values 
of the parameters, for comparatively large values we can see a clear 
difference in the quasinormal modes. So, in general, the quasinormal 
modes have a higher dependency on global monopoles than GUP. This 
property may be helpful to check the existence of a global monopoles on 
the black hole spacetime once we have a significant experimental 
result from LISA in the near future.

 It is worth to be mentioned that the GUP impacts on quasinormal modes of a 
Schwarzschild black hole have been studied previously \cite{gupbh}. However, in this study, we have 
considered three perturbations viz., scalar, electromagnetic and gravitational perturbations and 
compared the results in presence of bumblebee field, global monopole and cosmological constant. It 
clearly shows that the GUP impacts highly depend on the Lorentz violation, cosmological constant and 
presence of global monopoles. As we have mentioned in the introduction, quasinormal modes and greybody 
factors have been extensively studied in different modified theories of gravity including Rastall 
gravity, $f(R)$ gravity etc. We have seen that for different model parameters in different modified 
gravity theories, the quasinormal modes may show a different variation pattern. For example, in a recent 
study the authors studied the impacts of energy momentum conservation 
violating Rastall parameter on the 
quasinormal modes \cite{gogoi3}. It is seen that the impacts of general energy momentum conservation 
violation on quasinormal modes differ from that of Lorentz violation. We believe that such studies might 
be helpful to differentiate between 
different theories and bumblebee gravity in terms of quasinormal modes in the near future.

In the next part of study, we have studied the quantum thermodynamics of the black 
holes. Here at first we have obtained the Hawking temperature of the black holes 
within the framework of GUP for scalar, electromagnetic and gravitational 
perturbations. Then we have obtained the greybody factors of the black holes. We see 
that the Lorentz violation affects the greybody factor and greybody factor decreases 
with an increase in the parameter $\lambda$ value. Low values of greybody factor imply a low 
probability of the Hawking radiation to reach spatial infinity. So, future 
astrophysical observations may shed more light in this area and may provide support 
to the indirect evidence of Lorentz violation. On the other hand, the topological 
defects have an opposite impact on the greybody factors. We see that an increase 
in $\mu$ decreases the greybody factors and increases the probability of Hawking 
radiation to reach the spatial infinity. The GUP parameter $\beta$ also provides a 
similar effect on the greybody factors; however the effect is less in comparison to 
the previous case. An increase in the GUP parameter $\alpha$, on the other hand, 
decreases the probability of Hawking radiation to reach spatial infinity. So the 
study shows that the parameters $\alpha$ and $\lambda$ have a similar impact on the 
greybody factors.

Total Hawking radiation power emitted increases with an increase in GUP 
parameter $\alpha$ and Lorentz violation parameter $\lambda$. On the contrary, it 
decreases with increase the second GUP parameter $\beta$ and global monopole 
term $\mu$ and the peak of the distribution shifts towards low frequency range.

In the case of gravitational perturbation, the black hole is less sparse and in the case of 
scalar perturbation the black hole is highly sparse. In the case of all the three 
perturbations, with an increase in the Lorentz violation factor, the sparsity of the black 
holes decreases. On the other hand, the GUP parameters have a very very small impact 
on the sparsity of the black holes. With increase in the global monopole 
term, $\mu$, the sparsity of the black hole increases.

 The global monopole parameter does not imprint significant changes on the 
black hole sparsity in case of gravitational perturbation. Since the peak of 
the power spectrum for the gravitational perturbation is much higher and is 
at near the low frequency range, it results in the Hawking radiation to be 
less sparse in comparison to the scalar and electromagnetic perturbations.
Also due to this high value of power spectrum peak, the impacts of the black 
hole parameters including global monopole on the gravitational perturbation is 
not as large as we have observed for the other two perturbations, i.e.\
electromagnetic and scalar perturbations. This should be a universal feature 
of the sparsity of black holes as the gravitational perturbation around any 
black hole should be of a similar nature except some local effects introduced 
by the black hole type-parameters and its background. However, the effects due 
to the black hole background and the model parameters should be small like our 
case and hence such small variations represent the unique structure of a black 
hole solution.


Eventually, we have checked the area quantization of the black hole by using the adiabatic 
invariance method. It shows that the area quantization is affected by the presence of 
Lorentz violation. Therefore, we can conclude that the Hawking radiation power 
emitted is also quantized and the quantization is affected by the Lorentz violation. 
Recent studies show the impacts of area quantization of black holes  on GW 
echoes \cite{cardoso2019, datta2021, coates2022}. So, it can be possible in the near 
future to comment on the fate of quantum gravity and Lorentz violation provided we 
have sufficient astrophysical signatures.



\begin{thebibliography}{99}
\bibitem{kost2004}V. A. Kosteleck\'y, {\em Lorentz violation, and the standard model}, \href{https://doi.org/10.1103/PhysRevD.69.105009}{Phys. Rev. D {\bf 69}, 105009 (2004)} [arXiv:hep-th/0312310].

\bibitem{casana2018}R. Casana, A. Cavalcante, F. P. Poulis, and E. B. Santos, {\em Exact Schwarzschild-like Solution in a Bumblebee Gravity Model}, \href{https://doi.org/10.1103/PhysRevD.97.104001}{Phys. Rev. D {\bf 97}, 104001 (2018)} [arXiv:1711.02273].

\bibitem{bluhm2005}R. Bluhm and V. A. Kosteleck\'y, {\em Spontaneous Lorentz Violation, Nambu-Goldstone Modes, and Gravity}, \href{https://doi.org/10.1103/PhysRevD.71.065008}{Phys. Rev. D {\bf 71}, 065008 (2005)} [arXiv:hep-th/0412320].

\bibitem{bailey2006}Q. G. Bailey and V. A. Kosteleck\'y, {\em Signals for Lorentz Violation in Post-Newtonian Gravity}, \href{https://doi.org/10.1103/PhysRevD.74.045001}{Phys. Rev. D {\bf 74}, 045001 (2006)} [arXiv:gr-qc/0603030].

\bibitem{bailey2009}Q. G. Bailey, {\em Time Delay and Doppler Tests of the Lorentz Symmetry of Gravity}, \href{https://doi.org/10.1103/PhysRevD.80.044004}{Phys. Rev. D {\bf 80}, 044004 (2009)} [arXiv:0904.0278].

\bibitem{tso2011}R. Tso and Q. G. Bailey, {\em Light-Bending Tests of Lorentz Invariance}, \href{https://doi.org/10.1103/PhysRevD.84.085025}{Phys. Rev. D {\bf 84}, 085025 (2011)} [arXiv:1108.2071].

\bibitem{kost2009}V. A. Kosteleck\'y and J. D. Tasson, {\em Prospects for Large Relativity Violations in Matter-Gravity Couplings}, \href{https://doi.org/10.1103/PhysRevLett.102.010402}{Phys. Rev. Lett. {\bf 102}, 010402 (2009)} [arXiv:0810.1459].

\bibitem{maluf2013}R. V. Maluf, V. Santos, W. T. Cruz, and C. A. S. Almeida, {\em Matter-Gravity Scattering in the Presence of Spontaneous Lorentz Violation}, \href{https://doi.org/10.1103/PhysRevD.88.025005}{Phys. Rev. D {\bf 88}, 025005 (2013)} [arXiv:1304.2090].

\bibitem{maluf2014}R. V. Maluf, C. A. S. Almeida, R. Casana, and M. M. Ferreira, {\em Einstein-Hilbert Graviton Modes Modified by the Lorentz-Violating Bumblebee Field}, \href{https://doi.org/10.1103/PhysRevD.90.025007}{Phys. Rev. D {\bf 90}, 025007 (2014)} [arXiv:1402.3554].

\bibitem{santos2015}A. F. Santos, W. D. R. Jesus, J. R. Nascimento, and A. Yu. Petrov, {\em G\"odel Solution in the Bumblebee Gravity}, \href{http://dx.doi.org/10.1142/S021773231550011X}{Mod. Phys. Lett. A {\bf 30}, 1550011 (2015)} [arXiv:1407.5985].

\bibitem{kost2016}V. A. Kosteleck\'y, A. C. Melissinos, and M. Mewes, {\em Searching for Photon-Sector Lorentz Violation Using Gravitational-Wave Detectors}, \href{https://doi.org/10.1016/j.physletb.2016.08.001}{Phys. Lett. B {\bf 761}, 1 (2016)} [arXiv:1608.02592].

\bibitem{kost2016_2}V. A. Kosteleck\'y and M. Mewes, {\em Testing Local Lorentz Invariance with Gravitational Waves}, \href{https://doi.org/10.1016/j.physletb.2016.04.040}{Phys. Lett. B {\bf 757}, 510 (2016)} [arXiv:1602.04782].

\bibitem{kumar2021}S. Kumar Jha, H. Barman, and A. Rahaman, {\em Bumblebee Gravity and Particle Motion in Snyder Noncommutative Spacetime Structures}, \href{https://doi.org/10.1088/1475-7516/2021/04/036}{J. Cosmol. Astropart. Phys. {\bf 04}, 036 (2021)} [arXiv:2012.02642].

\bibitem{kanzi2019}S. Kanzi and \.I. Sakall\i, {\em GUP Modified Hawking Radiation in Bumblebee Gravity}, \href{https://doi.org/10.1016/j.nuclphysb.2019.114703}{Nuclear Physics B {\bf 946}, 114703 (2019)}.

\bibitem{jha2022}S. K. Jha, S. Aziz, and A. Rahaman, {\em Study of Einstein-Bumblebee Gravity with Kerr-Sen-like Solution in the Presence of a Dispersive Medium}, \href{https://doi.org/10.1140/epjc/s10052-022-10042-4}{Eur. Phys. J. C {\bf 82}, 106 (2022)} [arXiv:2103.17021].

\bibitem{gullu}\.I. G\"ull\"u and A. \"Ovg\"un, {\em Schwarzschild-like Black Hole with a Topological Defect in Bumblebee Gravity}, \href{https://doi.org/10.1016/j.aop.2021.168721}{Annals of Physics {\bf 436}, 168721 (2022)}.


\bibitem{Jusufi19}A. \"Ovg\"un, K. Jusufi, and \.{I} Sakall{\i}, {\em Exact Traversable Wormhole Solution in Bumblebee Gravity}, \href{https://doi.org/10.1103/PhysRevD.99.024042}{Phys. Rev. D {\bf 99}, 024042 (2019)} [arXiv:1804.09911 [gr-qc]].

\bibitem{Vishveshwara}C. V. Vishveshwara, {\em Stability of the Schwarzschild Metric}, \href{https://journals.aps.org/prd/pdf/10.1103/PhysRevD.1.2870}{Phys. Rev. D {\bf 1}, 2870 (1970)}.

\bibitem{Press}W. H. Press, {\em Long Wave Trains of Gravitational Waves from a Vibrating Black Hole}, \href{https://ui.adsabs.harvard.edu/link_gateway/1971ApJ...170L.105P/doi:10.1086/180849}{ApJ {\bf 170}, L105 (1971)}.

\bibitem{Chandrasekhar_qnms}S. Chandrasekhar and S. Detweiler, {\em The Quasi-Normal Modes of the Schwarzschild Black Hole}, \href{https://doi.org/10.1098/rspa.1975.0112}{Proc. R. Soc. Lond. A {\bf 344}, 441 (1975)}.

\bibitem{Ma}C. Ma, Y. Gui, W. Wang, F. Wang, {\em Massive scalar field quasinormal modes of a Schwarzschild black hole surrounded by quintessence}, \href{https://doi.org/10.2478/s11534-008-0056-7}{Cent. Eur. J. Phys. {\bf 6}, 194 (2008)} [arXiv:gr-qc/0611146]. 

\bibitem{gogoi1}D. J. Gogoi and U. D. Goswami, {\em A New f(R) Gravity Model and Properties of Gravitational Waves in It}, \href{https://doi.org/10.1140/epjc/s10052-020-08684-3}{Eur. Phys. J. C {\bf 80}, 1101 (2020)} [arXiv:2006.04011].

\bibitem{gogoi2}D. J. Gogoi and U. D. Goswami, {\em Gravitational Waves in $\mathbf {f(R)}$ Gravity Power Law Model}, \href{https://doi.org/10.1007/s12648-020-01998-8}{Indian J. Phys. {\bf 96}, 637 (2022)} [arXiv:1901.11277].

\bibitem{Liang_2017}D. Liang, Y. Gong, S. Hou and Y. Liu, {\em Polarizations of Gravitational Waves in $f(R)$ Gravity}, \href{https://journals.aps.org/prd/abstract/10.1103/PhysRevD.95.104034}{Phys. Rev. D {\bf 95}, 104034 (2017)} [arXiv:1701.05998].

\bibitem{qnm_bumblebee}R. Oliveira, D. M. Dantas, and C. A. S. Almeida, {\em Quasinormal Frequencies for a Black Hole in a Bumblebee Gravity}, \href{https://doi.org/10.1209/0295-5075/ac130c}{EPL {\bf 135}, 10003 (2021)} [arXiv:2105.07956].

\bibitem{gogoi3}D. J. Gogoi and U. D. Goswami, {\em Quasinormal Modes of Black Holes with Non-Linear-Electrodynamic Sources in Rastall Gravity}, \href{https://doi.org/10.1016/j.dark.2021.100860}{Physics of the Dark Universe {\bf 33}, 100860 (2021)} [arXiv:2104.13115].

\bibitem{Graca}J. P. M. Gra\c{c}a and I. P. Lobo, {\em Scalar QNMs for Higher Dimensional Black Holes Surrounded by Quintessence in Rastall Gravity}, \href{https://doi.org/10.1140/epjc/s10052-018-5598-2}{Eur. Phys. J. C {\bf 78}, 101 (2018)} [arXiv:1711.08714].

\bibitem{Zhang2}Y. Zhang, Y.X. Gui, F. Li, {\em Quasinormal modes of a Schwarzschild black hole surrounded by quintessence: electromagnetic perturbations}, \href{https://doi.org/10.1007/s10714-007-0434-2}{Gen. Relativ. Gravit. {\bf 39}, 1003 (2007)} [arXiv:gr-qc/0612010]. 

\bibitem{lopez2020}M. Bouhmadi-López, S. Brahma, C.-Y. Chen, P. Chen, and D. Yeom, {\em A Consistent Model of Non-Singular Schwarzschild Black Hole in Loop Quantum Gravity and Its Quasinormal Modes}, \href{https://doi.org/10.1088/1475-7516/2020/07/066}{J. Cosmol. Astropart. Phys. {\bf 07}, 066 (2020)} [arXiv:2004.13061].

\bibitem{Liang2018}J. Liang, {\em Quasinormal Modes of the Schwarzschild Black Hole Surrounded by the Quintessence Field in Rastall Gravity}, \href{https://doi.org/10.1088/0253-6102/70/6/695}{Commun. Theor. Phys. {\bf 70}, 695 (2018)}.

\bibitem{Hu}Y. Hu, C.-Y. Shao, Y.-J. Tan, C.-G. Shao, K. Lin, and W.-L. Qian, {\em Scalar Quasinormal Modes of Nonlinear Charged Black Holes in Rastall Gravity}, \href{https://doi.org/10.1209/0295-5075/128/50006}{EPL {\bf 128}, 50006 (2020)}.

\bibitem{hemawati2022}S. Giri, H. Nandan, L. K. Joshi, and S. D. Maharaj, {\em Geodesic Stability and Quasinormal Modes of Non-Commutative Schwarzschild Black Hole Employing Lyapunov Exponent}, \href{https://doi.org/10.1140/epjp/s13360-022-02403-5}{Eur. Phys. J.  Plus {\bf 137}, 181 (2022)}.

\bibitem{gogoi4}D. J. Gogoi, R. Karmakar, and U. D. Goswami, {\em Quasinormal Modes of Non-Linearly Charged Black Holes Surrounded by a Cloud of Strings in Rastall Gravity}, \href{https://arxiv.org/abs/2111.00854}{arXiv:2111.00854} (2021).

\bibitem{gupbh}M. A. Anacleto, J. A. V. Campos, F. A. Brito, and E. Passos, {\em Quasinormal Modes and Shadow of a Schwarzschild Black Hole with GUP}, \href{https://doi.org/10.1016/j.aop.2021.168662}{Annals of Physics {\bf 434}, 168662 (2021)} [arXiv:2108.04998].

\bibitem{Daghigh2012}R. G. Daghigh and M. D. Green, {\em Validity of the WKB Approximation in Calculating the Asymptotic Quasinormal Modes of Black Holes}, \href{https://doi.org/10.1103/PhysRevD.85.127501}{Phys. Rev. D {\bf 85}, 127501 (2012)} [arXiv:1112.5397 [gr-qc]].

\bibitem{Boonserm2018}P. Boonserm, T. Ngampitipan, and P. Wongjun, {\em Greybody Factor for Black Holes in DRGT Massive Gravity}, \href{https://doi.org/10.1140/epjc/s10052-018-5975-x}{Eur. Phys. J. C {\bf 78}, 492 (2018)} [arXiv:1705.03278 [gr-qc]].

\bibitem{Okyay2022}M. Okyay and A. \"Ovg\"un, {\em Nonlinear Electrodynamics Effects on the Black Hole Shadow, Deflection Angle, Quasinormal Modes and Greybody Factors}, \href{https://doi.org/10.1088/1475-7516/2022/01/009}{J. Cosmol. Astropart. Phys. 2022, 009 (2022)} [arXiv:2108.07766 [gr-qc]].

\bibitem{Boonserm2019}P. Boonserm, T. Ngampitipan, and P. Wongjun, {\em Greybody Factor for Black String in DRGT Massive Gravity}, \href{https://doi.org/10.1140/epjc/s10052-019-6827-z}{Eur. Phys. J. C {\bf 79}, 330 (2019)} [arXiv:1902.05215 [gr-qc]].

\bibitem{Javed2022}W. Javed, I. Hussain, and A. \"Ovg\"un, {\em Weak Deflection Angle of Kazakov-Solodukhin Black Hole in Plasma Medium Using Gauss-Bonnet Theorem and Its Greybody Bonding}, \href{https://doi.org/10.1140/epjp/s13360-022-02374-7}{Eur. Phys. J. Plus {\bf 137}, 148 (2022)} [arXiv:2201.09879 [gr-qc]].

\bibitem{Jusufi18}A. \"Ovg\"un and K. Jusufi, {\em Quasinormal Modes and Greybody Factors of $f(R)$ Gravity Minimally Coupled to a Cloud of Strings in $2+1$ Dimensions}, \href{https://doi.org/10.1016/j.aop.2018.05.013}{Annals of Physics {\bf 395}, 138 (2018)} [arXiv:1801.02555 [gr-qc]].

\bibitem{Daghigh2020}R. G. Daghigh, M. D. Green, J. C. Morey, and G. Kunstatter, {\em Scalar Perturbations of a Single-Horizon Regular Black Hole}, \href{https://doi.org/10.1103/PhysRevD.102.104040}{Phys. Rev. D {\bf 102}, 104040 (2020)} [arXiv:2009.02367 [gr-qc]].

\bibitem{Jusufi16}A. \"Ovg\"un and K. Jusufi, {\em Massive Vector Particles Tunneling From Noncommutative Charged Black Holes and Its GUP-Corrected Thermodynamics}, \href{https://doi.org/10.1140/epjp/i2016-16177-4}{Eur. Phys. J. Plus {\bf 131}, 177 (2016)} [arXiv:1512.05268 [gr-qc]].

\bibitem{Jusufi17}A. \"Ovg\"un and K. Jusufi, {\em The Effect of GUP to Massive Vector and Scalar Particles Tunneling From a Warped DGP Gravity Black Hole}, \href{https://doi.org/10.1140/epjp/i2017-11574-9}{Eur. Phys. J.  Plus {\bf 132}, 298 (2017)} [arXiv:1703.08073].



\bibitem{cardoso2019}V. Cardoso, V. F. Foit, and M. Kleban, {\em Gravitational Wave Echoes from Black Hole Area Quantization}, \href{https://doi.org/10.1088/1475-7516/2019/08/006}{J. Cosmol. Astropart. Phys. {\bf 08}, 006 (2019)} [arXiv:1902.10164].

\bibitem{datta2021}S. Datta and K. S. Phukon, {\em Imprint of Black Hole Area Quantization and Hawking Radiation on Inspiraling Binary}, \href{https://doi.org/10.1103/PhysRevD.104.124062}{Phys. Rev. D {\bf 104}, 124062 (2021)} [arXiv:2105.11140].

\bibitem{coates2022}A. Coates, S. H. V\"olkel, and K. D. Kokkotas, {\em On Black Hole Area Quantization and Echoes}, \href{https://doi.org/10.1088/1361-6382/ac4618}{Class. Quantum Grav. {\bf 39}, 045007 (2022)} [arXiv:2201.03245].

\bibitem{nw1}V. A. Kosteleck\'y and N. Russell, {\em Data Tables for Lorentz and C P T Violation}, \href{https://doi.org/10.1103/RevModPhys.83.11}{Rev. Mod. Phys. {\bf 83}, 11 (2011)}.
\bibitem{Seifert2010}M. D. Seifert, {\em Monopole Solution in a Lorentz-Violating Field Theory}, \href{https://journals.aps.org/prl/pdf/10.1103/PhysRevLett.105.201601}{Phys. Rev. Lett. {\bf 105}, 201601 (2010)}.
\bibitem{nw3} V. A. Kosteleck\'y and S. Samuel, {\em Gravitational Phenomenology in Higher-Dimensional Theories and Strings}, Phys. Rev. D {\bf 40}, 1886 (1989).
\bibitem{nw4}S. M. Carroll, J. A. Harvey, V. A. Kosteleck\'y, C. D. Lane, and T. Okamoto, {\em Noncommutative Field Theory and Lorentz Violation}, \href{https://journals.aps.org/prl/pdf/10.1103/PhysRevLett.87.141601}{Phys. Rev. Lett. {\bf 87}, 141601 (2001)}.

\bibitem{Kibble1976}T. W. B. Kibble, {\em Topology of Cosmic Domains and Strings}, \href{https://doi.org/10.1088/0305-4470/9/8/029}{J. Phys. A: Math. Gen. {\bf 9}, 1387 (1976)}.
\bibitem{Guth1981}A. H. Guth, {\em Inflationary Universe: A Possible Solution to the Horizon and Flatness Problems}, \href{https://journals.aps.org/prd/pdf/10.1103/PhysRevD.23.347}{Phys. Rev. D {\bf 23}, 347 (1981)}.
\bibitem{Durrer2002}R. Durrer, M. Kunz, and A. Melchiorri, {\em Cosmic Structure Formation with Topological Defects}, \href{https://doi.org/10.1016/S0370-1573(02)00014-5}{Physics Reports {\bf 364}, 1 (2002)}.


\bibitem{LGUP1}A. N. Tawfik, H. Magdy, and A. F. Ali, {\em Lorentz Invariance Violation and Generalized Uncertainty Principle}, \href{https://doi.org/10.1134/S1547477116010179}{Phys. Part. Nuclei Lett. {\bf 13}, 59 (2016)}.

\bibitem{LGUP2} G. Lambiase and F. Scardigli, {\em Lorentz Violation and Generalized Uncertainty Principle}, \href{https://doi.org/10.1103/PhysRevD.97.075003}{Phys. Rev. D {\bf 97}, 075003 (2018)}.


\bibitem{Bertolami2005}O. Bertolami and J. P\'aramos, {\em Vacuum Solutions of a Gravity Model with Vector-Induced Spontaneous Lorentz Symmetry Breaking}, \href{https://doi.org/10.1103/PhysRevD.72.044001}{Phys. Rev. D {\bf 72}, 044001 (2005)}.

\bibitem{Anacleto_01}M. A. Anacleto, F. A. Brito, J. A. V. Campos, and E. Passos, {\em Quantum-Corrected Scattering and Absorption of a Schwarzschild Black Hole with GUP}, \href{https://doi.org/10.1016/j.physletb.2020.135830}{Phys. Lett. B {\bf 810}, 135830 (2020)} [arXiv:2003.13464].

\bibitem{Gangopadhyay}S. Gangopadhyay, A. Dutta, and M. Faizal, {\em Constraints on the Generalized Uncertainty Principle from Black-Hole Thermodynamics}, \href{https://doi.org/10.1209/0295-5075/112/20006}{EPL {\bf 112}, 20006 (2015)} [arXiv:1501.01482].

\bibitem{Ali_gup}A. F. Ali, S. Das, and E. C. Vagenas, {\em Discreteness of Space from the Generalized Uncertainty Principle}, \href{https://doi.org/10.1016/j.physletb.2009.06.061}{Phys. Lett. B {\bf 678}, 497 (2009)} [arXiv:0906.5396].

\bibitem{Tawfik}A. N. Tawfik and A. M. Diab, {\em Generalized Uncertainty Principle: Approaches and Applications}, \href{https://doi.org/10.1142/S0218271814300250}{Int. J. Mod. Phys. D {\bf 23}, 1430025 (2014)} [arXiv:1410.0206].

\bibitem{Tawfik2}A. N. Tawfik and E. A. El Dahab, {\em Corrections to Entropy and Thermodynamics of Charged Black Hole Using Generalized Uncertainty Principle}, \href{https://doi.org/10.1142/S0217751X1550030X}{Int. J. Mod. Phys. A {\bf 30}, 1550030 (2015)} [arXiv:1501.01286].

\bibitem{maluf2021}R. V. Maluf and J. C. S. Neves, {\em Black Holes with a Cosmological Constant in Bumblebee Gravity}, \href{https://doi.org/10.1103/PhysRevD.103.044002}{Phys. Rev. D {\bf 103}, 044002 (2021)} [arXiv:2011.12841].

\bibitem{chandrasekhar}S. Chandrasekhar, {\em The mathematical theory of black holes}, Oxford University Press, Oxford (1992).

\bibitem{chen19}C.-Y. Chen and P. Chen, {\em Gravitational perturbations of nonsingular black holes in conformal gravity}, \href{https://doi.org/10.1103/PhysRevD.99.104003}{Phys. Rev. D {\bf 99}, 104003 (2019)} [arXiv:1902.01678].

\bibitem{Schutz}B. F. Schutz and C. M. Will, {\em Black Hole Normal Modes - A Semi analytic Approach}, \href{https://ui.adsabs.harvard.edu/link_gateway/1985ApJ...291L..33S/doi:10.1086/184453}{The Astrophysical Journal {\bf 291}, L33 (1985)}.

\bibitem{Will_wkb}S. Iyer and C. M. Will, {\em Black-Hole Normal Modes: A WKB Approach. I. Foundations and Application of a Higher-Order WKB Analysis of Potential-Barrier Scattering}, \href{https://doi.org/10.1103/PhysRevD.35.3621}{Phys. Rev. D {\bf 35}, 3621 (1987)}.

\bibitem{Konoplya_wkb}R. A. Konoplya, {\em Quasinormal Behavior of the D -Dimensional Schwarzschild Black Hole and the Higher Order WKB Approach}, \href{https://doi.org/10.1103/PhysRevD.68.024018}{Phys. Rev. D {\bf 68}, 024018 (2003)} [arXiv:gr-qc/0303052].

\bibitem{Maty_wkb}J. Matyjasek and M. Telecka, {\em Quasinormal Modes of Black Holes. II. Pad\'e Summation of the Higher-Order WKB Terms}, \href{https://doi.org/10.1103/PhysRevD.100.124006}{Phys. Rev. D {\bf 100}, 124006 (2019)} [arXiv:1908.09389].

\bibitem{gundlach} C. Gundlach, R. H. Price and J. Pullin, {\em Late time behavior of stellar collapse and explosions: 2. Nonlinear evolution}, \href{https://doi.org/10.1103/PhysRevD.49.890}{Phys. Rev. D {\bf 49}, 890 (1994)} [arXiv:gr-qc/9307010].

\bibitem{levenberg1944}K. Levenberg, {\em A Method for the Solution of Certain Non-Linear Problems in Least Squares}, \href{https://www.jstor.org/stable/43633451}{Quart. Appl. Math. {\bf 2}, 164 (1944)}.

\bibitem{marq1963}D. W. Marquardt, {\em An Algorithm for Least-Squares Estimation of Nonlinear Parameters}, \href{https://www.jstor.org/stable/2098941}{Journal of the Society for Industrial and Applied Mathematics {\bf 11}, 431 (1963)}.

\bibitem{berti2007}E. Berti, V. Cardoso, J. A. Gonz\'alez, and U. Sperhake, {\em Mining Information from Binary Black Hole Mergers: A Comparison of Estimation Methods for Complex Exponentials in Noise}, \href{https://doi.org/10.1103/PhysRevD.75.124017}{Phys. Rev. D {\bf 75}, 124017 (2007)} [arXiv:gr-qc/0701086].

\bibitem{Ferrari2008}V. Ferrari and L. Gualtieri, {\em Quasi-Normal Modes and Gravitational Wave Astronomy}, \href{https://link.springer.com/article/10.1007/s10714-007-0585-1}{Gen. Relativ. Gravit. {\bf 40}, 945 (2008)}.

\bibitem{Yamamoto22}K. Yamamoto, C. Vorndamme, O. Hartwig, M. Staab, T. S. Schwarze, and G. Heinzel, {\em Experimental Verification of Intersatellite Clock Synchronization at LISA Performance Levels}, \href{https://doi.org/10.1103/PhysRevD.105.042009}{Phys. Rev. D {\bf 105}, 042009 (2022)}.

\bibitem{ghez_sgr}A. M. Ghez, S. Salim, S. D. Hornstein, A. Tanner, J. R. Lu, M. Morris, E. E. Becklin, and G. Duchene, {\em Stellar Orbits around the Galactic Center Black Hole}, \href{https://doi.org/10.1086/427175}{ApJ {\bf 620}, 744 (2005)}.

\bibitem{miao2017}Y.-G. Miao and Z.-M. Xu, {\em Hawking Radiation of Five-Dimensional Charged Black Holes with Scalar Fields}, \href{https://doi.org/10.1016/j.physletb.2017.07.023}{Phys. Lett. B {\bf 772}, 542 (2017)}.

\bibitem{gray2016}F. Gray, S. Schuster, A. Van–Brunt, and M. Visser, {\em The Hawking Cascade from a Black Hole Is Extremely Sparse}, \href{https://doi.org/10.1088/0264-9381/33/11/115003}{Class. Quantum Grav. {\bf 33}, 115003 (2016)}.

\bibitem{visser1999}M. Visser, {\em Some General Bounds for One-Dimensional Scattering}, \href{https://doi.org/10.1103/PhysRevA.59.427}{Phys. Rev. A {\bf 59}, 427 (1999)} [arXiv:quant-ph/9901030].

\bibitem{boon2008}P. Boonserm and M. Visser, {\em Bounding the Bogoliubov Coefficients}, \href{https://doi.org/10.1016/j.aop.2008.02.002}{Annals of Physics {\bf 323}, 2779 (2008)} [arXiv:0801.0610].

\bibitem{boon2008_2}P. Boonserm and M. Visser, {\em Bounding the Greybody Factors for Schwarzschild Black Holes}, \href{https://doi.org/10.1103/PhysRevD.78.101502}{Phys. Rev. D {\bf 78}, 101502 (2008)} [arXiv:0806.2209].

\bibitem{boon2014}P. Boonserm, T. Ngampitipan, and M. Visser, {\em Bounding the Greybody Factors for Scalar Field Excitations on the Kerr-Newman Spacetime}, \href{https://doi.org/10.1007/JHEP03%282014%29113}{J. High Energ. Phys. {\bf 2014}, 113 (2014)} [arXiv:1401.0568].

\bibitem{chowdhury2020}A. Chowdhury and N. Banerjee, {\em Greybody Factor and Sparsity of Hawking Radiation from a Charged Spherical Black Hole with Scalar Hair}, \href{https://doi.org/10.1016/j.physletb.2020.135417}{Phys. Lett. B {\bf 805}, 135417 (2020)}.

\bibitem{hod2016}S. Hod, {\em The Hawking Cascades of Gravitons from Higher-Dimensional Schwarzschild Black Holes}, \href{https://doi.org/10.1016/j.physletb.2016.03.002}{Phys. Lett. B {\bf 756}, 133 (2016)} [arXiv:1605.08440].

\bibitem{hod2015}S. Hod, {\em The Hawking Evaporation Process of Rapidly-Rotating Black Holes: An Almost Continuous Cascade of Gravitons}, \href{https://doi.org/10.1140/epjc/s10052-015-3554-y}{Eur. Phys. J. C {\bf 75}, 329 (2015)} [arXiv:1506.05457].

\bibitem{Majhi2011}B. R. Majhi and E. C. Vagenas, {\em Black hole spectroscopy via adiabatic invariance}, \href{
https://doi.org/10.1016/j.physletb.2011.06.025}{Phys. Lett.  B {\bf 701}, 623 (2011)} [arXiv:1106.2292].

\bibitem{qqjiang2012}Q.-Q. Jiang and Y. Han, {\em On black hole spectroscopy via adiabatic invariance}, \href{
https://doi.org/10.1016/j.physletb.2012.10.031}{Phys. Lett.  B {\bf 718}, 584 (2012)} [arXiv:1210.4002].

\bibitem{Shahjalal2019}M. Shahjalal, {\em Area and entropy quantization of quantum-corrected Schwarzschild black hole surrounded by quintessence}, \href{https://doi.org/10.1142/S0217751X1950091X}{Int. J. Mod. Phys. A {\bf 34}, 1950091 (2019)}.

\bibitem{umetsu2010}K. Umetsu, {\em Hawking radiation from Kerr–Newman black hole and tunneling mechanism}, \href{https://doi.org/10.1142/S0217751X10050251}{Int. J. Mod. Phys. A {\bf 25}, 4123 (2010)} [arXiv:0907.1420 [hep-th]].

\bibitem{gibbons1978}G. W. Gibbons and M. J. Perry, {\em Black holes and thermal Green functions}, \href{https://www.jstor.org/stable/79482}{Proc. R. Soc. {\bf 358}, 467 (1978)}.




\end{thebibliography}
\end{document}